\documentclass[fleqn,usenatbib]{mnras}

\usepackage[T1]{fontenc}
\usepackage{graphicx}	% Including figure files
\usepackage{amsmath}	% Advanced maths commands
\usepackage{amssymb}	% Extra maths symbols
\usepackage{rotating}
\usepackage{textalpha}
\usepackage{pdflscape}
\usepackage{soul}

\title[Age calibration from star clusters]{
Open cluster age calibration from colour-magnitude morphological indices using  {\textit{Gaia}} DR3 data}

\author[F. A. Ferreira et al.]{
F. A. Ferreira$^{1}$\thanks{E-mail: filipe1906@ufmg.br},
J. F. C. Santos Jr.$^{1}$,
W. J. B. Corradi$^{1,4}$,
M. S. Angelo$^{3}$, \newauthor
\ and F. F. S. Maia$^{2}$
\\
$^{1}$Universidade Federal de Minas Gerais, Departamento de F\'isica, Av. Ant\^onio Carlos 6627, 31270-901, Brazil \\
$^{2}$Universidade Federal do Rio de Janeiro, Instituto de F\'isica, 21941-972, Brazil 
\\
$^{3}$Centro Federal de Educa\c c\~ao Tecnol\'ogica de Minas Gerais, Av. Monsenhor Luiz de Gonzaga, 103, 37250-000, Brazil\\
$^{4}$Laborat\'orio Nacional de Astrof{\'{\i}}sica, R. Estados Unidos, 154, 37504-364, Itajub\'a, MG, Brazil
}

\date{Accepted XXX. Received YYY; in original form ZZZ}

\pubyear{2025}

\begin{document}
\label{firstpage}
\pagerange{\pageref{firstpage}--\pageref{lastpage}}
\maketitle

% Abstract of the paper
\begin{abstract}
Star clusters are crucial for understanding how stars evolve. Their colour-
magnitude diagrams show the effects of stellar evolution of approximately coeval objects with the same
chemical composition. Furthermore, the determination of their astrophysical parameters
(age, distance, colour excess and metallicity) together with their spatial distribution provides information about the structure and the evolution of
the Galaxy itself. Using data from the \textit{Gaia} DR3 and 2MASS catalogues, we develop methodologies for characterizing open clusters. Precise membership lists, mean astrometric parameters and radii are obtained. Using photometric data from both data sources, we carried out new age calibrations that rely on morphological indices based on colour ($\Delta BR$) and magnitude ($\Delta G$)
differences between the red clump and the turnoff for a sample of 34 open clusters with ages covering the interval $8.3 < \log[t({\rm yr})] < 9.9$. A set of age calibration functions based on \textit{Gaia} morphological age indices are determined for the first time. We demonstrate their accuracy, obtaining a mean residual of 0.06 dex in $\log[t(yr)]$. Our results also show that stellar evolution models tend to predict the difference $\Delta G$. However, they typically overestimate the difference $\Delta BR$ for objects younger than $\log[t({\rm yr})] = 8.8$.

\end{abstract}

% Select between one and six entries from the list of approved keywords.
% Don't make up new ones.

%%%%%%%%%%%%%%%%%%%%%%%%%%%%%%%%%%%%%%%%%%%%%%%%%%

%%%%%%%%%%%%%%%%% BODY OF PAPER %%%%%%%%%%%%%%%%%%

\section{Introduction}
\label{sect:intro}

Open clusters (OCs) are important objects to study the history and structure of the Galactic disc. Young OCs reveal how stars form in embedded environments as well as the recent disc history \citep{Lada:2003}. On the other hand, the older OCs are the fingerprints of the chemical and dynamical evolution of the Galactic disc (\citealp[e.g.][]{friel1995,2007A&A...476..217C}; \citealp{2016A&A...585A.150N} [hereafter N2016]; \citealp{2020A&A...640A...1C} [hereafter C2020]).

%Open clusters (OCs) are important objects to study the history and structure of the Galactic disc. Young OCs reveal how stars form in embedded environments as well as the recent disc history \citep{Lada:2003}. On the other hand, the older OCs are the fingerprints of the chemical and dynamical evolution of the Galactic disc (\citealp[e.g.][]{friel1995,2007A&A...476..217C}; \cite[][hereafter N2016]{2016A&A...585A.150N}; \citealp{2020A&A...640A...1C}).

The Galactic OCs present a vast range of ages ($\log[t({\rm yr})] \lesssim 10.0$) and typically solar metallicities ($-0.5 \lesssim [Fe/H] \lesssim 0.5$) (\citealt{kharchenko2013}; N2016; \citealt{10.1093/mnras/stab770}[hereafter D2021]). Their colour-magnitude diagrams (CMDs) contain the fiducial positions of evolving stars of different masses according to their evolutionary stages at a given age, allowing us to study in detail the stellar evolution.

%therefore their CMDs are import laboratories of stellar evolution due presenting empirically the fiducial positions of evolving stars of different masses since their birth to their final evolutionary stages.

The separation between fiducial OCs members from field stars, especially in high stellar density environments, like the Galactic disc, is a challenging task. However, due to the high precision of the astrometric and photometric data provided by the most recent releases of the Gaia catalogue (DR2: \citealp{Lindegren:2018,Evans:2018}; EDR3: \citealp{2021A&A...649A...1G}, DR3: \citealp{2023A&A...674A...1G}) it has been possible to disentangle both cluster and field populations, allowing a drastic increase of works providing precise memberlists of OCs (\citealp[]{cjv18,2019A&A...624A...8A,2021MNRAS.500.4338A};D2021). Combined with efficient methodologies, a number of new objects has been discovered (\citealp[]{2019MNRAS.483.5508F,sla19,10.1093/mnras/staa1684,cjl20,2021MNRAS.502L..90F,2022arXiv220908504H}; \citealp{2023A&A...673A.114H} [hereafter H2023].

%(DR2: Gaia Collaboration et al. 2018; EDR3: Gaia Collaboration et al. 2021; DR3, Gaia Collaboration et al. 2022) 
%\textit{Gaia} DR2 catalogue \citep{Lindegren:2018,Evans:2018}
%\cite{2021A&A...649A...1G} GAIA EDR3
%\cite{2023A&A...674A...1G} GAIA DR3

%The CMD is the most important tool to obtain astrophysical parameters from OCs, so that an observational HR diagram can be constructed by correcting a CMD from extinction, distance modulus and colour excess. Such diagrams of intermediate age and old stellar populations present two main structures: the main sequence (MS), constituted by objects at the Hidrogen burning phase and the Red Giant Branch (RGB), for evolved stars.

%By taking advantage of the resulting cleaner and decontaminated CMDs, their stellar populations can be studied. An important feature, known as the Red Clump (RC), characteristic of intermediate age and older stellar populatoons passing through the evolutionary stage of Helium burning in the core, can be easily noticed. The RC is a concentration of cold and luminous stars in the Red Giant Branch (RGB) phase in the CMD. Although the RC is not often observed in the CMD of very young clusters, because the He-burning phase is rapid in the more massive stars, is has been reported for clusters within the age interval of .

%In this sense, the RC has been widely used to determine distances of vcluster within the milky way and the neighring gales as weel a age indicator
By taking advantage of the resulting decontaminated CMDs from such precise member star lists, the stellar populations from OCs can be better studied. An important feature, known as the Red Clump (RC), characteristic of intermediate age and older stellar populations whose more massive stars are passing through the evolutionary stage of Helium burning in the core, can be easily noticed (see Figs 2, 3 and 10 from \citealp{2018A&A...616A..10G}). The RC is a concentration of cold and luminous stars which occurs after the Red Giant Branch (RGB) phase in the CMD, and is the longest subsequent evolutionary stage in the life of a star, after the Main Sequence (MS). The RC is not often observed in the CMD of very young clusters, because the He-burning phase is rapid in the more massive stars, thereby it has been reported for clusters within the age interval of $8.3 \lesssim \log[t({\rm yr})] \lesssim 10.0$  \citep{2002AJ....123.1603G,2007A&A...463..559V,2019MNRAS.486.5600O}.

%After the Main Sequence (MS) phase, the longest subsequent evolutionary stage in the life of a star is the Helium burning phase in the core. Observationally, in the HR diagram, this evolutionary phase is characterized by a concentration of cold and luminous stars in the Red Giant Branch  (RGB) phase, which feature is often referred as the Red Clump (RC). It is easy to notice the presence of a RC in observational HR diagrams of stellar populations and CMDs of evolved star clusters \citep{2018A&A...616A..10G}. But this feature is not often observed in very young star clusters, because the Helium burning phase is rapid in more massive stars, which form less frequently: althoughg the RC has already been observed in Galactic OCs CMDs within the age interval of $8.3 \lesssim \log[t({\rm yr})] \lesssim 10.0$  \citep{2002AJ....123.1603G,2007A&A...463..559V,2019MNRAS.486.5600O}.
%On the other hand, it is easy to notice the presence of a RC in observational HR diagrams of stellar populations and CMDs of evolved star clusters \citep{2018A&A...616A..10G}. 

The RC is widely used as a standard candle, because although the colours of stars in this evolutionary phase depend strongly on age and metallicity, for low-mass stars ($M\lesssim 1.9 M_{\odot}$) the absolute magnitude do not  show large variations, especially in the infrared. In this way, by establishing a reliable RC mean magnitude value for old populations ($\log[t({\rm yr})] \gtrsim 9.0$), we are capable to determine distances to structures within our Galaxy and to neighboring galaxies \citep{2002AJ....123.1603G,2007A&A...463..559V,2013Ap&SS.344..417B,2016ARA&A..54...95G,2019MNRAS.486.5600O}. % (Grocholski \& Sarajedini 2002, van Helshoecht \& Groenewegen 2007, Bilir et al. 2013, Girardi 2016, Onozato et al. 2019.

For intermediate age and older clusters $8.3 \lesssim \log[t({\rm yr})] \lesssim 10.0$, the average position of the RC stars in the CMD can also be used as an age indicator. Indices based on CMD morphology are common in the literature. For example, the difference in magnitude $\Delta V$ between the MS turnoff (bluest point on the MS) and the mean magnitude of the RC tend to increase with age for star populations older than 1 Gyr. On the other hand the difference in colour $\Delta (B-V)$ of the same regions tend to decrease with age \citep{1985ApJ...291..595A}. A morphological age index called MAR is also defined by \cite{1985ApJ...291..595A}, taking the ratio $\Delta V$/$\Delta (B-V)$.

 \cite{1994AJ....107.1079P} defined the morphological age indices $\delta V $ and $\delta 1$. The index $\delta V$ is the difference between the inflection point of the main sequence or the base of the giant branch and the mean magnitude value of the RC, while $\delta 1$ is the colour difference between the point one magnitude brighter than the turnoff and the base of the giant branch. The index $\delta V $ established in \cite{1994AJ....107.1079P} was later calibrated with the ages and metallicities of star clusters in \cite{2004A&A...414..163S}.

Such indices based on the CMD morphology are common in the literature and are well-known age indicators for Galactic clusters \citep{1985ApJ...291..595A,1994AJ....107.1079P,2004A&A...414..163S,
10.1111/j.1365-2966.2009.16106.x,2009A&A...508.1279B,2015NewA...34..195O} and Magellanic Clouds clusters \citep{1997AJ....114.1920G,2014AJ....147...71P}. Those age determination techniques are independent of distance and reddening, but it is necessary the consistent identification of those key evolutionary features in the CMDs.
%(Anthony-Twarog \& Twarog 1985 , Phelps et al. 1994, Salaris et al. 2004, Piatti et al. 2010, Beletsky et al. 2009, Oralhan et al. 2015)
%(Geisler et al. 1997, Parisi et al. 2014)
 
In this work, we used data from the 2MASS \citep{Skrutskie:2006} and \textit{Gaia} DR3 catalogues to establish morphological age indices for a set of Galactic OCs, extending the morphological age indices for OCs as young as $\log[t({\rm yr})]=8.3$. We adopted the magnitude difference between the turnoff point and the RC, an index widely employed as an age indicator, both in the visible ($\Delta V$, \citealt{1994A&A...287..761C}) and in the infrared ($\Delta K$, \citealt{2009A&A...508.1279B,2013AJ....146...64Z}). We also present age calibrations with those indices, using the 2MASS index $\Delta K$ as a benchmark.

This paper is structured as follows. In Section \ref{sect:data} the data is presented. In Section \ref{sect:sample} the OC sample selection is described. In Section \ref{sect:method} the analysis procedures are discussed, including membership assessment and determination of mean astrometric parameters. The CMDs properties of our OC sample are explored in Section \ref{sect:CMD}. The main results are presented in Section \ref{sect:result} and the concluding remarks are given in Section \ref{sect:concl}.

\section{Data}
\label{sect:data}
\subsection{\textit{Gaia}}

%\textit{Gaia} DR2 catalogue \citep{Lindegren:2018,Evans:2018}

%\cite{2021A&A...649A...1G} GAIA EDR3

%\cite{2023A&A...674A...1G} GAIA DR3

%Cantat-Gaudin et al. (2018)

The \textit{Gaia DR3} catalogue provides positions, proper motions in right ascension and declination, parallaxes and photometry ($G$, $G_{BP}$, and $G_{RP}$ passbands) for nearly two billion sources. The Gaia@AIP (https://gaia.aip.de/) online services have been used to extract \textit{Gaia} DR3 data for each OC in circular regions centred on the coordinates presented in D2021 catalogue. For a first guess, we adopted 
an extraction radius of 1, 2 and 3 degrees for OCs of heliocentric distances $d>2\,{\rm kpc}$, $1\,{\rm kpc} < d < 2\,{\rm kpc}$ and $d <  1\,{\rm kpc}$, respectively. Those sizes were large enough to restrict the studied OCs and an adjacent comparison star field. Confirmed OCs in D2021 present mean tidal radius ($r_{t}$) of $9.85\,{\rm pc}$ with standard deviation of $5.21\,{\rm pc}$ \citep{kharchenko2013}. Assuming a typical superior limit in OC size of $r_{t} \sim 15\,{\rm pc}$, an OC would appear to have apparent radii of $\sim 26$ arcmin at $2\,{\rm kpc}$, $\sim 52$ arcmin at $1\,{\rm kpc}$ and $\sim 104$ arcmin at $500\,{\rm pc}$.
This means that those extracting radii are capable to encompass a typical OC at those ranges of distances.

%The paper Gaia Early Data Release 3: Parallax bias vs. magnitude, colour, and position by Lindegren et al. (2021) provides a tentative recipe which allows one to correct the parallax of a given Gaia (E)DR3 source for the zero point bias. The correction is provided separately for 5- and 6-parameter astrometric solutions and is given as a function of source magnitude, colour, and celestial position. This correction, which applies equally to Gaia EDR3 and to Gaia DR3 astrometry, is implemented as a Python code which can be found here: 

%https://gitlab.com/icc-ub/public/gaiadr3_zeropoint

To extract the data with the corrected version of the photometric flux excess factor \citep{2021A&A...649A...3R}, we adopted the query examples presented in Gaia EDR3 documentation (Appendix B, \citeauthor{2021A&A...649A...1G} \citeyear{2021A&A...649A...1G}). We also adopted a filter to keep only stars with $G < 19$ to avoid less informative sources,
which is also the nominal magnitude limit to ensure astrometric and photometric completeness \citep{2021A&A...649A...2L}.

In order to correct the original data for the parallax zero-point bias, we applied the available recipe presented in \cite{2021A&A...649A...2L}, which is applied equally to Gaia EDR3 and to Gaia DR3 astrometry. This correction is provided separately for sources with available 5- and 6-parameter astrometric solutions and is given as a function of the source magnitude, colour, and celestial position. No further corrections in flux or G-band magnitudes have been applied, since they are already
implemented in DR3.% \citep{2023A&A...674A...1G}.
%https://gitlab.com/icc-ub/public/gaiadr3_zeropoint

We applied quality filters limiting our database to remove spurious astrometric and photometric solutions by keeping sources consistent with the two following equations:

\begin{equation}
|C^{*}|<5 \sigma_{C^{*}}
\end{equation}
\begin{equation}
 RUWE<1.4,
\end{equation}

\noindent where $C^{*}$ is the corrected value of the BP and RP flux excess factor, $\sigma_{C^{*}}$ is given by equation 18 of \cite{2021A&A...649A...3R} and RUWE is the renormalised unit weight error \citep{2021A&A...649A...2L}.

%Following the prescriptions informed in the Gaia documentation 3 , we applied the available recipes 4 accompanying the third data release in order to correct the original data for parallax zero-point (Lindegren et al. 2021) and to provide a corrected version of the photometric flux excess factor (E(BP/RP ); Riello et al. 2021). No corrections in flux or G-band magnitudes have been applied, since they are already
%implemented in DR3 (Gaia Collaboration et al. 2022).

%a minimum of 5 transits for a source (matched transits > 5) and 9 visibility periods (visibility periods used > 9), thus allowing astrometric and photometric completeness (Fabricius et al. 2021).

\subsection{2MASS}
 
 The Two Micron All Sky Survey (2MASS) provides positions and near-infrared photometry ($J$, $H$, and $K$ passbands) for nearly 470 million sources, covering the entire sky. For each target, we have used the Vizier service to extract data from the 2MASS catalogue for stars inside the circular regions quoted above. To ensure quality to our sample, we only extracted stars with 2MASS JHKs photometric quality flag $'AAA'$. We then crossmatched \textit{Gaia} catalogue with 2MASS by selecting 2MASS sources within 1 arcsec from \textit{Gaia} sources. The photometric depth of our catalogues is governed by \textit{Gaia} photometry, in other words, if a star from \textit{Gaia} do not have a counterpart in 2MASS survey, this star is not excluded.

\section{The OC sample}
\label{sect:sample}

It is beyond the scope of this work to determine the astrophysical parameters of the OCs (age, distance, color excess, and metallicity). Therefore, we decided to use the OC catalogue published by N2016, which contains a homogenized sample of 172 OCs with averaged ages and metallicities, based on [Fe/H] abundances from numerous individual studies, as well as their associated uncertainties. To select our OC sample, we restricted their catalogue to objects closer than 3 kpc from the Sun according to D2021 and for which metallicities were taken from high resolution spectroscopy. For the remaining sample, we kept OCs for which we could visually identify in the CMD a concentration of stars around the RC position and for which the turnoff points are brighter than $G = 19$, resulting in a final sample of 34 OCs, covering $8.3<logt[yr]<9.9$ and $-0.44<[Fe/H]<0.37$.

The catalogue presented in N2016 is widely adopted, especially as a reference for metallicity \citep{2019A&A...623A.108B,2020MNRAS.495.2673C,2021ApJ...919...52Z,2023AJ....165...53I} and recent works still adopt their age values \citep{2020MNRAS.495.2673C}. To ensure the quality of their data as age references for the range of ages and distances explored in this work, we compared the selected sample of OCs with recent catalogs based on \textit{Gaia} data (C2020; D2021; H2023). The comparison between the age values reported by these catalogs is presented in Table \ref{Tab:logt_compara_literatura}. The age values in N2016 do not show significant offsets from recent literature values, and their comparison with recent catalogs exhibits similar correlations and residuals, as observed when catalogs based on \textit{Gaia} photometry are compared with each other. The only discrepant age value found was for the OC NGC2354, whose age value was not included in the comparisons. A discussion concerning this particular OC is presented in Appendix \ref{sect:ngc_2354}.

%To select our OC sample, We restricted their sample to objects closer than 3 kpc from the Sun according to D2021 and for which metallicities were taken from high resolution spectroscopy. For the remaining sample, we kept OCs for which we could visually identify in the CMD a concentration of stars around the RC position and for which the turnoff points are brighter than $G = 19$, resulting in a final sample of 34 OCs, covering $8.3<logt[yr]<9.9$ and $-0.44<[Fe/H]<0.37$. 

We have also compiled distances and colour excesses from D2021. Positions, ages, metallicities, distances and colour excesses of our sample are presented in Table \ref{Tab:clusters_prop2}.

 \begin{table}
\caption{Comparision between the adopted age values from N2016 with recent catalogues based on \textit{Gaia} data. The parameter $\Delta log[t]$ represents the average difference between the first source and the second source. The mean residuals and the correlation between them are also reported.}
  \small
\begin{tabular}{|l|r|r|r|r|}
\hline
   \multicolumn{1}{|c|}{Sources} &
   \multicolumn{1}{c|}{$\Delta log[t]$} &
  \multicolumn{1}{c|}{Mean residuals}& 
      \multicolumn{1}{c|}{Correlation}      
      \\
      \hline
   \multicolumn{1}{|c|}{} &
   \multicolumn{1}{c|}{$dex$} &
  \multicolumn{1}{c|}{$dex$}& 
      \multicolumn{1}{c|}{}      
      \\
   %\multicolumn{1}{c|}{$Method$} \\
\hline
   N2016 x C2020  & $-0.05  $ & $0.10  $ & $0.95 $ \\
    \hline
    N2016 x D2021 &$-0.07  $ & $0.09  $ & $0.97 $ \\
      \hline
      N2016 x H2023 &$ 0.03$ & $0.17  $  & $0.82  $   \\
  \hline
  H2023 x C2020 & $ -0.07$ & $0.15  $  & $0.79  $   \\
    \hline
    H2023x D2021 &$ -0.1$ & $0.13  $  & $0.84  $   \\
     \hline
    C2020 x D2021&$ -0.02$ & $0.07  $  & $0.97  $   \\
  \hline
\end{tabular}
   \label{Tab:logt_compara_literatura}
\end{table}

\begin{table*}
\small
%\centering
\caption{Properties of the investigated clusters from the literature.}
\label{Tab:clusters_prop2}
\begin{tabular}{|l|r|r|r|r|r|r|r|r|r|r|}
\hline
  \multicolumn{1}{|c|}{OC} &
  \multicolumn{1}{c|}{$\alpha_{J2000}$} &
  \multicolumn{1}{c|}{$\delta_{J2000}$} &
  \multicolumn{1}{c|}{$\log[t({\rm yr})]$} &
  \multicolumn{1}{c|}{$\sigma_{\log[t({\rm yr})]}$} &
  \multicolumn{1}{c|}{$[Fe/H]$} &
  \multicolumn{1}{c|}{$\sigma_{[Fe/H]}$} &
  \multicolumn{1}{c|}{$d$} &
  \multicolumn{1}{c|}{$\sigma_{d}$} &
  \multicolumn{1}{c|}{$E(B-V)$} &
  \multicolumn{1}{c|}{$\sigma_{E(B-V)}$} \\
    \multicolumn{1}{|c|}{} &
  \multicolumn{1}{c|}{degrees} &
  \multicolumn{1}{c|}{degrees} &
  \multicolumn{1}{c|}{dex} &
  \multicolumn{1}{c|}{dex} &
  \multicolumn{1}{c|}{dex} &
  \multicolumn{1}{c|}{dex} &
  \multicolumn{1}{c|}{pc} &
  \multicolumn{1}{c|}{pc} &
  \multicolumn{1}{c|}{mag} &
  \multicolumn{1}{c|}{mag} \\
\hline
  NGC 188 & 11.749 & 85.243 & 9.80 & 0.17 & 0.11 & 0.04 & 1859 & 36 & 0.075 & 0.008\\
  NGC 752 & 29.120 & 37.760 & 9.23 & 0.18 & -0.03 & 0.06 & 441 & 4 & 0.051 & 0.024\\
  NGC 1245 & 48.699 & 47.253 & 9.03 & 0.09 & 0.02 & 0.03 & 2763 & 97 & 0.268 & 0.010\\
  NGC 1817 & 78.153 & 16.695 & 8.91 & 0.17 & -0.11 & 0.03 & 1582 & 40 & 0.233 & 0.012\\
  NGC 2099 & 88.048 & 32.568 & 8.56 & 0.25 & 0.02 & 0.05 & 1299 & 22 & 0.297 & 0.016\\
  Trumpler 5 & 99.107 & 9.454 & 9.64 & 0.21 & -0.44 & 0.07 & 3260 & 122 & 0.652 & 0.007\\
  Collinder 110 & 99.681 & 2.100 & 9.09 & 0.24 & 0.03 & 0.02 & 1991 & 80 & 0.530 & 0.021\\
  NGC 2354 & 108.520 & -25.725 & 9.14$^{*}$& 0.06 $^{*}$ & -0.18 & 0.02 & 1258 & 42 & 0.171 & 0.019\\
  NGC 2355 & 109.269 & 13.766 & 8.91 & 0.08 & -0.05 & 0.08 & 1794 & 41 & 0.104 & 0.006\\
  NGC 2360 & 109.447 & -15.623 & 9.04 & 0.27 & -0.03 & 0.06 & 1054 & 21 & 0.132 & 0.020\\
  NGC 2423 & 114.310 & -13.880 & 8.90 & 0.19 & 0.08 & 0.05 & 924 & 14 & 0.095 & 0.021\\
  NGC 2420 & 114.603 & 21.576 & 9.37 & 0.13 & -0.05 & 0.02 & 2435 & 46 & 0.038 & 0.005\\
  NGC 2447 & 116.150 & -23.872 & 8.60 & 0.17 & -0.05 & 0.01 & 1004 & 10 & 0.037 & 0.012\\
  NGC 2477 & 118.054 & -38.505 & 8.93 & 0.1 & 0.07 & 0.03 & 1351 & 47 & 0.384 & 0.025\\
  NGC 2527 & 121.280 & -28.150 & 8.81 & 0.12 & -0.10 & 0.04 & 630 & 8 & 0.075 & 0.018\\
  NGC 2539 & 122.665 & -12.830 & 8.70 & 0.11 & -0.02 & 0.08 & 1243 & 50 & 0.069 & 0.002\\
  NGC 2660 & 130.665 & -47.203 & 9.12 & 0.09 & 0.04 & 0.03 & 2642 & 105 & 0.470 & 0.048\\
  NGC 2682 & 132.822 & 11.839 & 9.54 & 0.15 & 0.03 & 0.05 & 865 & 18 & 0.041 & 0.013\\
  IC 2714 & 169.370 & -62.710 & 8.45 & 0.14 & 0.02 & 0.06 & 1229 & 16 & 0.393 & 0.013\\
  NGC 3960 & 177.650 & -55.684 & 8.99 & 0.10 & -0.04 & 0.10 & 2074 & 114 & 0.347 & 0.027\\
  NGC 4337 & 186.022 & -58.121 & 9.24 & 0.11 & 0.12 & 0.05 & 2416 & 99 & 0.418 & 0.021\\
  NGC 4349 & 186.070 & -61.875 & 8.52 & 0.27 & -0.07 & 0.06 & 1656 & 36 & 0.420 & 0.019\\
  Collinder 261 & 189.56 & -68.400 & 9.86 & 0.17 & 0.00 & 0.04 & 2806 & 119 & 0.322 & 0.035\\
  NGC 5822 & 225.900 & -54.300 & 8.95 & 0.11 & 0.08 & 0.08 & 796 & 27 & 0.155 & 0.019\\
  NGC 6134 & 246.950 & -49.150 & 8.96 & 0.16 & 0.11 & 0.07 & 1055 & 51 & 0.411 & 0.024\\
  NGC 6253 & 254.770 & -52.715 & 9.59 & 0.13 & 0.34 & 0.11 & 1674 & 72 & 0.256 & 0.019\\
  IC 4651 & 261.179 & -49.917 & 9.27 & 0.18 & 0.12 & 0.04 & 920 & 15 & 0.113 & 0.017\\
  NGC 6583 & 273.960 & -22.150 & 9.00 & 0.10 & 0.37 & 0.04 & 2162 & 132 & 0.592 & 0.041\\
  IC 4756 & 279.670 & 5.520 & 8.82 & 0.08 & 0.02 & 0.04 & 472 & 1 & 0.204 & 0.012\\
  NGC 6705 & 282.790 & -6.255 & 8.28 & 0.19 & 0.12 & 0.09 & 1888 & 65 & 0.470 & 0.016\\
  Ruprecht 147 & 289.100 & -16.350 & 9.37 & 0.04 & 0.16 & 0.08 & 305 & 0 & 0.094 & 0.016\\
  NGC 6811 & 294.350 & 46.390 & 8.86 & 0.11 & 0.03 & 0.01 & 1102 & 13 & 0.069 & 0.011\\
  NGC 6819 & 295.332 & 40.192 & 9.32 & 0.09 & 0.09 & 0.01 & 2444 & 54 & 0.157 & 0.016\\
  NGC 7789 & 359.330 & 56.720 & 9.18 & 0.06 & 0.05 & 0.07 & 1907 & 30 & 0.328 & 0.008\\
\hline
\end{tabular}
 \vspace{1ex}
 
     {\raggedright $*$ Age calculated based on recent literature parameters (see Sect \ref{sect:ngc_2354}). \par}
%$*$ Age calculated based on recent literature parameters (See \ref{section:apendiz_2354}).\\
\end{table*}

\begin{figure*}
\includegraphics[width=0.42\linewidth]{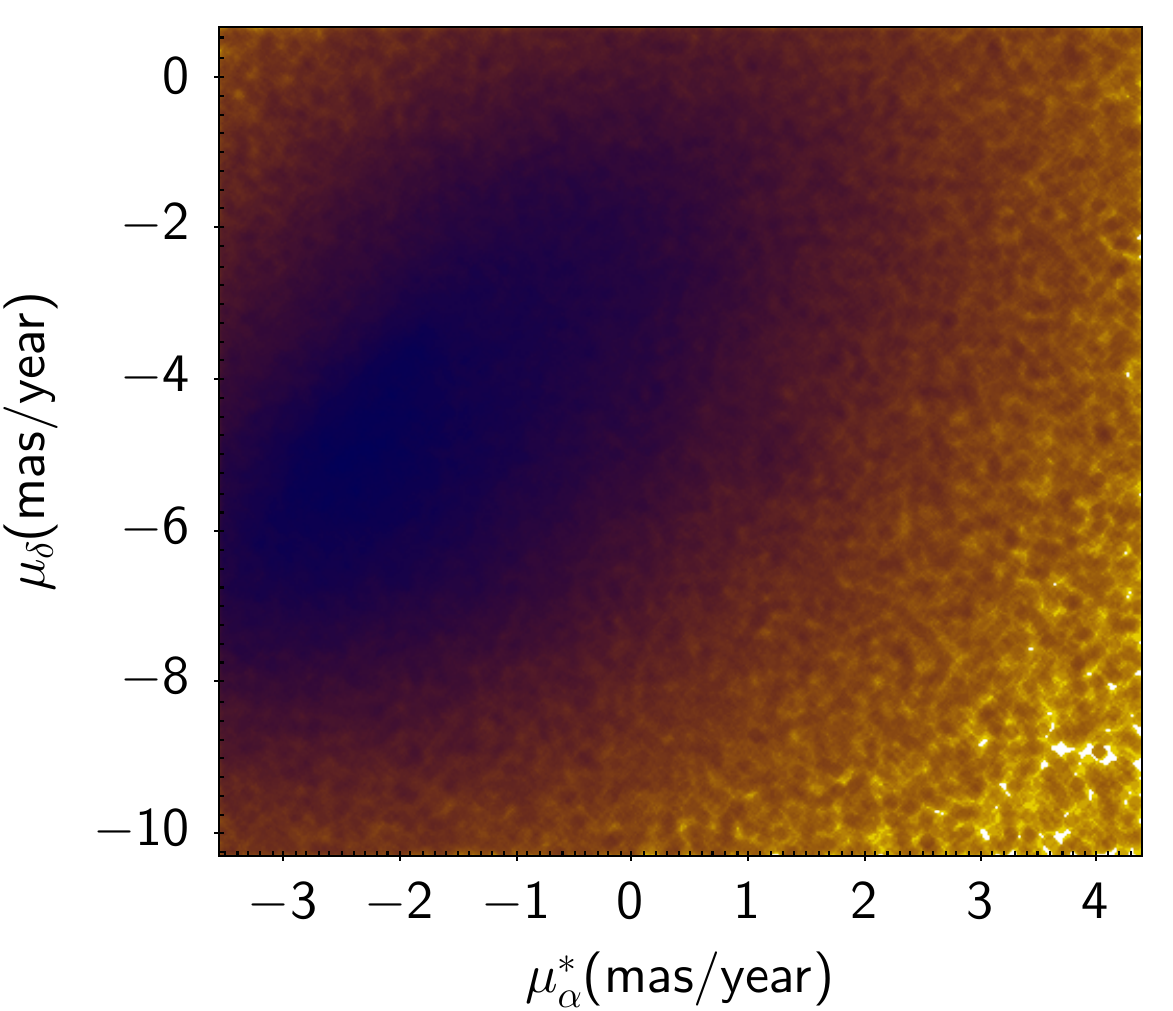}
\includegraphics[width=0.42\linewidth]{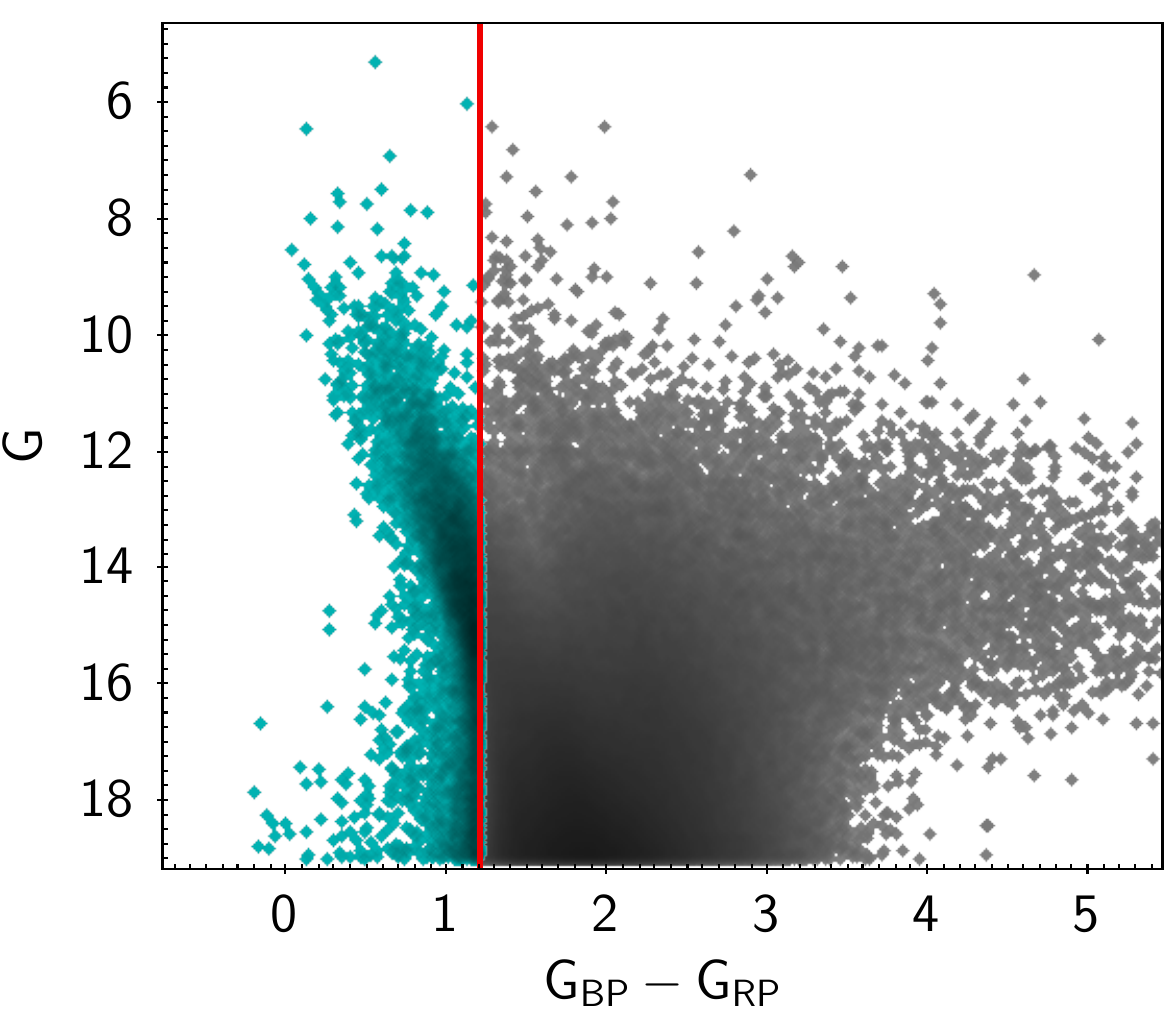}
\includegraphics[width=0.45\linewidth]{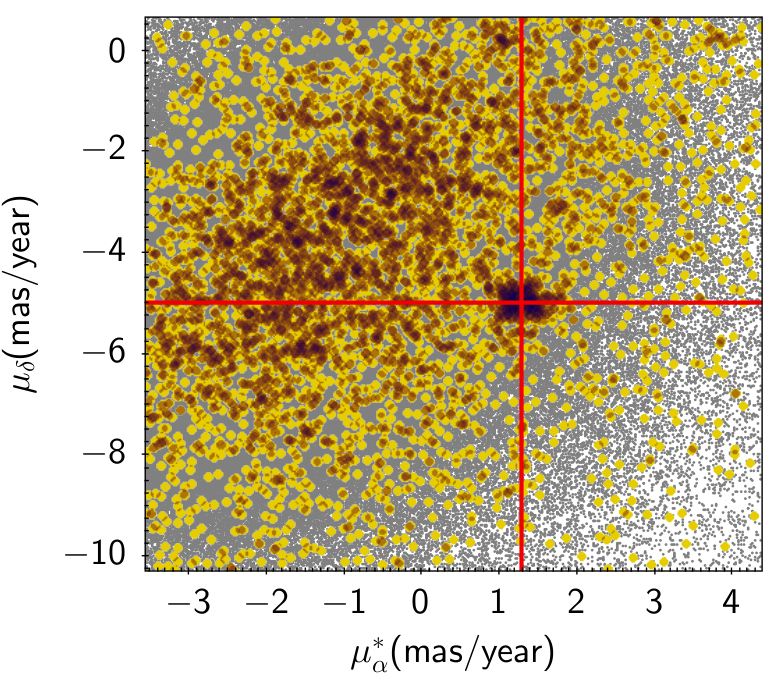}
\includegraphics[width=0.45\linewidth]{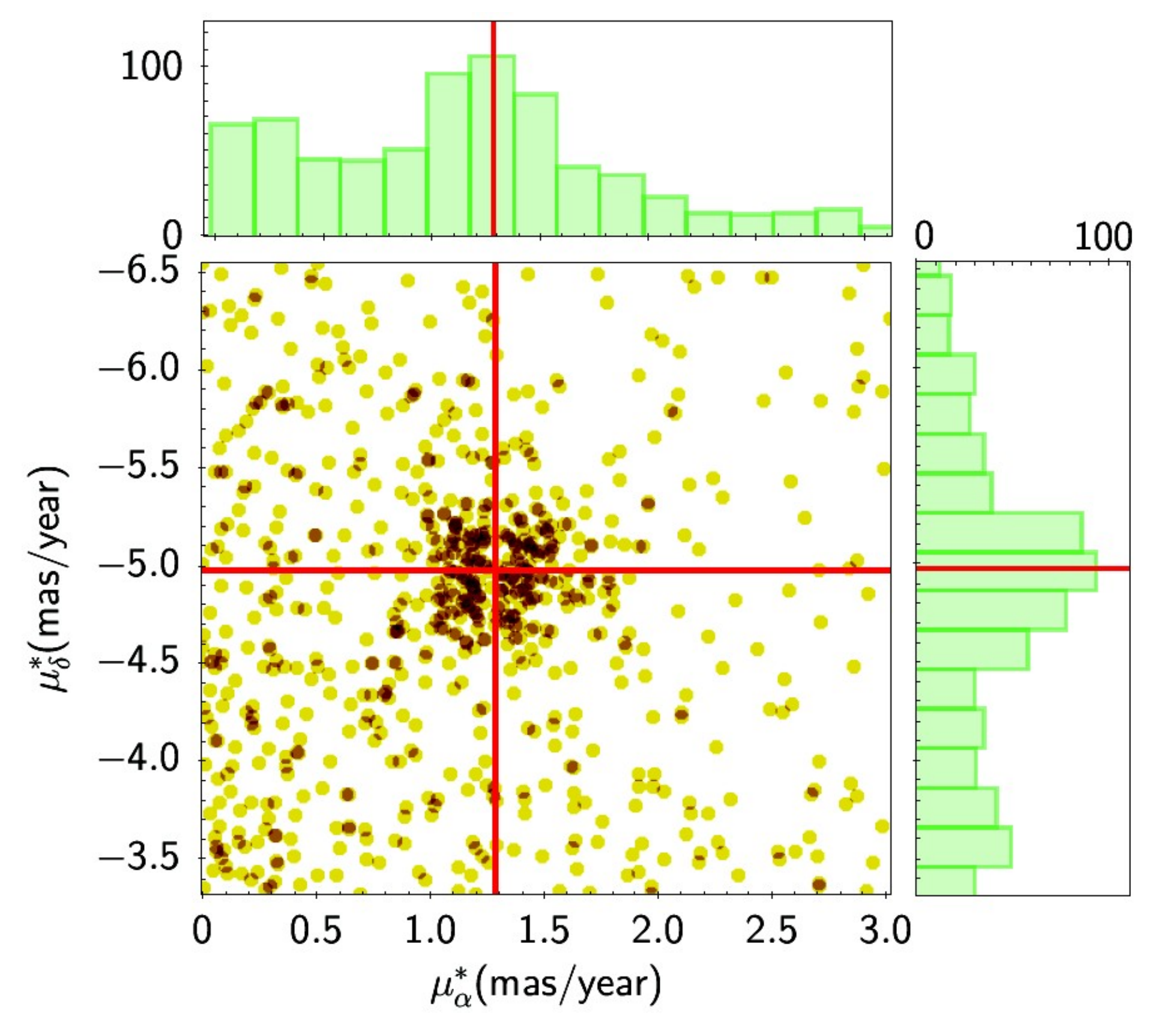}
\caption{Data inspection for the OC IC 4756. Top left: VPD of all data within 2 degrees radii from its centre, containing 629323 stars. Top right: CMD built from the entire sample (grey dots), where the colour filter is represented by the red line, selecting the cyan sample, which corresponds to 8402 stars. Bottom left: VPD with the sample filtered by colour (yellow and brown samples) plotted over the entire sample (grey dots), the red lines mark the proper motion peak. Bottom right: The same VPD with the colour filtered sample, but with the histograms represented where the proper motion peak values were determined.}
\label{fig:pmra_detect}
\end{figure*}

%\caption{Data inspection from OC IC 4756. Top left: VPD of all data within 2 degrees radii from its centre, containing 629323 stars. Top right: A CMD built from the entire sample (grey dots), where the colour filter is represented by the red line $G_{BP}-G_{RP}<1.20$, selecting the cyan sample, which corresponds to remaining 8402 stars. Bottom left: VPD with the sample filtered by colour (yellow sample) plotted over the entire sample (gray dots), the red lines mark the proper motion peak in $\mu_{\alpha}^{*}=1.28$ and $\mu_{\delta}=-4.98$. Bottom right: The same VPD with the colour filtered sample, but with the histograms represented where the proper motion peak values were calculated.}

\section{Methodology}
 \label{sect:method}
To assess memberships and remove the field population from the clusters sample, we have employed proper motion and parallax selections of members. For this purpose, we have developed a methodology based on \cite{2019MNRAS.483.5508F} that includes:

%the use of colour filters, the building of radial density profiles (RDPs) and the selection of members based on Gaussian fittings over the proper motion and parallax distributions.

\begin{enumerate}

%\item Data extraction centered on the object coordinates according to \cite{dias2002} and application of quality filters on the data;
\item Preliminary analysis of Vector Point Diagrams (VPDs) to determine the mode of the cluster's proper motion distribution for both components;
\item Construction of a proper motion mask around the determined proper motion mode, to create a subsample almost free of contamination from field stars;
%\item Application of the proper motion mask on the base sample;
   \item Determination of the cluster centre and radius by building radial density profiles (RDPs);
   \item Gaussian fittings over the proper motion components and parallaxes distributions and filters restricting stars based on such distributions.
 %   \item Restriction of stars with parallax values.
\end{enumerate}

\subsection{The clusters proper motion detection}
To find the OCs signatures in the VPD, we used the same method adopted in \citeauthor{2019MNRAS.483.5508F}(\citeyear{2019MNRAS.483.5508F}; \citeyear{10.1093/mnras/staa1684}; \citeyear{2021MNRAS.502L..90F}), where a colour filter is applied on the sample to discard very reddened field stars and maximize the contrast between cluster and field population.
We started with a colour threshold value $G_{BP}-G_{RP}<2.5$. For the cases of more distant and highly reddened clusters, we increased this threshold value and, for cases of less reddened ones, this value is decreased, in order to make the initial detection of the cluster evident as an overdensity in the VPD. We then computed the mode of the proper motions in right ascension ($\mu_{\alpha}^{*}$) and declination ($\mu_{\delta}$).

In Fig. \ref{fig:pmra_detect} we show how we identified the proper motion signature for the OC IC\,4756. The top left panel shows a VPD for all stars within 2 degrees from its centre, containing 629323 stars, showing that without any filter, we are not capable to find the cluster. In the top right panel, a colour filter is represented by the red line $G_{BP}-G_{RP}<1.20$, selecting the cyan sample, which corresponds to 8402 stars. In the bottom left panel, the VPD with the sample filtered by colour (yellow and brown samples) is plotted over the entire sample (gray dots), exhibiting an overdensity of stars corresponding to the cluster, where the red lines mark its proper motion modes at $\mu_{\alpha}^{*}=1.28$ \,{\rm mas\,yr$^{-1}$} and $\mu_{\delta}=-4.98$ \,{\rm mas\,yr$^{-1}$}. The same VPD with the colour filtered sample, including histograms where the proper motion modes are indicated, is shown in the bottom right panel.

%To find the OCs signatures in VPD, we use the same method adopted in \cite{2019MNRAS.483.5508F}, where a colour filter is applied to discart very reddened field stars and maximize the contrast between clusters and field population. Therefore, from our base sample, we performed a color cut, constructed VPDs and initially looked for overdensities in them. We compute the peak values of the proper motions in right assencion and declination in this procedure.

%Initially we started with a color threshold value $G_{BP}-G_{RP}<2.5$ that preserves the majority of stars in our clusters and, for the cases of more distant and severely contaminated clusters, we increased this threshold value, in order to make the initial detection of the cluster in the VPD. We compute the peak values of the proper motions in right assencion and declination in this procedure.

\subsection{The proper motion filter}

In order to estimate how the dispersion in proper motion components of real OCs behave (on average) as function of distance, we used the recent catalogue published by H2023, where the OCs members were obtained by using HDBSCAN algorithm over \textit{Gaia} DR3 data. We built a sample by limiting the OCs to distances between 40\,pc and 6\,kpc and number of members larger than 200. The catalogue used provides, for each OC, its position, astrophysical parameters, and the mean astrometric parameters and dispersions presented by its probable members. We then separated the OCs into intervals of 50\,pc and calculated the mean proper motion dispersion values for the entire group of OCs within each distance interval. In this procedure, we established a small catalog of distances and the expected OC proper motion dispersions at those distances.

\begin{figure*}
\includegraphics[width=0.48\linewidth]{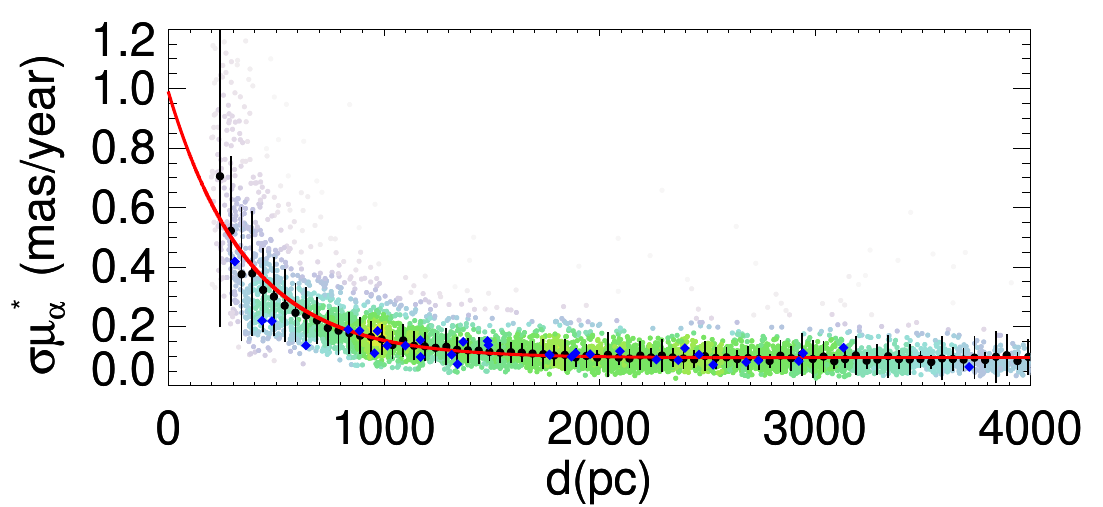}
\includegraphics[width=0.48\linewidth]{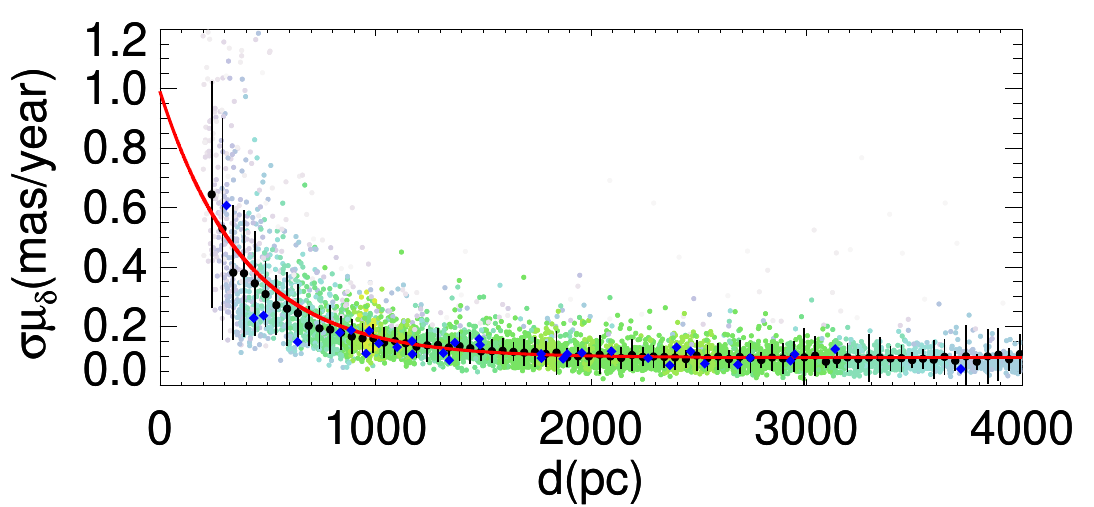}
\caption{Mean values of dispersion in $\mu_{\alpha}^{*}$ (top) and $\mu_{\delta}$ (bottom) for a set of 1229 OCs as function of the heliocentric distance. The red lines represent the best exponential fit, the coloured symbols represent the density of OCs and the black filled circles represent the local mean value of proper motion dispersion. Error bars represent the standard deviation in each bin and the blue filled dots represent our sample of OCs.}
\label{fig:pmra_mean_expo}
\end{figure*}

As shown in Fig. \ref{fig:pmra_mean_expo}, we note that the nearest clusters tend to exhibit higher dispersion values, due to the apparent random motion of the stars being greater than the proper motion uncertainties. On the other hand, clusters located at distances farther than 2\,kpc tend (on average) to exhibit an approximately fixed value, which shows that beyond this limit the physical dispersion of proper motions tends to be negligible in face of astrometric errors. In this way, we adopted an exponential fitting over these data to reproduce this behavior for both proper motion components (Fig. \ref{fig:pmra_mean_expo}).

%At this point we have defined our working sample for the radial profile procedure: best quality Gaia data within the square proper motion box and 300 from the objects centres

\indent This procedure was carried out to construct proper motion masks of sizes that fit the clusters proper motion spread for different distances. From the mean values $\sigma_{rep}$ of proper motion dispersion, we adopted boxes with sides equal to 20 times this value, that is, we limited the samples of stars within 10\,$\sigma_{rep}$ around the mode of the proper motion distributions for both components (Fig \ref{fig:pmra_box}). The adopted sizes of the proper motion masks were:

\small
\begin{equation}
 L_{pmra} = 20\times[0.895\times exp(-0.0027\times D_{cluster}) + 0.094] 
 \end{equation}
 \begin{equation}
 L_{pmde} = 20\times[0.810\times exp(-0.0025\times D_{cluster}) + 0.095]  
\end{equation}

\normalsize
 
\noindent where $D_{cluster}$ is the distance (in pc) according to D2021 and $L_{pmra}$ and $L_{pmde}$ represent the sizes of the mask in proper motion units in right ascension and declination, respectively. For subsequent analyses, we will abandon any colour filters, as they obviously exclude very cool low main sequence stars from nearby clusters and possible RGB stars for some older clusters. This procedure defines our database. We restricted the OCs proper motion space by employing the box-shaped filter centered on the modal values of $\mu_{\alpha}^{*}$ and $\mu_{\delta}$, as shown in the top panels of Fig. \ref{fig:pmra_box}. %In the following procedures, we showed a set of examples of how our methodology performed on the open cluster NGC7789.

\subsection{Centre and radial density profile}

 After establishing the size of the proper motion mask for our OCs sample, we applied it to the original database to reduce the contamination by field stars. We constructed frequency histograms with the equatorial coordinates of the stars and estimated the centre values of the clusters in both coordinates through Gaussian fittings (taking the average values of the fit). This average value is used as a first guess for the central coordinates to build the RDPs, however the final values are based on the best RDP constructed.

  \begin{figure*}
\includegraphics[width=0.32\linewidth]{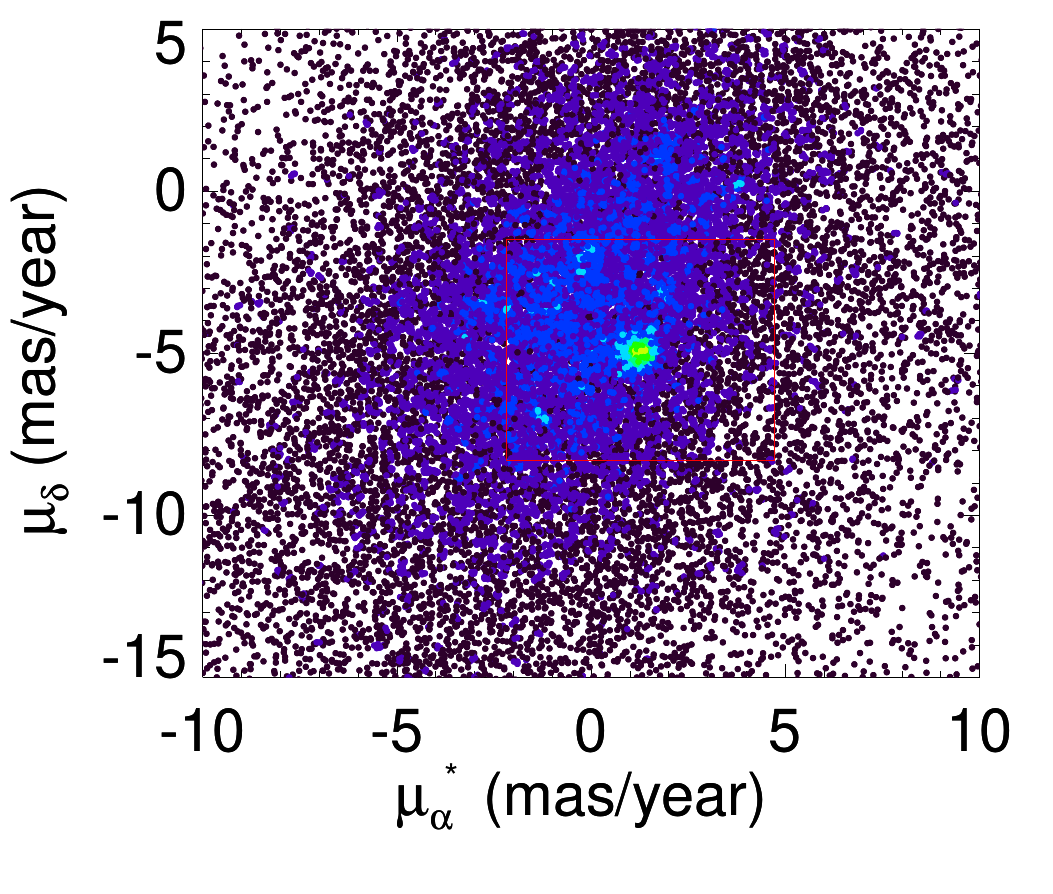}
\includegraphics[width=0.32\linewidth]{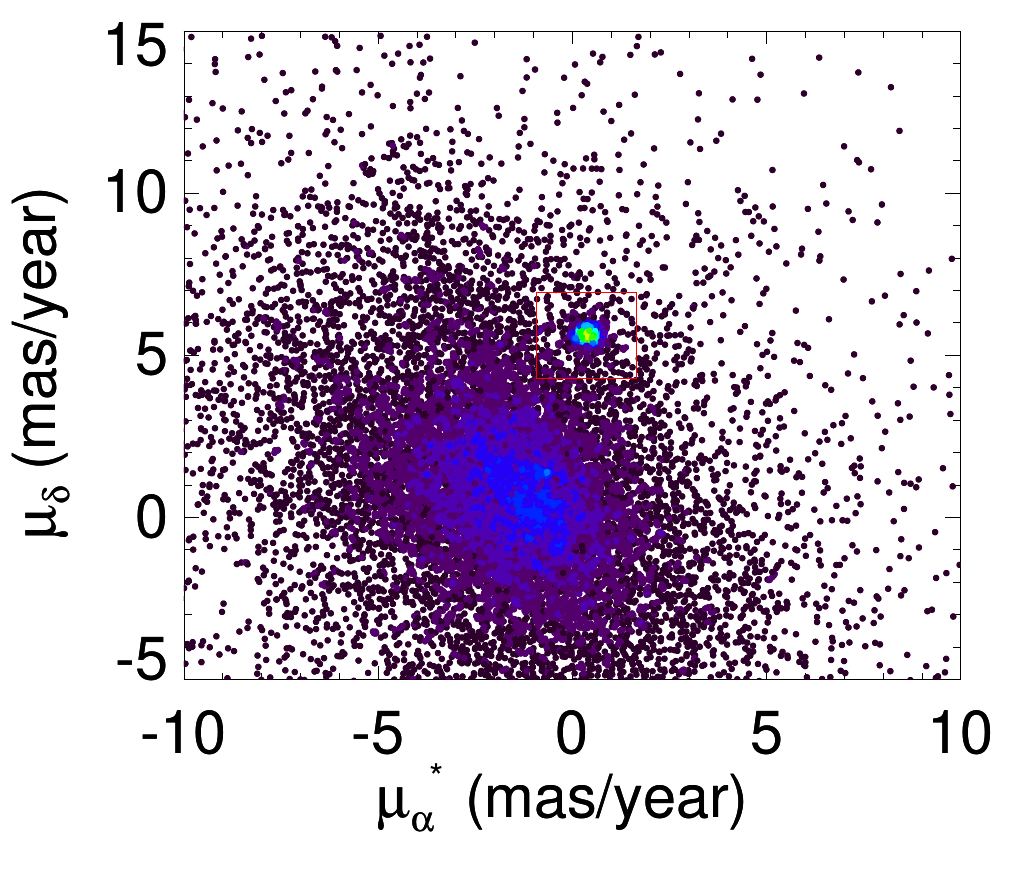}
\includegraphics[width=0.32\linewidth]{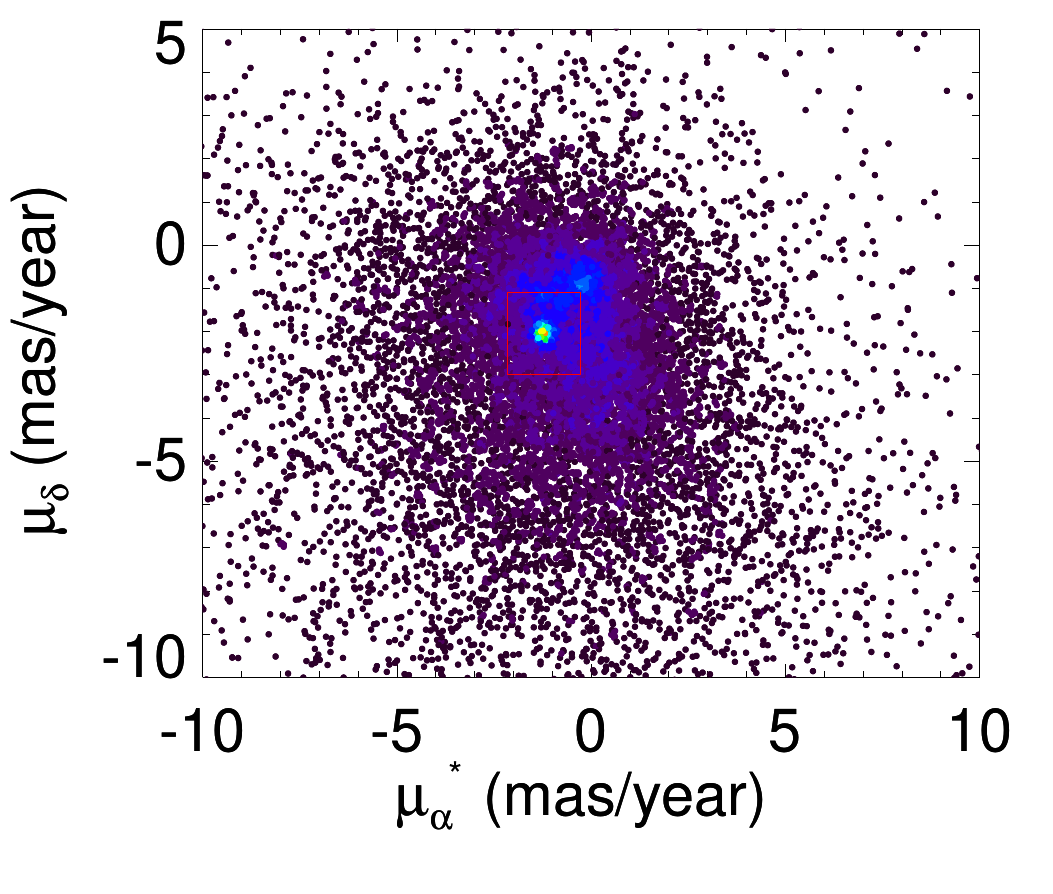}
\includegraphics[width=0.32\linewidth]{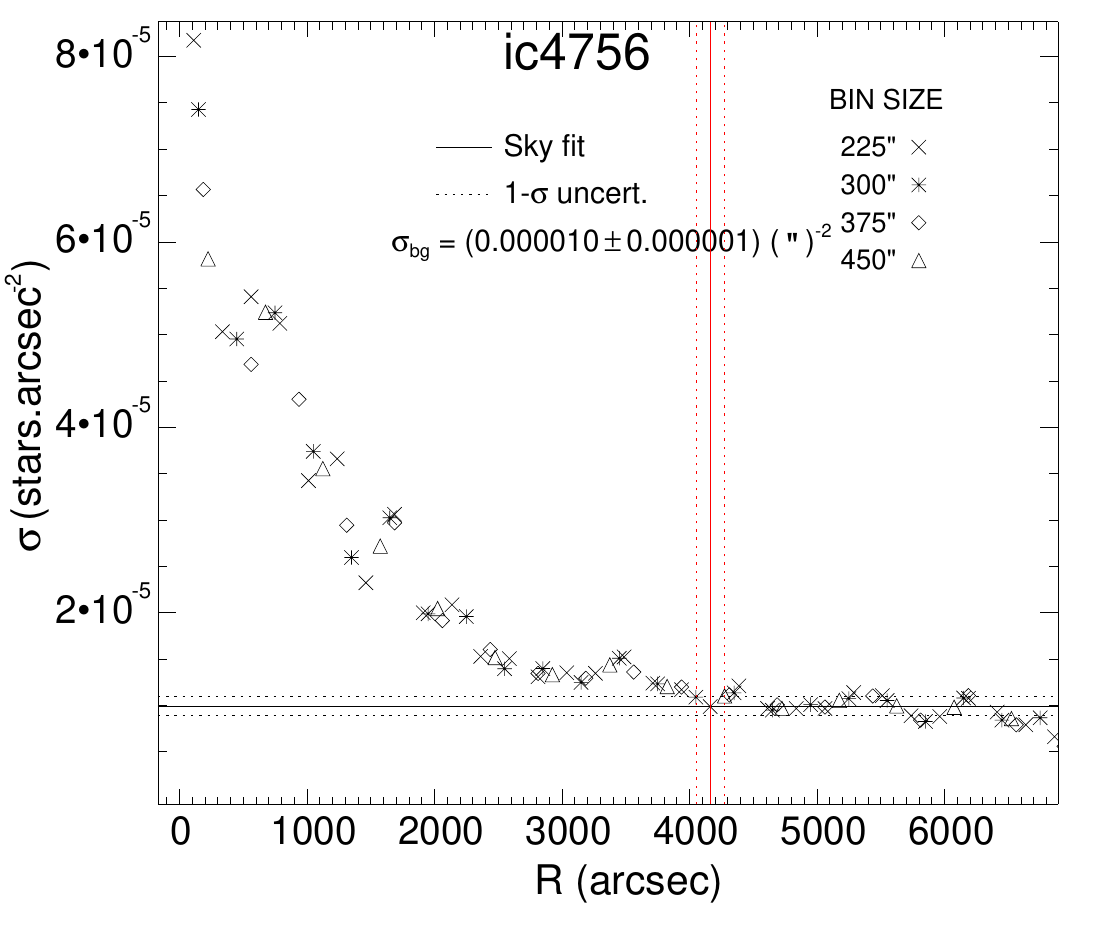}
\includegraphics[width=0.32\linewidth]{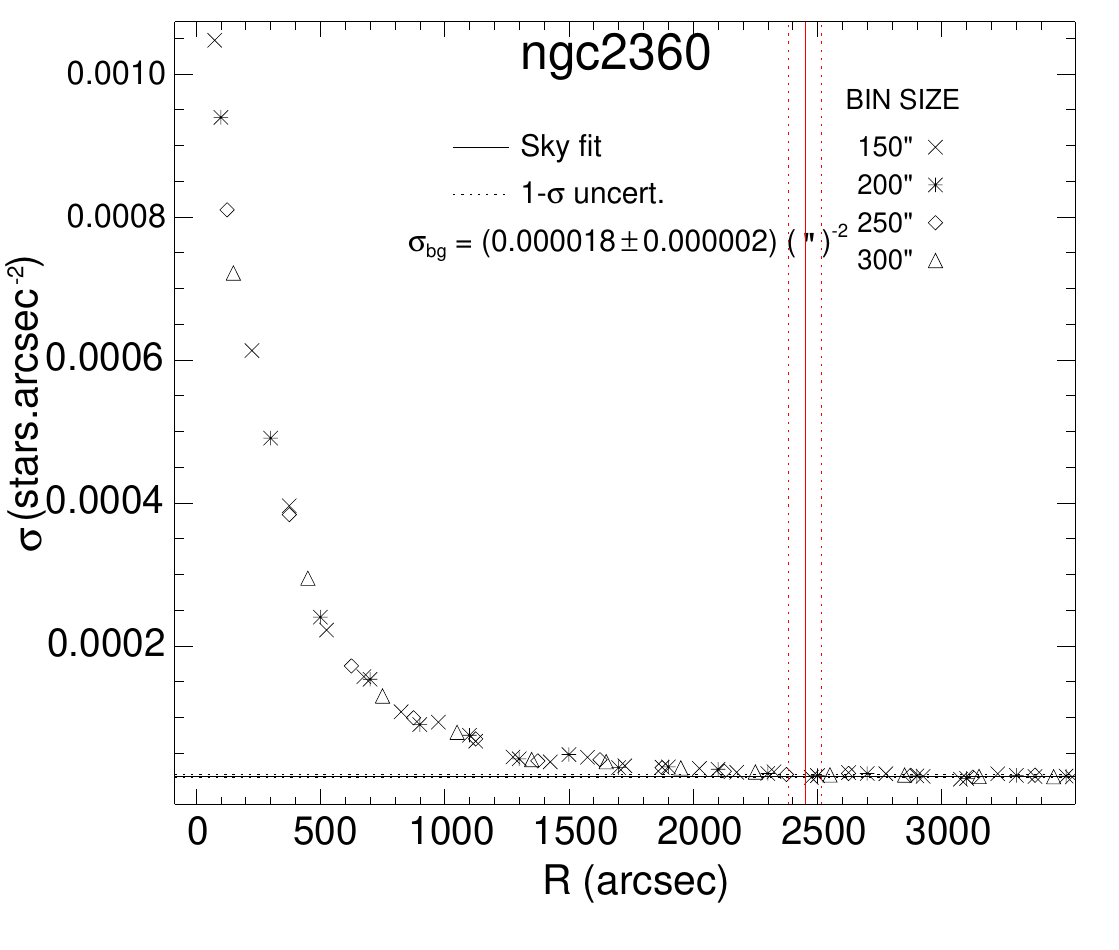}
\includegraphics[width=0.32\linewidth]{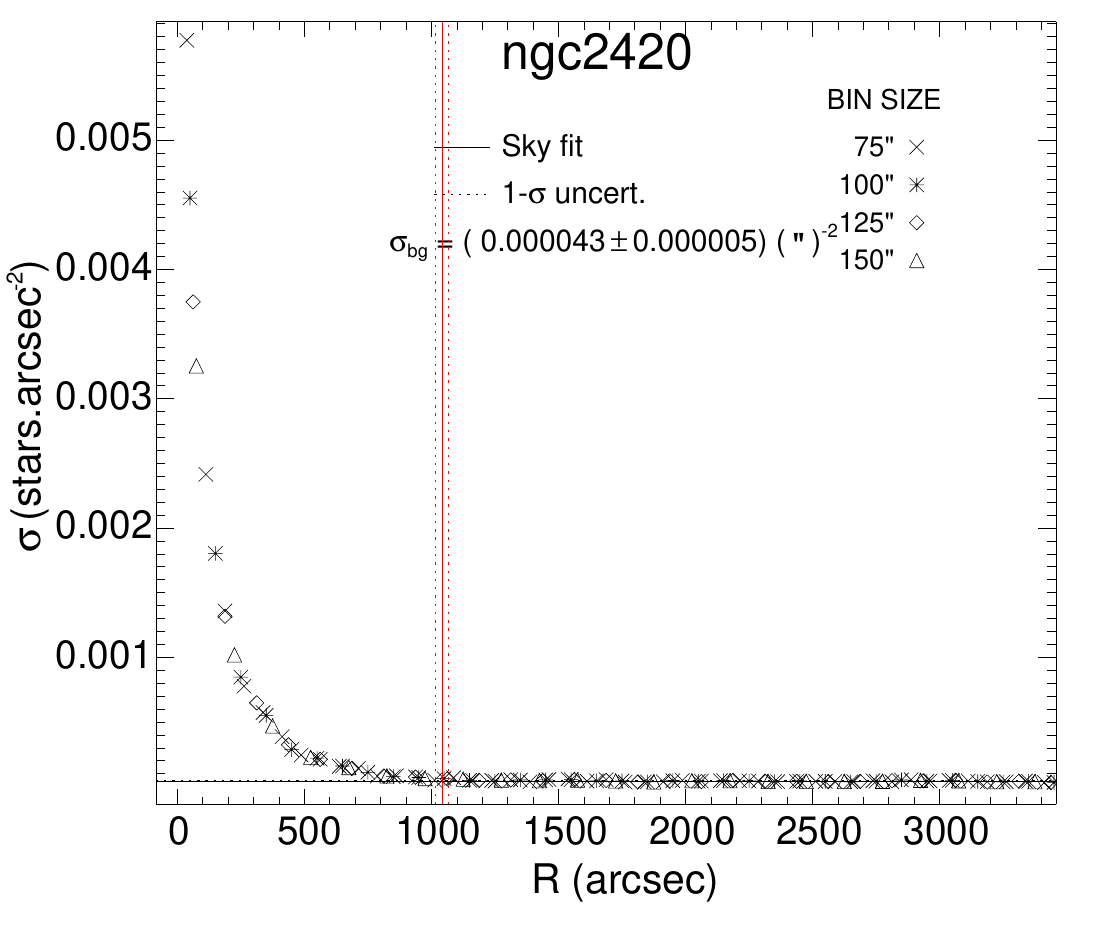}
\caption{Top: Proper motion mask applied to the VPD for OCs of different distances: IC4756 (left), NGC 2360 (middle) and NGC 2420 (right).Bottom: Radial density profiles of the same OCs: IC4756 (left), NGC2360 (middle), NGC2420 (right) with their limiting radius (vertical red line) and mean background density level (horizontal black line) indicated.}
\label{fig:pmra_box}
\end{figure*}

We built the RDPs by counting stars within concentric rings with same thickness as function of the distance to the cluster centre. We repeated this procedure for 4 ring thicknesses (75, 100, 125 and 150 arcseconds, except for the closest clusters, for which larger bin sizes were used) in order to mitigate binning effects on the density distribution. We also computed the density values of the background far from the centre of the cluster. Then, we determined the value of the cluster's limiting radius ($r_{lim}$) as the distance at which the density level reaches the mean value computed for the sky background (bottom panels, Fig. \ref{fig:pmra_box}).

In this procedure, we applied small variations in the coordinates of the centre so that the density profile had a well-defined central maximum. The radius containing $50\%$ of the members ($r_{50}$) was also determined. The OCs Ruprecht\,147, NGC\,752, IC\,4756, NGC\,2527 and IC\,4651 have their spatial distributions significantly affected by field stars due their projection towards dense star fields and/or due to the fact that they are poor stellar concentrations. Therefore, to build the RDPs for those objects, we adopted a subsample with parallaxes above $0.9$\,mas to increase the contrast with the field population. Taking into account the expected dispersion in parallax, this selection did not remove members from the respective OCs.% and was effective to enhance contrast between member and field stars.

\subsection{Two-dimensional proper motion filter and parallax filter}

\indent In order to obtain precise member star lists, we established a filter capable of better predicting the  morphology of the distribution of stars in the VPD, since the box-shaped masks determined previously tend to encompass regions substantially larger than the dispersions expected for the clusters.

%At this point, the work sample has stars projected into the same region of space, as there are stars with parallax values very different from the average value of the clusters and stars with different values of proper motion of the values of the clusters, since the square masks of proper motion tend to encompass regions substantially larger than the dispersions expected for the clusters.

For this purpose, we constructed a two-dimensional histogram of the VPD with the samples restricted by the limiting radius determined for the cluster. Then we also constructed two-dimensional histograms for an adjacent concentric annular stellar field with the same area as the cluster: the internal radius of the control field is 1.3 times greater than the limiting radius. In order to remove the contribution of field stars in the proper motion space, we subtracted the histograms and performed 2D Gaussian fittings over the resulting ones. Subsequently, those stars with proper motion components outside 3-$\sigma$ of the 2D Gaussian mean values, considered as proper motion outliers, were removed from the sample.

To discard remaining probable field stars with discrepant parallaxes from the average OC parallax, we performed an one-dimensional Gaussian fitting to the parallax distribution of the sample filtered by proper motion, limiting them to 3-$\sigma$ of the average value (Fig. \ref{fig:ajuste_gauss_pm}).

\subsection{Memberlists}
\label{sect:member}
\begin{figure*}
%\includegraphics[width=0.31\linewidth]{fig/dados_ajuste_mov_proprio_2cngc7789.eps}
%\includegraphics[width=0.31\linewidth]{fig/dados_cortes_pm_gauss_1ngc7789.eps}
%%\includegraphics[width=0.32\linewidth]{fig/membership/dados_ajuste_mov_proprio_2c_ngc2099.pdf}
%%\includegraphics[width=0.32\linewidth]{fig/membership/dados_ajuste_mov_proprio_2c_ngc2420.pdf}
%%\includegraphics[width=0.32\linewidth]{fig/membership/dados_ajuste_mov_proprio_2c_ngc6134.pdf}
%%\includegraphics[width=0.32\linewidth]{fig/membership/dados_cortes_pm_gauss_1_ngc2099.pdf}
%%\includegraphics[width=0.32\linewidth]{fig/membership/dados_cortes_pm_gauss_1_ngc2420.pdf}
%%\includegraphics[width=0.32\linewidth]{fig/membership/dados_cortes_pm_gauss_1_ngc6134.pdf}
%%\includegraphics[width=0.32\linewidth]{fig/membership/dados_cortes_pm_gauss_plx_2_ngc2099.pdf}
%%\includegraphics[width=0.32\linewidth]{fig/membership/dados_cortes_pm_gauss_plx_2_ngc2420.pdf}
%%\includegraphics[width=0.32\linewidth]{fig/membership/dados_cortes_pm_gauss_plx_2_ngc6134.pdf}
%------------------------
\includegraphics[width=0.32\linewidth]{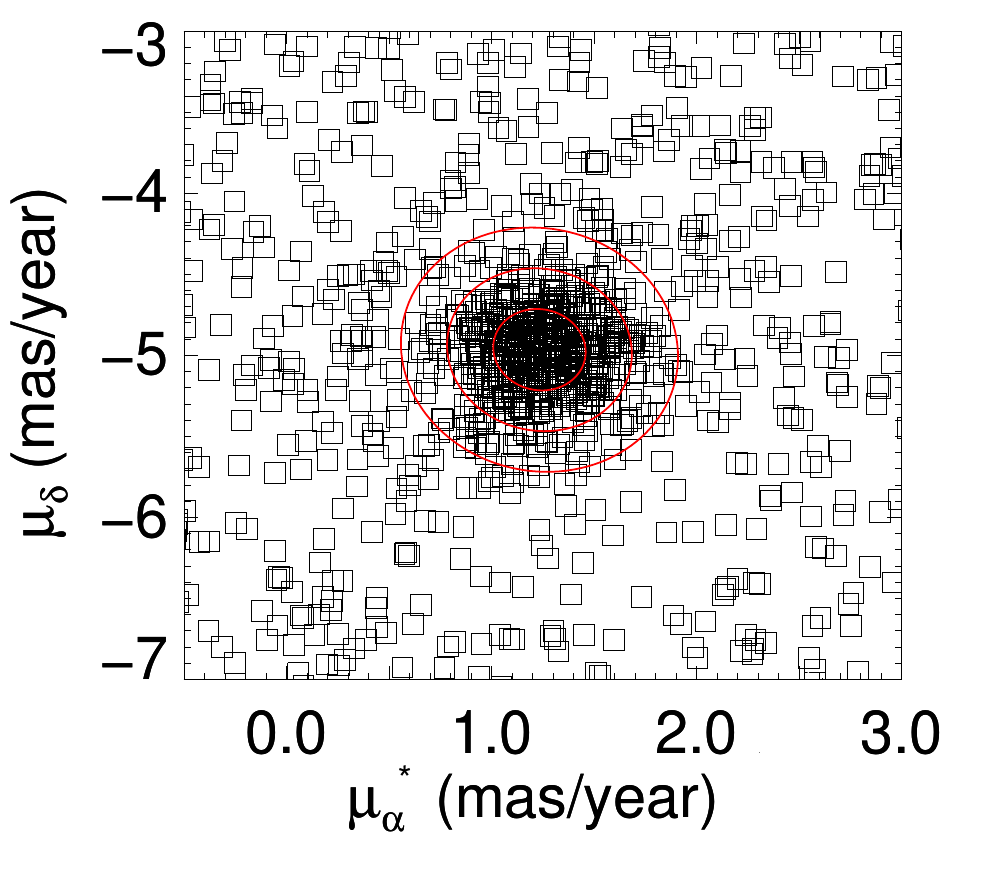}
\includegraphics[width=0.30\linewidth]{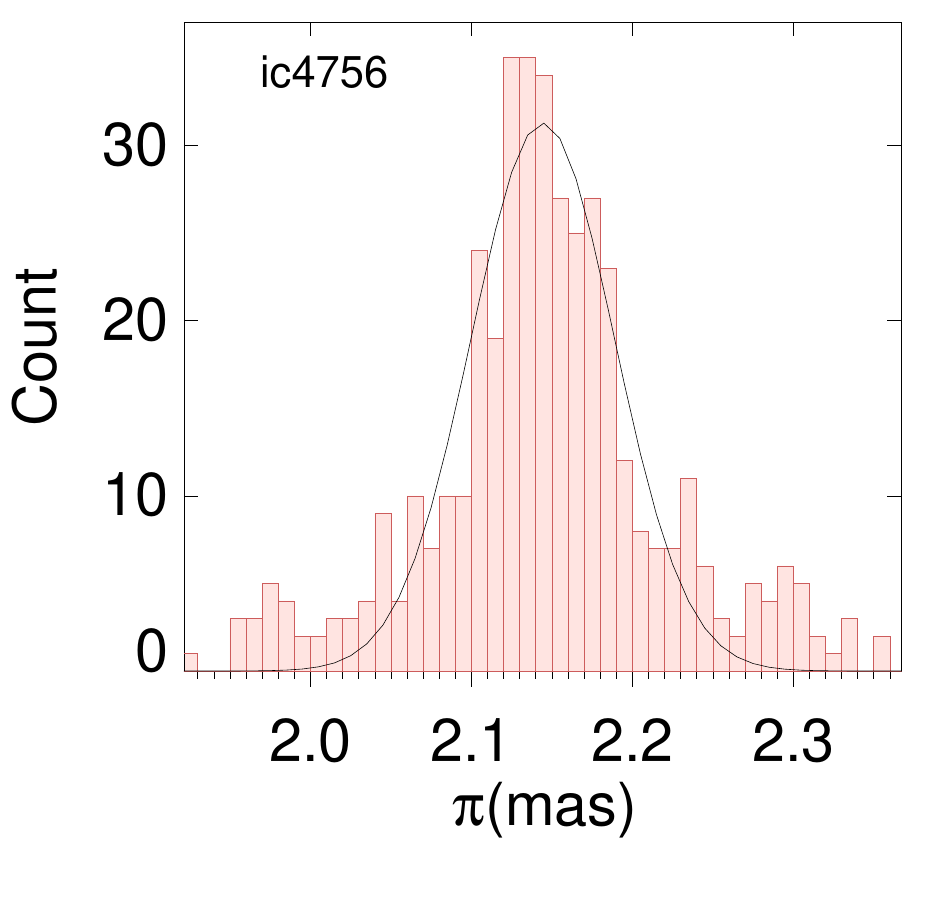}
\includegraphics[width=0.32\linewidth]{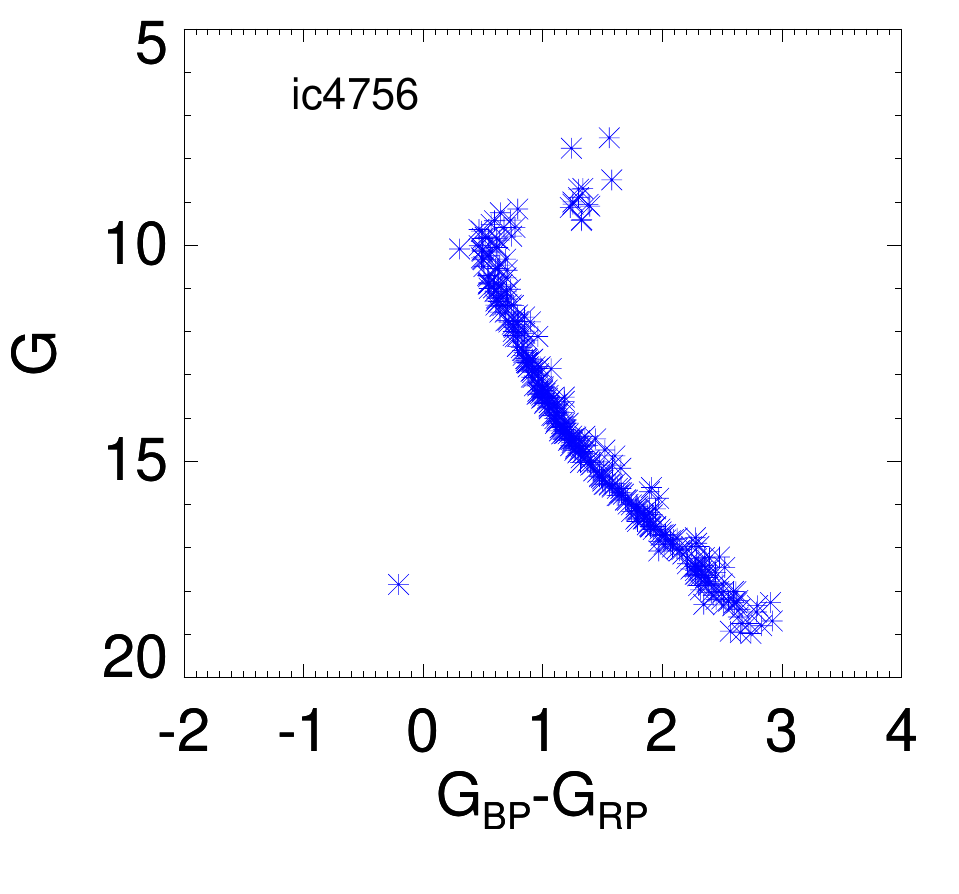}
\includegraphics[width=0.32\linewidth]{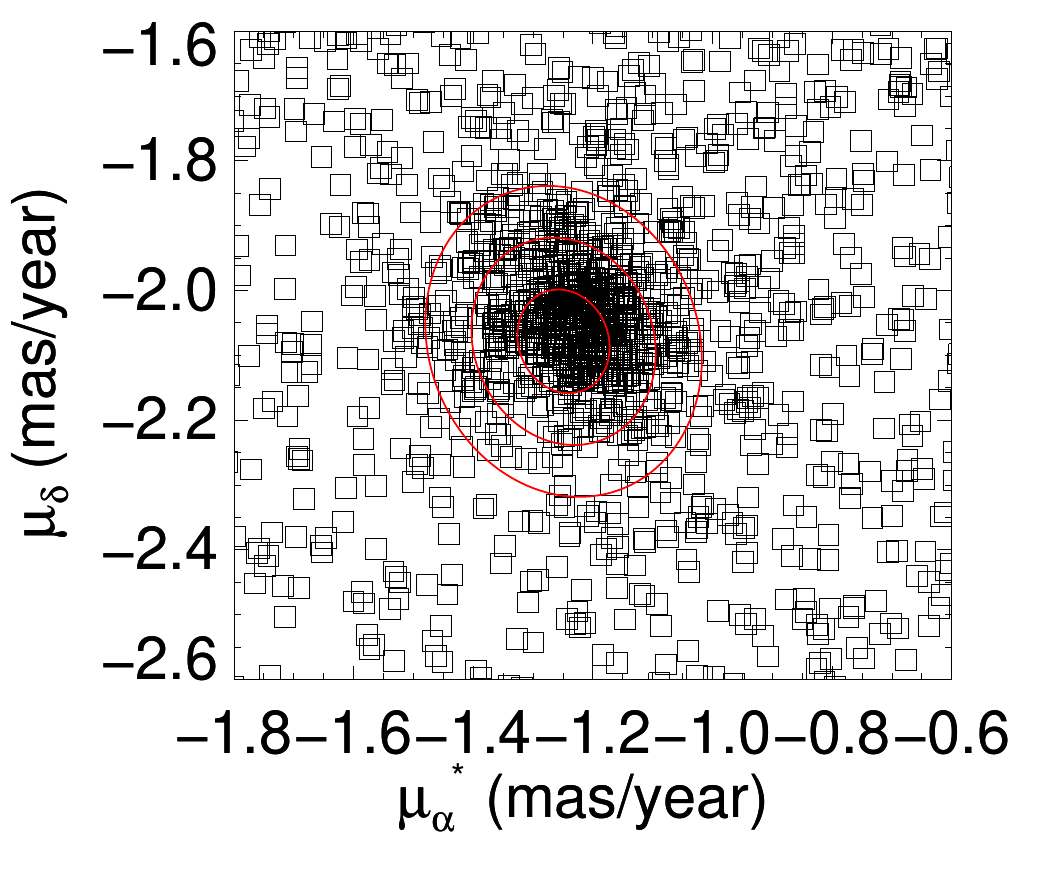}
\includegraphics[width=0.30\linewidth]{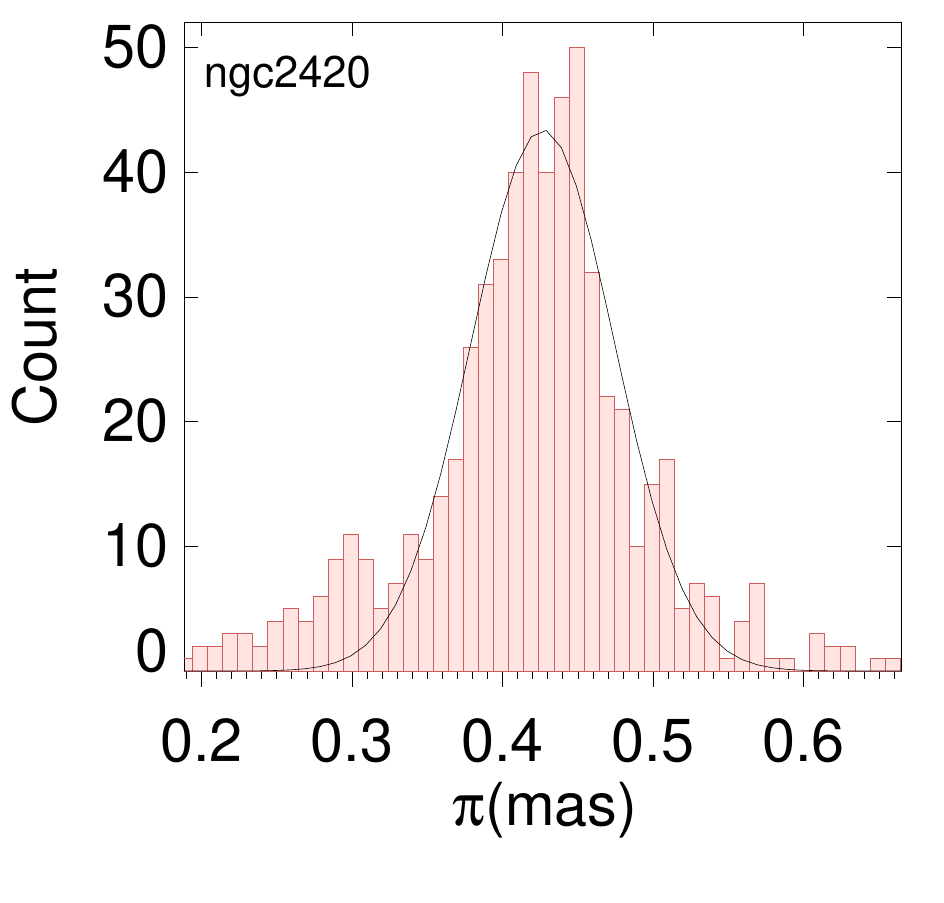}
\includegraphics[width=0.32\linewidth]{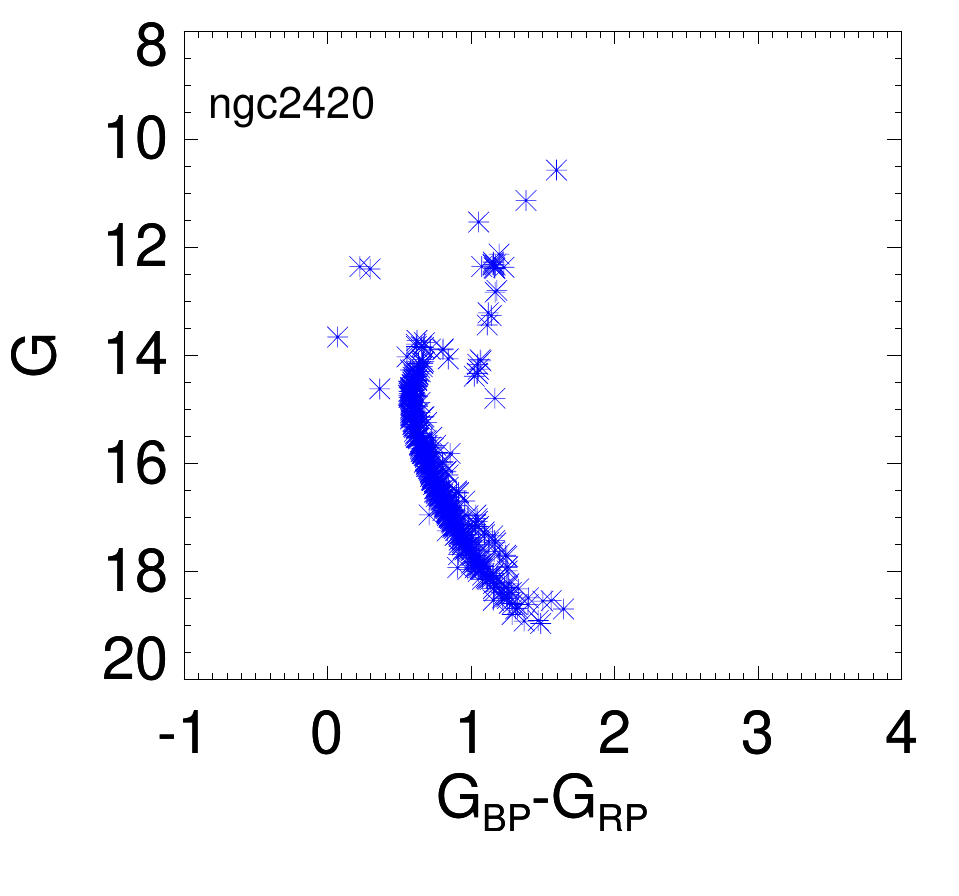}
\caption{Left panels: Projection of the fitted 2D Gaussian function over the VPD, where the red ellipses represent the two-dimensional dispersion in the 1, 2 and 3-$\sigma$ levels. Middle panels: one-dimensional Gaussian fit over the parallax values for the proper motion filtered sample. Right panels: CMD with \textit{Gaia} passsbands built from the final memberlist restricted by proper motion and parallax. In the examples, the OCs represented are: IC 4756 (top) and NGC 2420 (bottom).}
\label{fig:ajuste_gauss_pm}
\end{figure*}

\begin{figure}
%------------------------
%\includegraphics[width=0.32\linewidth]{fig/membership/comparision/pmra_nos_alf-crop.pdf}
%\includegraphics[width=0.32\linewidth]{fig/membership/comparision/pmde_nos_alf-crop.pdf}
%\includegraphics[width=0.32\linewidth]{fig/membership/comparision/plx_nos_alf-crop.pdf}
%\includegraphics[width=0.32\linewidth]{fig/membership/comparision/pmra_nos_hunt-crop.pdf}
%\includegraphics[width=0.32\linewidth]{fig/membership/comparision/pmde_nos_hunt-crop.pdf}
%\includegraphics[width=0.32\linewidth]{fig/membership/comparision/plx_nos_hunt-crop.pdf}
\includegraphics[width=0.49\linewidth]{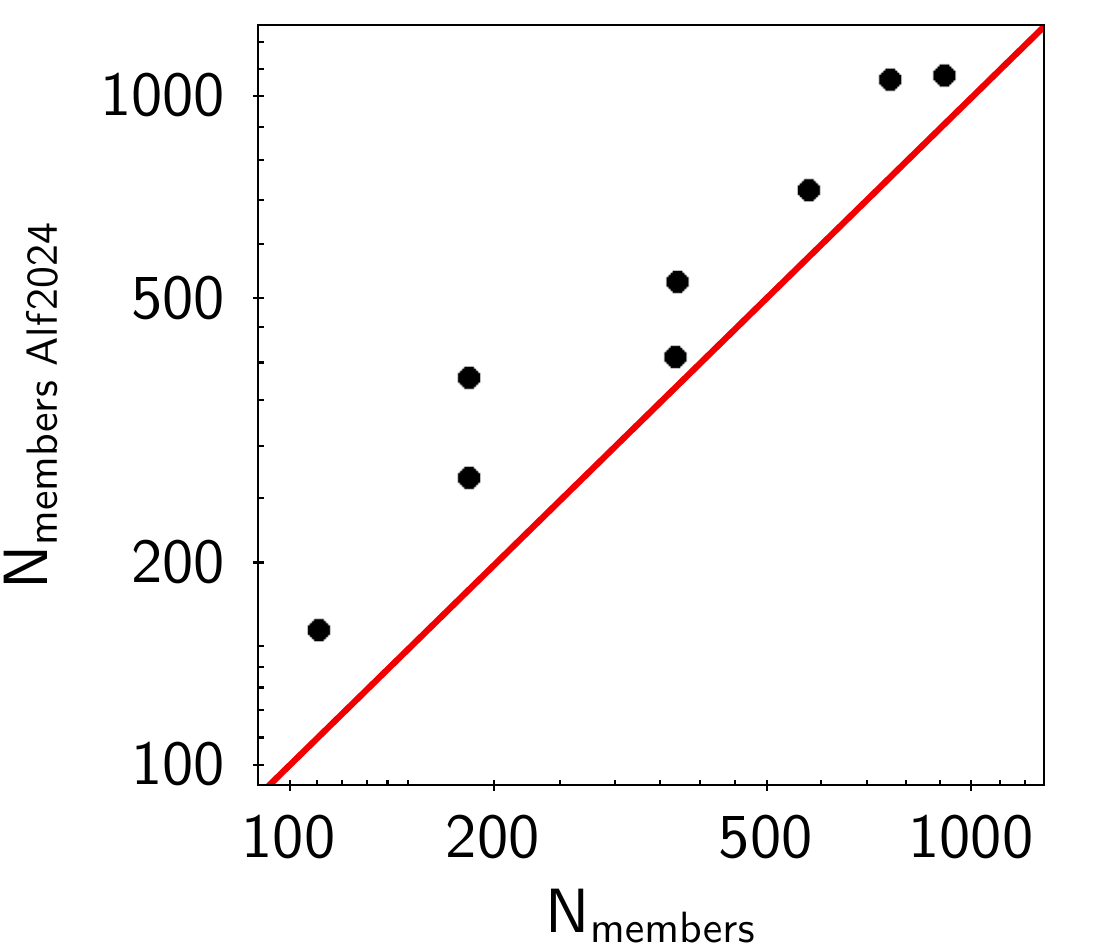}
\includegraphics[width=0.49\linewidth]{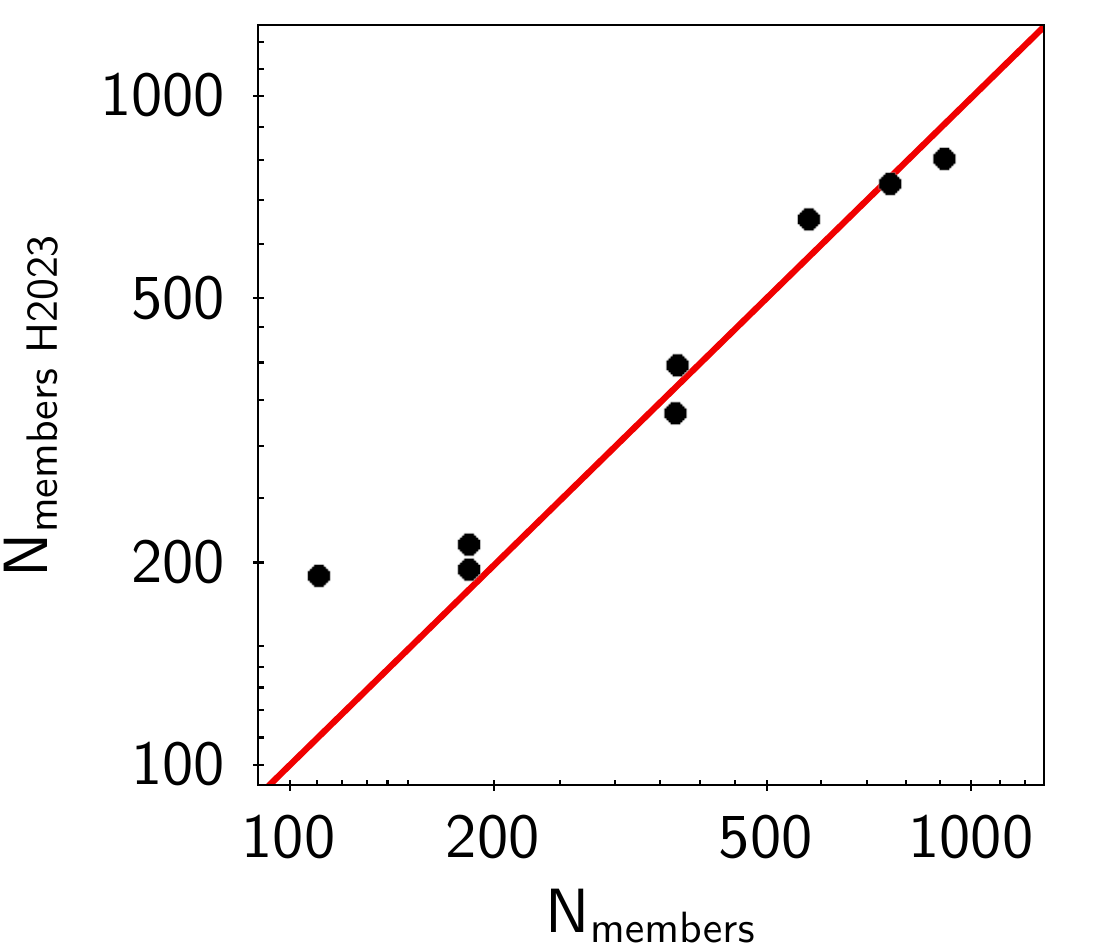}

\caption{Comparision between the number of members obtained in this work (x-axis) with the literature (y-axis) for a sample of OCs within 1kpc. Left panel: comparision with Alf2024 results. Right panel: comparision with H2023 results.}
\label{fig:membership_compara}
\end{figure}

\indent Our OCs memberlists were obtained through the filters mentioned above. The panels in Fig. \ref{fig:ajuste_gauss_pm} show the members proper motions and parallaxes distributions and the cleaned CMD.  We compared our OCs memberlists with those of recent works (H2023; \citealp[]{2024A&A...689A..18A} [hereafter Alf2024]) that used \textit{Gaia} DR3 data to perform membership determination using similar techniques (HDBSCAN). Alf2024 provides memberlists for OCs within 1 kpc. We compared our OCs with those present in both catalogues, resulting in 8 OCs that span a considerable range of sizes and numbers of members. The Table \ref{Tab:pmra_dec_plx_literatura} exhibits a comparison between the mean astrometric parameters derived in the studies for coincidental OCs. The mean astrometric parameters derived in this work are in good agreement with the literature, with no significant offsets in the compared values. We also compared the number of member stars in our study ($N_{members}$) with those reported in the literature, as shown in Fig. \ref{fig:membership_compara}. Our number of members differs from that given by Alf2024, which did not apply a spatial restriction to their OCs, resulting in clusters that appear to encompass more members. In contrast, H2023 reported the total number of members within the tidal radius, which is in good agreement with our results (see Fig. \ref{fig:membership_compara}).

The Galactic coordinates $l$ and $b$, $r_{lim}$, $r_{50}$, proper motions in right ascension and declination ($\mu_{\alpha}^{*}$ and $\mu_{\delta}$), parallaxes ($\varpi$) and their dispersions and the number of probable members $N$ are presented in Table \ref{Tab:members}. Tables containing memberlists and all parameters determined for our sample of OCs are available electronically through
Vizier\footnote{http://cdsarc.u-strasbg.fr/vizier/cat/J/MNRAS/{\bf vol/page}}.

%Comparisons between the mean astrometric parameters 
%,μα cos δ, μδ  obtained in this work (x-axis) and in CJV2018 (y-axis). The outlier is
%the OC Ruprecht 26 (see text for details). The rightmost plot in the second line compares the number of member stars determined for each cluster. 32 OCs are
%common to both studies. In all panels, the dashed line is the identity locus.

%\caption{Top panels: Projection of the fitted function over the VPD, where the red ellipses represent the two-dimensional Gaussian dispersion in 1, 2 and 3-$\sigma$. Middle panels: one-dimensional Gaussian fitting over the parallax remaining sample. Bottom panels: CMD with \textit{Gaia} passsbands built from the final memberlist restricted by proper motion and parallax. In the examples, the OCs represented are: NGC 2099 (left column), NGC 2420 (middle column) and NGC 6134 (right column).}

%N_nos_alf-crop.pdf     pmde_nos_alf-crop.pdf   r50_nos_alf-crop.pdf
%N_nos_hunt-crop.pdf    pmde_nos_hunt-crop.pdf  r50_nos_hunt-crop.pdf
%plx_nos_alf-crop.pdf   pmra_nos_alf-crop.pdf
%plx_nos_hunt-crop.pdf  pmra_nos_hunt-crop.pdf

  \begin{table}
\caption{Mean differences ($\Delta \mu_{\alpha}^{*}$, $\Delta \mu_{\delta}$ and $\Delta \varpi$) between our OCs astrometric parameters ($\mu_{\alpha}^{*}$, $\mu_{\delta}$ and $\varpi$) and 8 coincidental OCs from the literature. The standard deviation is also calculated.}
  \small
\begin{tabular}{|l|r|r|r|}
\hline
   \multicolumn{1}{|c|}{Source} &
   \multicolumn{1}{c|}{$\Delta \mu_{\alpha}^{*}$} &
  \multicolumn{1}{c|}{$\Delta \mu_{\delta}$}& 
    \multicolumn{1}{c|}{$\Delta \varpi$} \\
   %\multicolumn{1}{c|}{$Method$} \\
\hline
   \multicolumn{1}{|c|}{} &
   \multicolumn{1}{c|}{mas/year} &
  \multicolumn{1}{c|}{mas/year}& 
    \multicolumn{1}{c|}{mas} \\
   %\multicolumn{1}{c|}{$Method$} \\
\hline
  H2023 & $0.005 \pm 0.032$ & $-0.005 \pm 0.014 $  & $-0.003 \pm 0.006 $   \\
  \hline
  Alf2024  & $0.005 \pm 0.022 $ & $0.007 \pm 0.010 $ & $0.004 \pm 0.006 $ \\
  \hline
\end{tabular}
   \label{Tab:pmra_dec_plx_literatura}
\end{table}

\begin{table*}
\small
\caption{Astrometric parameters obtained from our OCs memberlists.}
\begin{tabular}{|l|r|r|r|r|r|r|r|r|r|r|r|}
\hline
  \multicolumn{1}{|c|}{Cluster} &
  \multicolumn{1}{c|}{$l$} &
  \multicolumn{1}{c|}{$b$} &
  \multicolumn{1}{c|}{$r_{lim}$} &
  \multicolumn{1}{c|}{$r_{50}$} &
  \multicolumn{1}{c|}{$\mu_{\alpha}^{*}$} &
  \multicolumn{1}{c|}{$\sigma_{\mu_{\alpha}^{*}}$} &
  \multicolumn{1}{c|}{$\mu_{\delta}$} &
  \multicolumn{1}{c|}{$\sigma_{\mu_{\delta}}$} &
  \multicolumn{1}{c|}{$\varpi$} &
  \multicolumn{1}{c|}{$\sigma_{\varpi}$} &
  \multicolumn{1}{c|}{N} \\
  \multicolumn{1}{|c|}{} &
  \multicolumn{1}{c|}{degrees} &
  \multicolumn{1}{c|}{degrees} &
  \multicolumn{1}{c|}{arcmin} &
  \multicolumn{1}{c|}{degress} &
  \multicolumn{1}{c|}{mas/year} &
  \multicolumn{1}{c|}{mas/year} &
  \multicolumn{1}{c|}{mas/year} &
  \multicolumn{1}{c|}{mas/year} &
  \multicolumn{1}{c|}{mas} &
  \multicolumn{1}{c|}{mas} &
  \multicolumn{1}{c|}{ } \\
\hline
  NGC 188 & 122.832 & 22.372 & 26.0 & 0.109 & -2.323 & 0.095 & -1.018 & 0.114 & 0.549 & 0.031 & 820\\
  NGC 752 & 136.884 & -23.343 & 79.7 & 0.399 & 9.779 & 0.225 & -11.83 & 0.224 & 2.307 & 0.041 & 183\\
  NGC 1245 & 146.653 & -8.908 & 15.7 & 0.078 & 0.470 & 0.088 & -1.662 & 0.068 & 0.335 & 0.046 & 641\\
  NGC 1817 & 186.201 & -13.021 & 22.1 & 0.166 & 0.424 & 0.082 & -0.934 & 0.078 & 0.61 & 0.036 & 419\\
  NGC 2099 & 177.611 & 3.08 & 29.6 & 0.123 & 1.882 & 0.155 & -5.617 & 0.140 & 0.712 & 0.043 & 1541\\
  Trumpler 5 & 202.816 & 1.000 & 24.0 & 0.120 & -0.617 & 0.154 & 0.271 & 0.144 & 0.336 & 0.098 & 3593\\
  Collinder 110 & 209.612 & -1.868 & 27.5 & 0.138 & -1.097 & 0.093 & -2.045 & 0.096 & 0.472 & 0.050  & 1030\\
  NGC 2354 & 238.391 & -6.825 & 26.9 & 0.112 & -2.862 & 0.085 & 1.860 & 0.090 & 0.812 & 0.026 & 234\\
  NGC 2355 & 203.385 & 11.830 & 26.9 & 0.090 & -3.841 & 0.077 & -1.065 & 0.083 & 0.563 & 0.042 & 356\\
  NGC 2360 & 229.799 & -1.403 & 40.8 & 0.170 & 0.376 & 0.138 & 5.623 & 0.136 & 0.955 & 0.037 & 715\\
  NGC 2423 & 230.507 & 3.563 & 43.3 & 0.253 & -0.750 & 0.116 & -3.586 & 0.120 & 1.098 & 0.030 & 367\\
  NGC 2420 & 198.107 & 19.642 & 17.4 & 0.072 & -1.221 & 0.077 & -2.048 & 0.074 & 0.427 & 0.047 & 550\\
  NGC 2447 & 240.063 & 0.146 & 37.4 & 0.156 & -3.572 & 0.130 & 5.086 & 0.140 & 1.028 & 0.032 & 657\\
  NGC 2477 & 253.547 & -5.817 & 33.7 & 0.168 & -2.430 & 0.163 & 0.899 & 0.170 & 0.722 & 0.029 & 2548\\
  NGC 2527 & 246.108 & 1.882 & 34.7 & 0.231 & -5.566 & 0.134 & 7.338 & 0.146 & 1.603 & 0.029 & 183\\
  NGC 2539 & 233.721 & 11.114 & 25.0 & 0.146 & -2.329 & 0.105 & -0.538 & 0.100 & 0.798 & 0.033 & 421\\
  NGC 2660 & 265.934 & -3.011 & 13.5 & 0.045 & -2.740 & 0.090 & 5.202 & 0.083 & 0.371 & 0.044 & 559\\
  NGC 2682 & 215.653 & 31.909 & 37.5 & 0.219 & -10.964 & 0.184 & -2.922 & 0.179 & 1.193 & 0.04 & 912\\
  IC 2714 & 292.397 & -1.776 & 25.3 & 0.147 & -7.585 & 0.145 & 2.691 & 0.158 & 0.774 & 0.036 & 1001\\
  NGC 3960 & 294.377 & 6.174 & 15.2 & 0.076 & -6.520 & 0.085 & 1.876 & 0.083 & 0.453 & 0.039 & 472\\
  NGC 4337 & 299.316 & 4.559 & 22.1 & 0.074 & -8.855 & 0.064 & 1.500 & 0.079 & 0.415 & 0.04 & 373\\
  NGC 4349 & 299.737 & 0.828 & 23.6 & 0.118 & -7.845 & 0.134 & -0.265 & 0.144 & 0.557 & 0.047 & 1052\\
  Collinder 261 & 301.712 & -5.560 & 17.4 & 0.087 & -6.369 & 0.129 & -2.682 & 0.132 & 0.375 & 0.067 & 2374\\
  NGC 5822 & 321.525 & 3.730 & 55.0 & 0.321 & -7.485 & 0.185 & -5.491 & 0.171 & 1.239 & 0.034 & 577\\
  NGC 6134 & 334.922 & -0.200 & 25.8 & 0.129 & 2.145 & 0.174 & -4.446 & 0.159 & 0.912 & 0.048 & 752\\
  NGC 6253 & 335.454 & -6.254 & 18.2 & 0.091 & -4.555 & 0.119 & -5.288 & 0.113 & 0.624 & 0.043 & 778\\
  IC 4651 & 340.093 & -7.884 & 30.6 & 0.178 & -2.438 & 0.194 & -5.049 & 0.180 & 1.100 & 0.036 & 758\\
  NGC 6583 & 9.273 & -2.545 & 8.7 & 0.043 & 1.326 & 0.082 & 0.095 & 0.074 & 0.443 & 0.036 & 182\\
  IC 4756 & 36.407 & 5.345 & 69.4 & 0.405 & 1.267 & 0.215 & -4.955 & 0.238 & 2.144 & 0.046 & 370\\
  NGC 6705 & 27.329 & -2.786 & 23.1 & 0.135 & -1.550 & 0.192 & -4.172 & 0.187 & 0.426 & 0.112 & 3323\\
  Ruprecht 147 & 20.919 & -12.779 & 77.8 & 0.519 & -0.869 & 0.309 & -26.701 & 0.468 & 3.299 & 0.064 & 111\\
  NGC 6811 & 79.221 & 11.998 & 37.4 & 0.156 & -3.349 & 0.096 & -8.805 & 0.101 & 0.905 & 0.027 & 287\\
  NGC 6819 & 73.985 & 8.479 & 22.1 & 0.074 & -2.896 & 0.097 & -3.867 & 0.108 & 0.400 & 0.038 & 1754\\
  NGC 7789 & 115.524 & -5.371 & 35.9 & 0.150 & -0.915 & 0.126 & -1.958 & 0.131 & 0.507 & 0.036 & 3233\\
\hline\end{tabular}
\label{Tab:members}
\end{table*}

\section{RC and turnoff positions}
\label{sect:CMD}

The position of the RC in the CMD is often obtained by applying a simple filter over the stars distribution to take its mean value in magnitude and/or colour \citep{2002AJ....123.1603G,2009A&A...508.1279B,2007A&A...463..559V,2019MNRAS.486.5600O}. We determined the mean values and dispersion of colour ($G_{BP}-G_{RP}$ and $J-K$) and magnitudes ($G$ and $K$) of the RC by restricting the sample in a box with width of 1 mag for $G$, 0.4 mag for $G_{BP}-G_{RP}$, 1.4 mag for $K$ and 0.4 mag for $J-K$. Those sizes were chosen due to the visual dispersion of the stars in the RC of both visible and infrared CMDs, encompassing the region defined by RC stars. As uncertainties in the colour and magnitude of the RC were adopted the corresponding standard deviation from the mean.

We considered the bluest point of the MS as the turnoff position. However, some OCs of our sample present blue straggler stars and/or remaining field stars in the final memberlist that have colour indices bluer than the expected turnoff position. In order to remove those isolated stars, we identified the number of nearest neighbours of each star from the CMD within squared boxes of width 0.05 mag. We then removed stars with few close neighbours and, from the remaining sample, measured the colour and magnitude values of the bluest star left in the MS.

\begin{figure}
\includegraphics[width=0.48\linewidth]{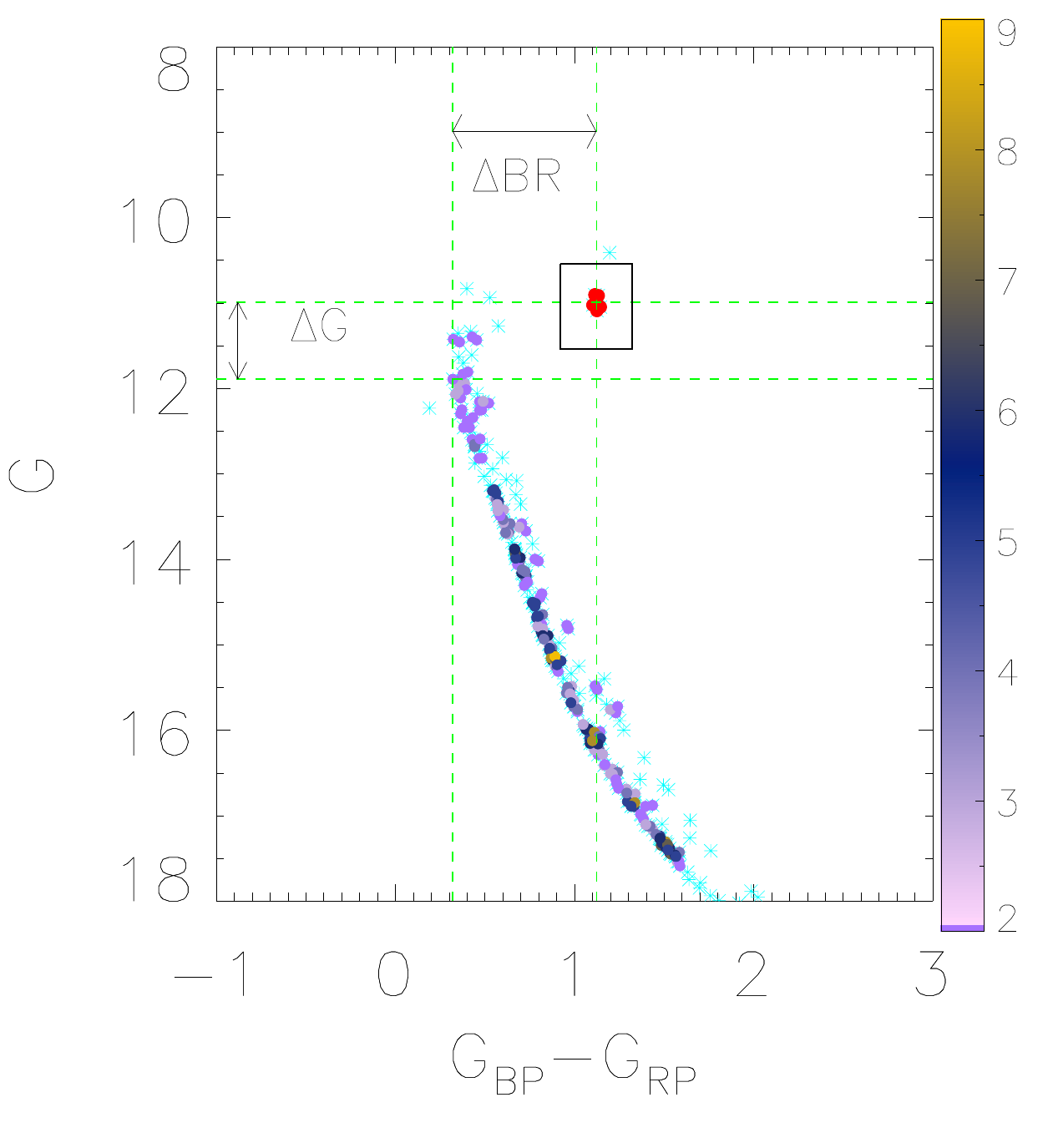}
\includegraphics[width=0.48\linewidth]{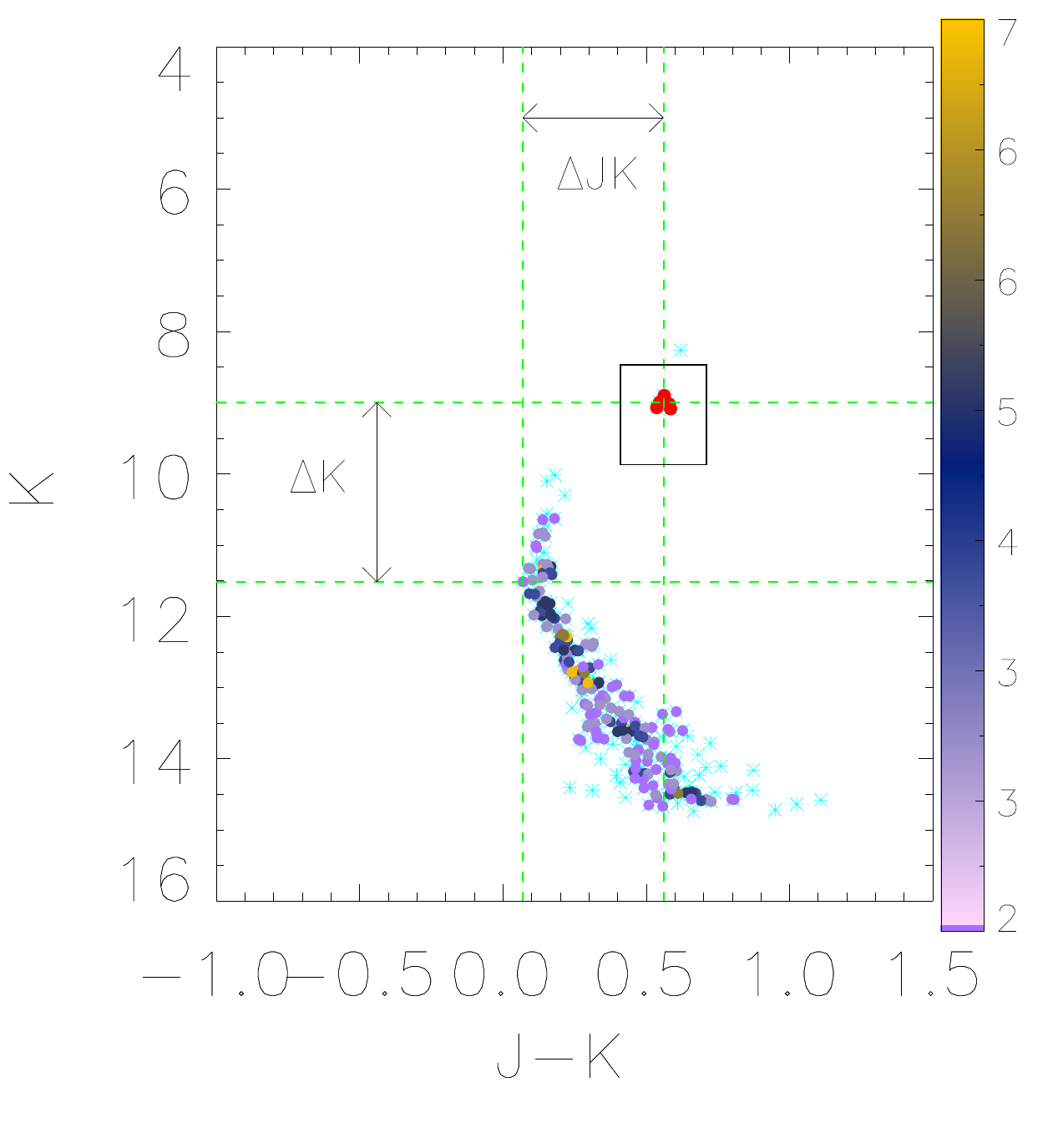}
\includegraphics[width=0.48\linewidth]{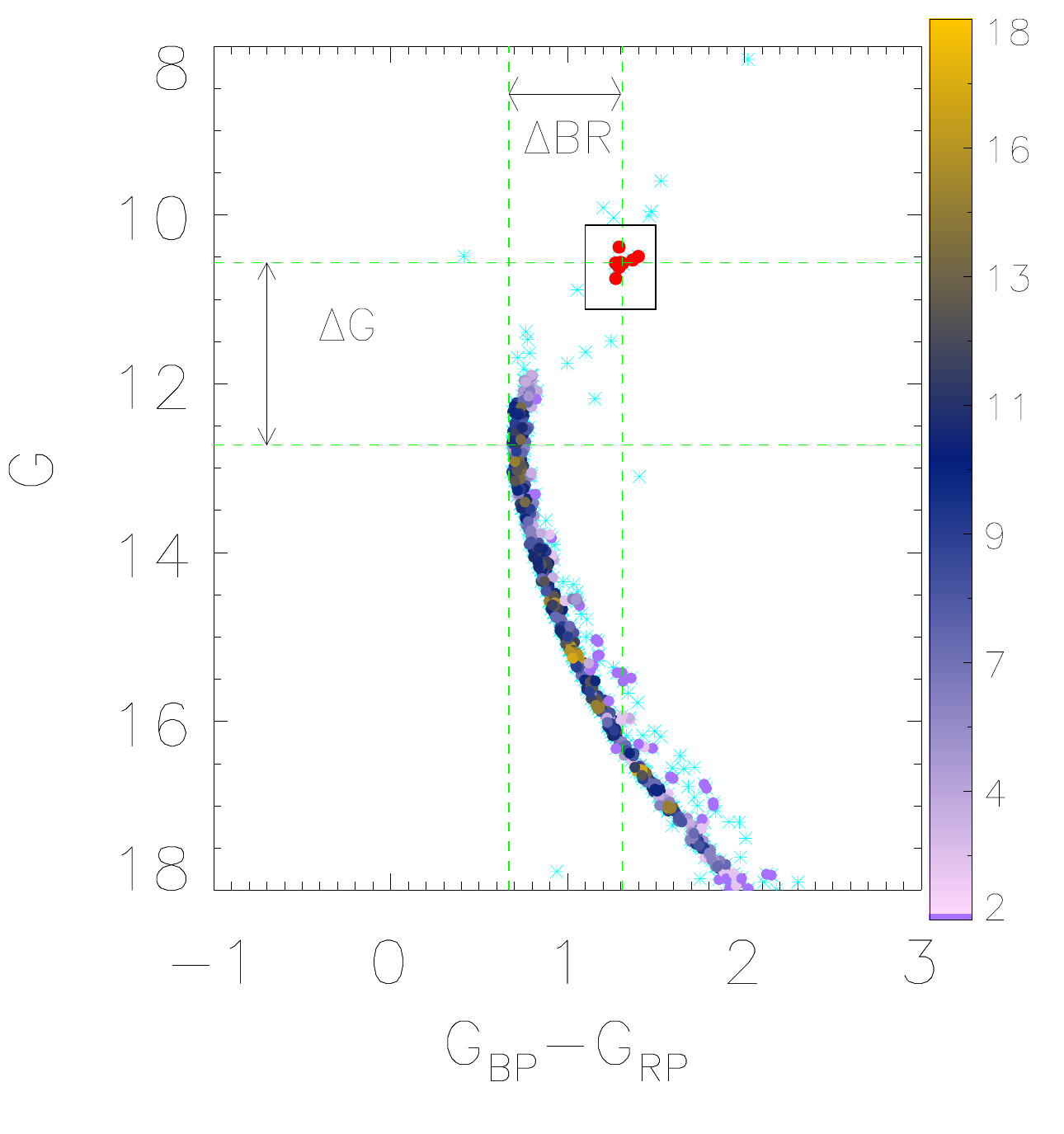}
\includegraphics[width=0.48\linewidth]{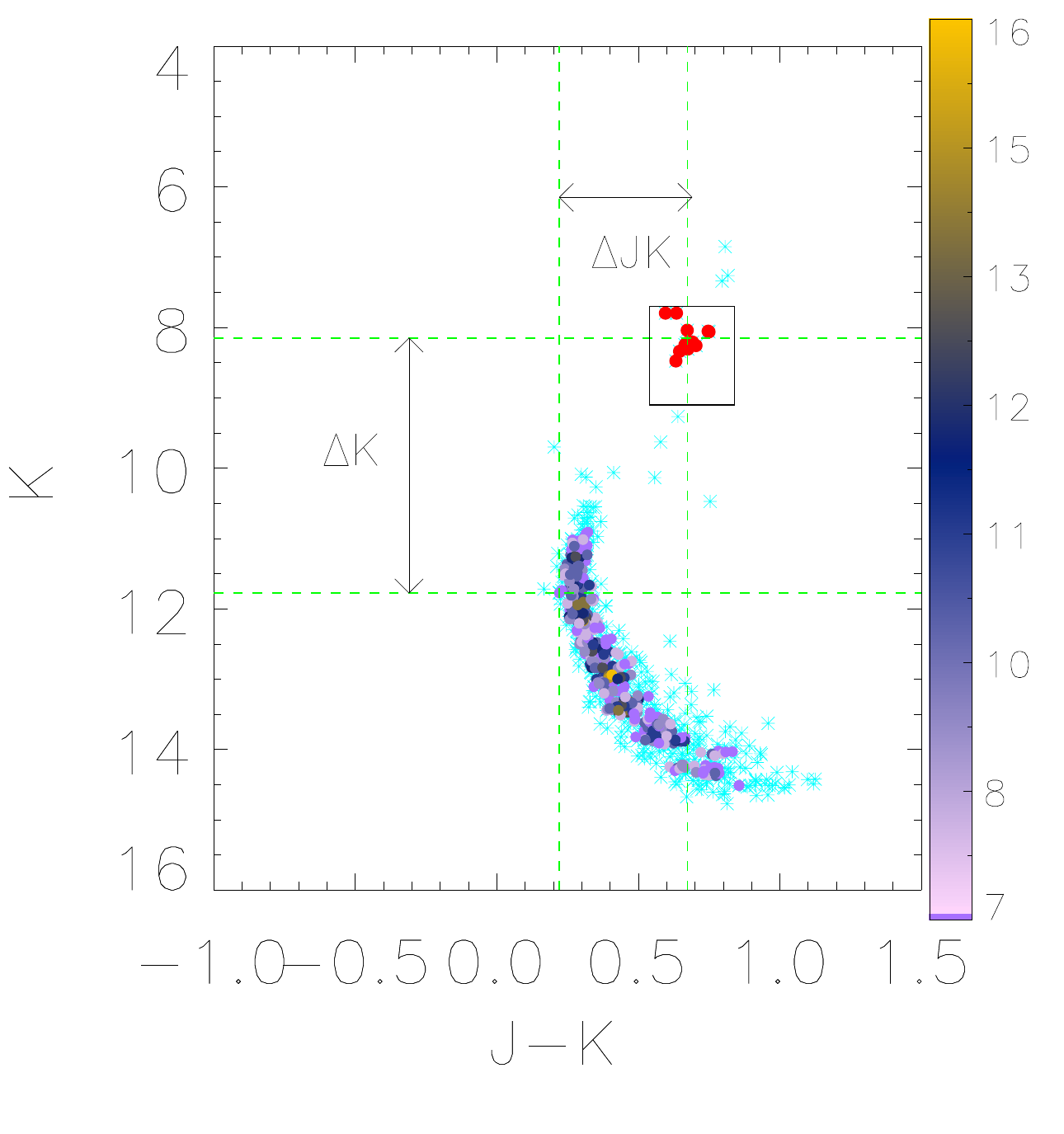}
\includegraphics[width=0.48\linewidth]{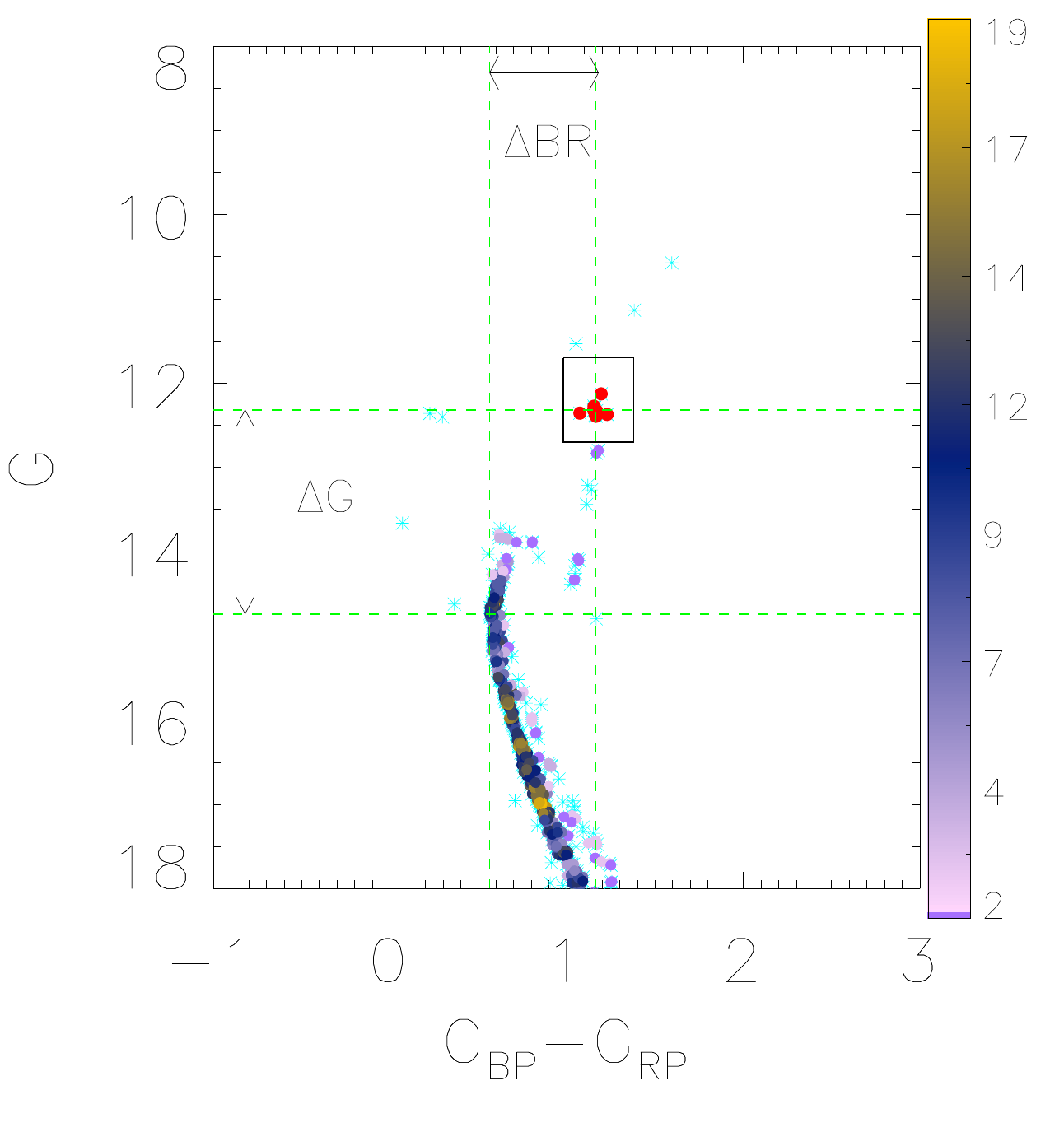}
\includegraphics[width=0.48\linewidth]{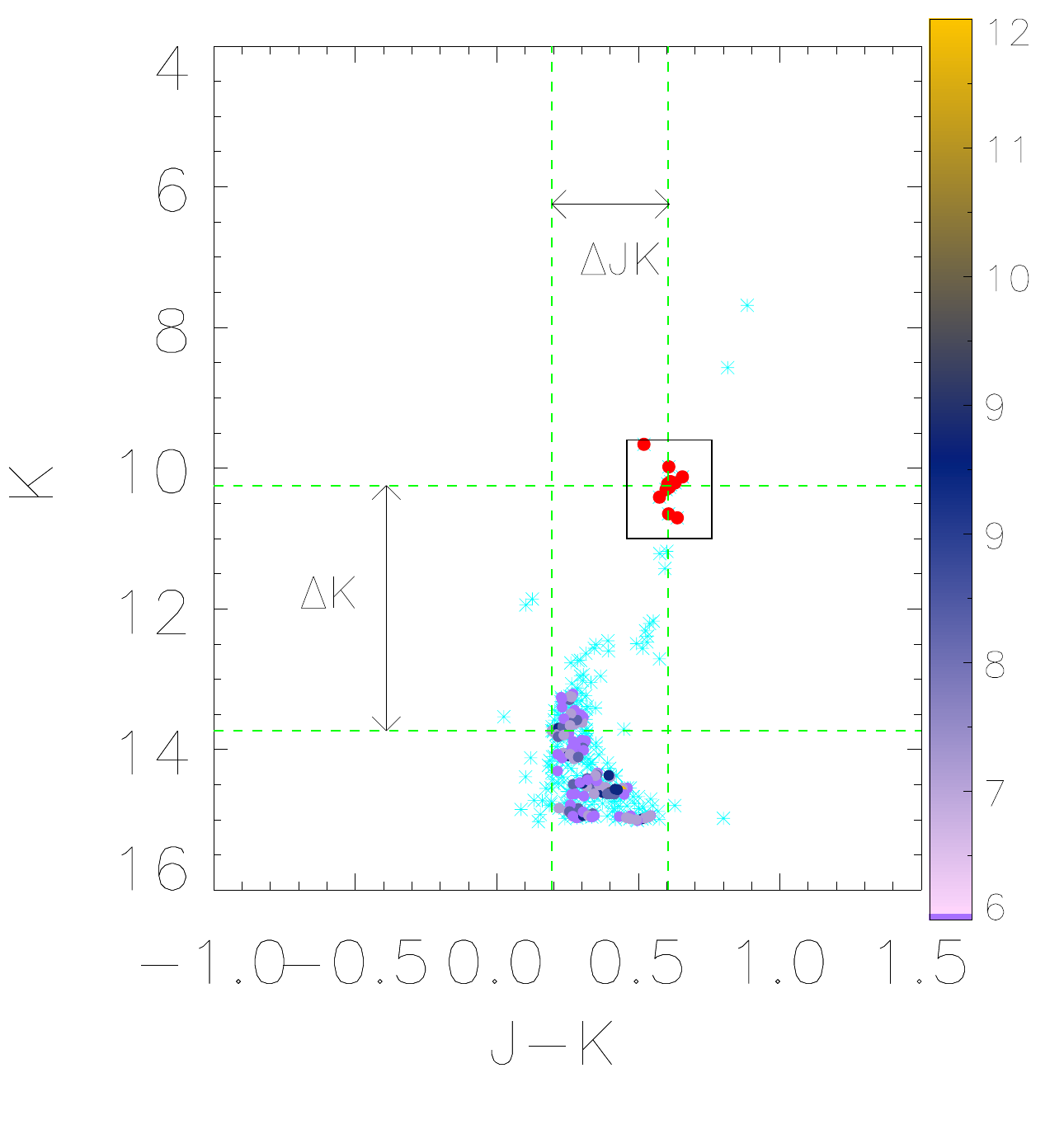} 
\caption{The method used to determine the colour and magnitude for the RC and turnoff positions. Left: CMDs with \textit{Gaia} passbands of the OCs NGC 6811 (top), IC 4651 (middle) and NGC 2420 (bottom). Right: 2MASS CMDs for the same OCs. Cyan symbols represent excluded stars with few neighbours, the blue and yellow symbols represent the density and the colour bar the number of close neighbours. The black box represents the region used to select RC candidates and the red filled circles identify the RC stars, whose mean magnitude and colour were measured. Green dotted lines mark the RC and turnoff positions. The measured indices are also indicated.}
\label{fig:rc_demarca}
\end{figure}

For CMDs with \textit{Gaia} filters, the removal of stars with only one close neighbour worked for most of our sample, and for only four OCs we had to remove stars with more than three close neighbours. On the other hand, the 2MASS photometry presents larger errors and the removal of stars with one, two or three close neighbours worked for most of our sample, although in some cases we had to elevate this cutoff value (for five OCs we had to remove stars with more than 6 neighbours). To provide uncertainties, we adopted the standard deviation value of colour and magnitude of the 5 bluest remaining MS stars. For 2MASS data, we did not determine the turnoff position for the OCs NGC 6819, Trumpler 5 and Collinder 261 due the observational limit, which is brighter than the turnoff of such distant and old objects. The scheme used to identify the turnoff positions and RC average magnitudes is summarized in Fig \ref{fig:rc_demarca}.

%\caption{The figure represent the method used to calculate the colour and magnitude for RC and turnoff stars. Left column: CMD of \textit{Gaia} passbands for OCs NGC6811 (top) and IC4651 (bottom). Right: The same, but for 2MASS photometry. Cyan symbols represent excluded stars with few neighbors, the coloured symbols represent the density and the colour bar the number of close neighbors. The black box represent the region used to select RC candidates and the red filled circles the RC stars used to calculate the mean magnitude and colour. Green dotted lines represent the RC and turnoff positions.}

  We also applied a similar procedure for these same evolutionary regions to the PARSEC isochrones \citep{bressan2012}, with ages and metallicities representative of our OC sample. Initially, we selected
stars where the isochrone table \textit{label} identifier was equal to 1 (MS) and 4, 5 and 6 (He-burning stars = RC). The turnoff position was taken as the bluest point of the MS, however we rejected the blue hook-like structure, that represents stars at the end of the main sequence, in order to keep consistency with the age range analysed.% In order to demarcate the position of the RC, we took the point of maximum effective temperature for stars with \textit{label} between 4 and 6 (bottom left panel, Fig \ref{fig:rc_demarca_modelos}).

\begin{figure}
\includegraphics[width=0.32\linewidth]{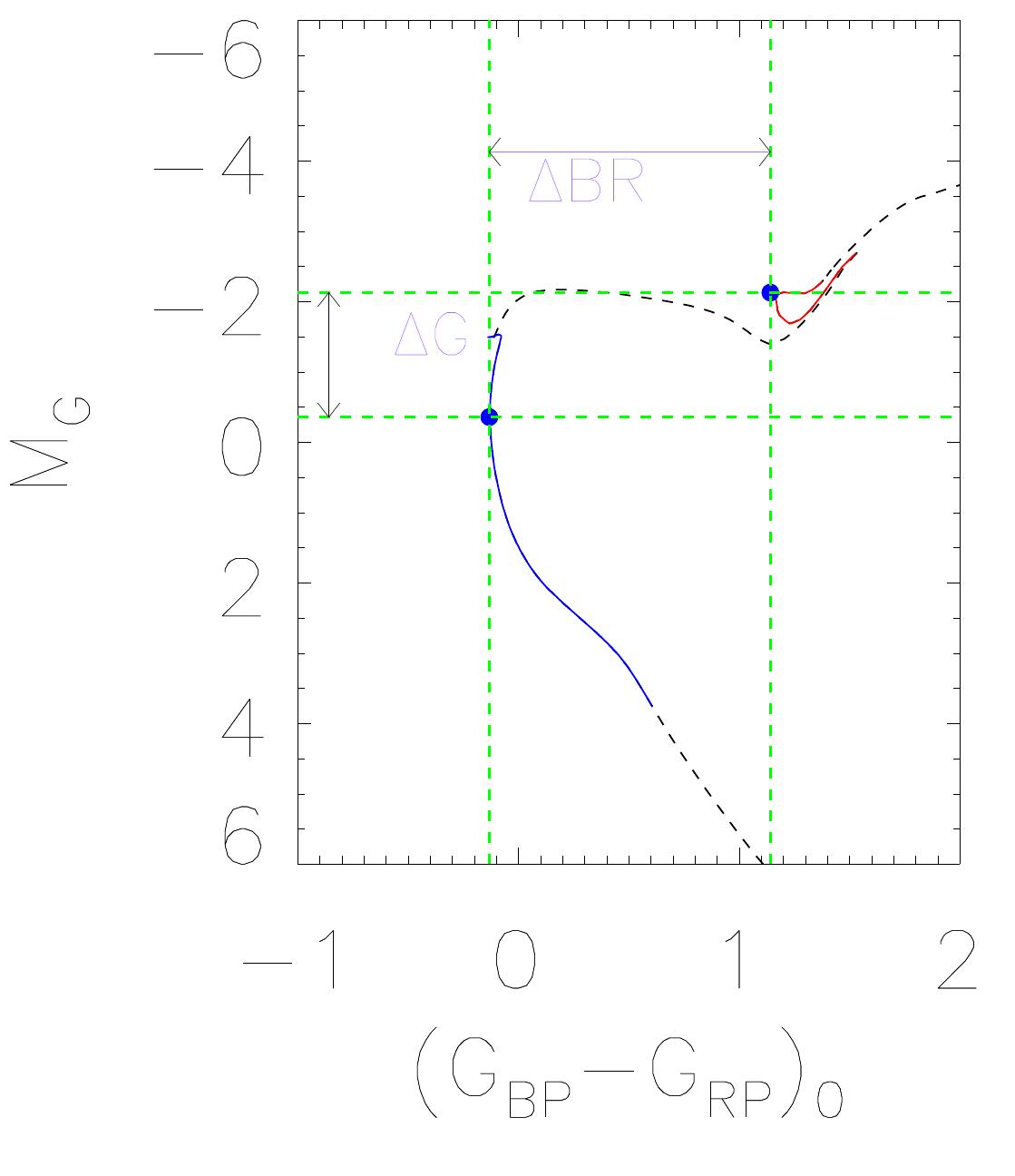}
\includegraphics[width=0.32\linewidth]{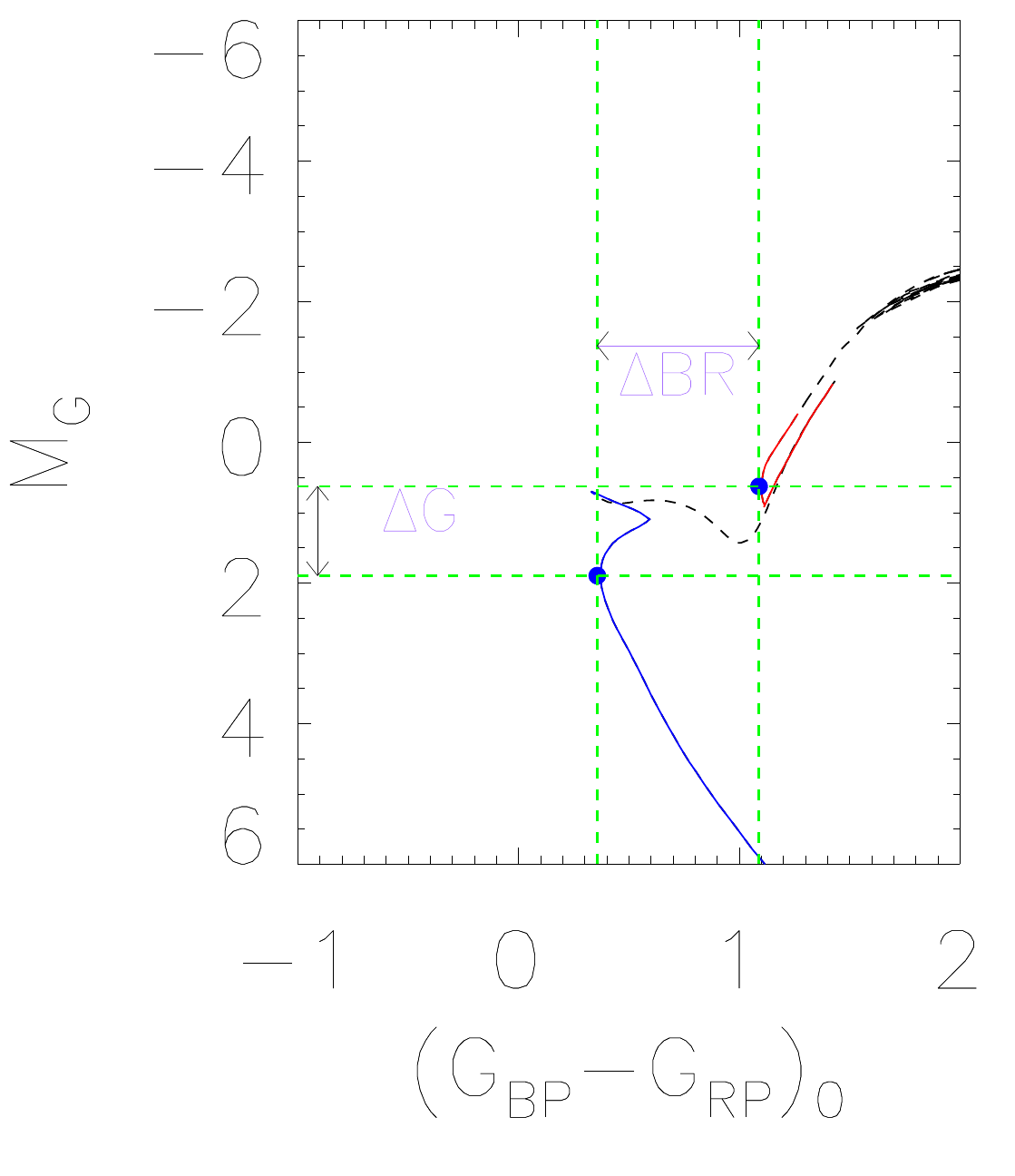}
\includegraphics[width=0.32\linewidth]{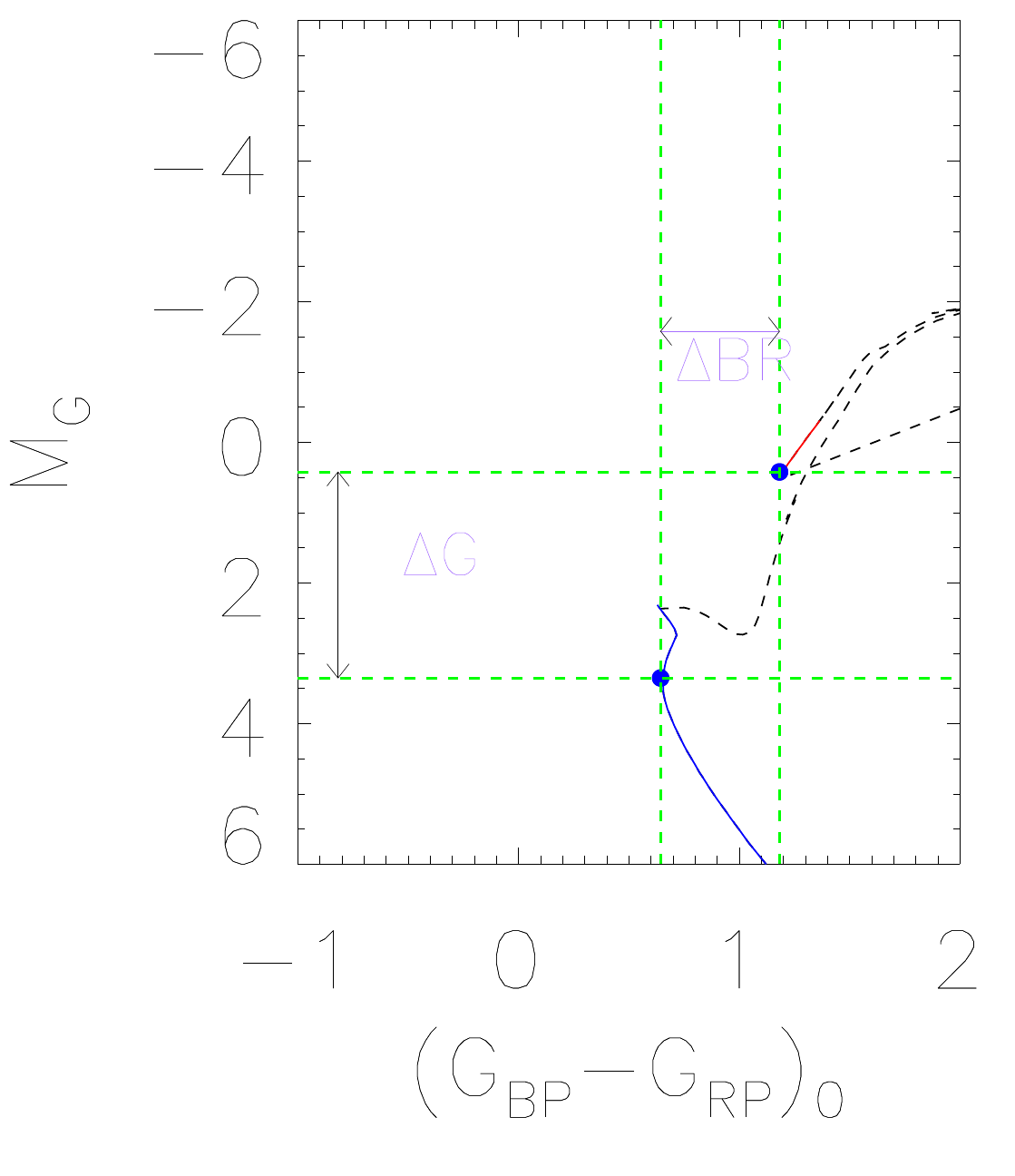}

\includegraphics[width=0.45\linewidth]{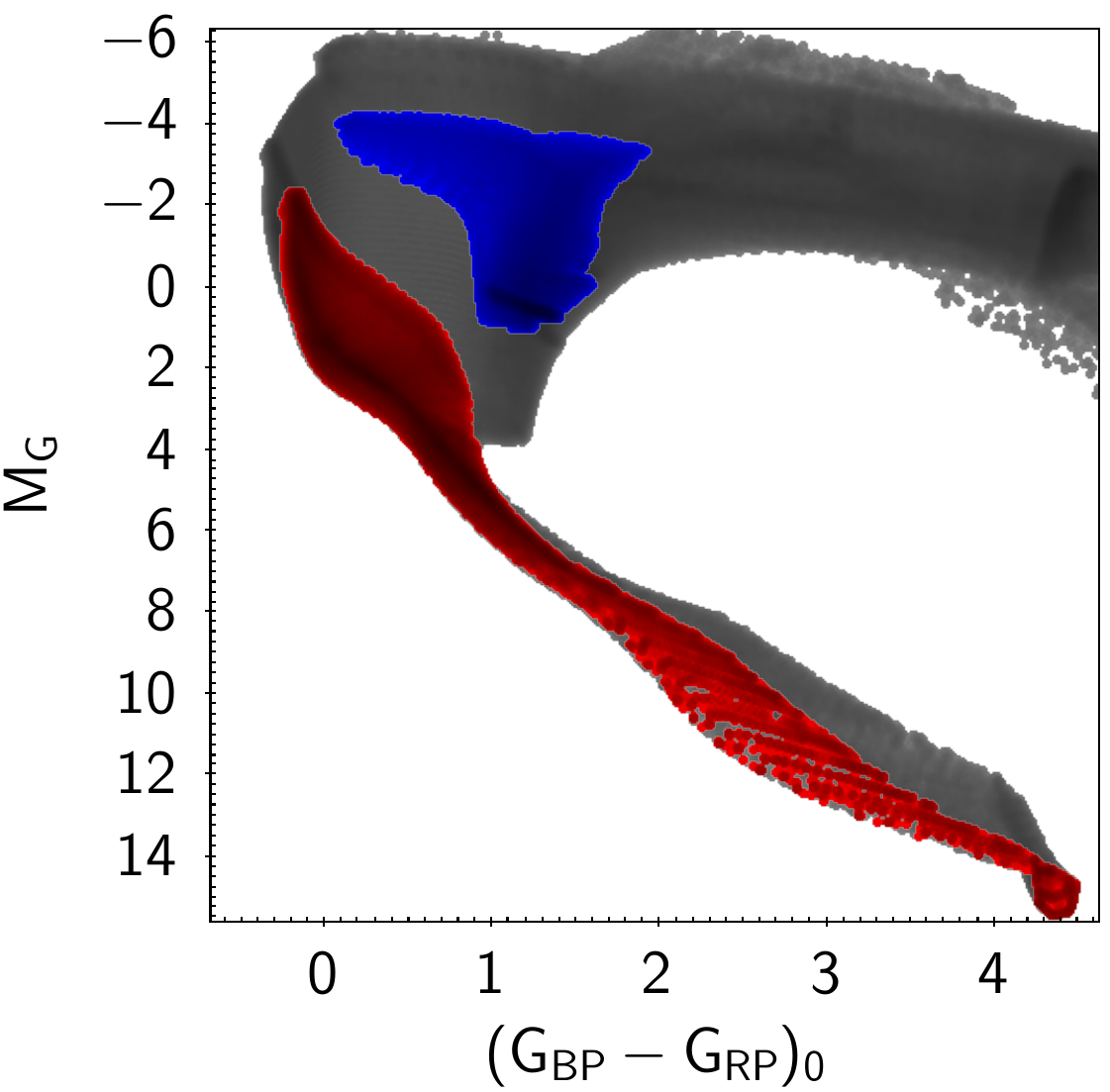}
\includegraphics[width=0.51\linewidth]{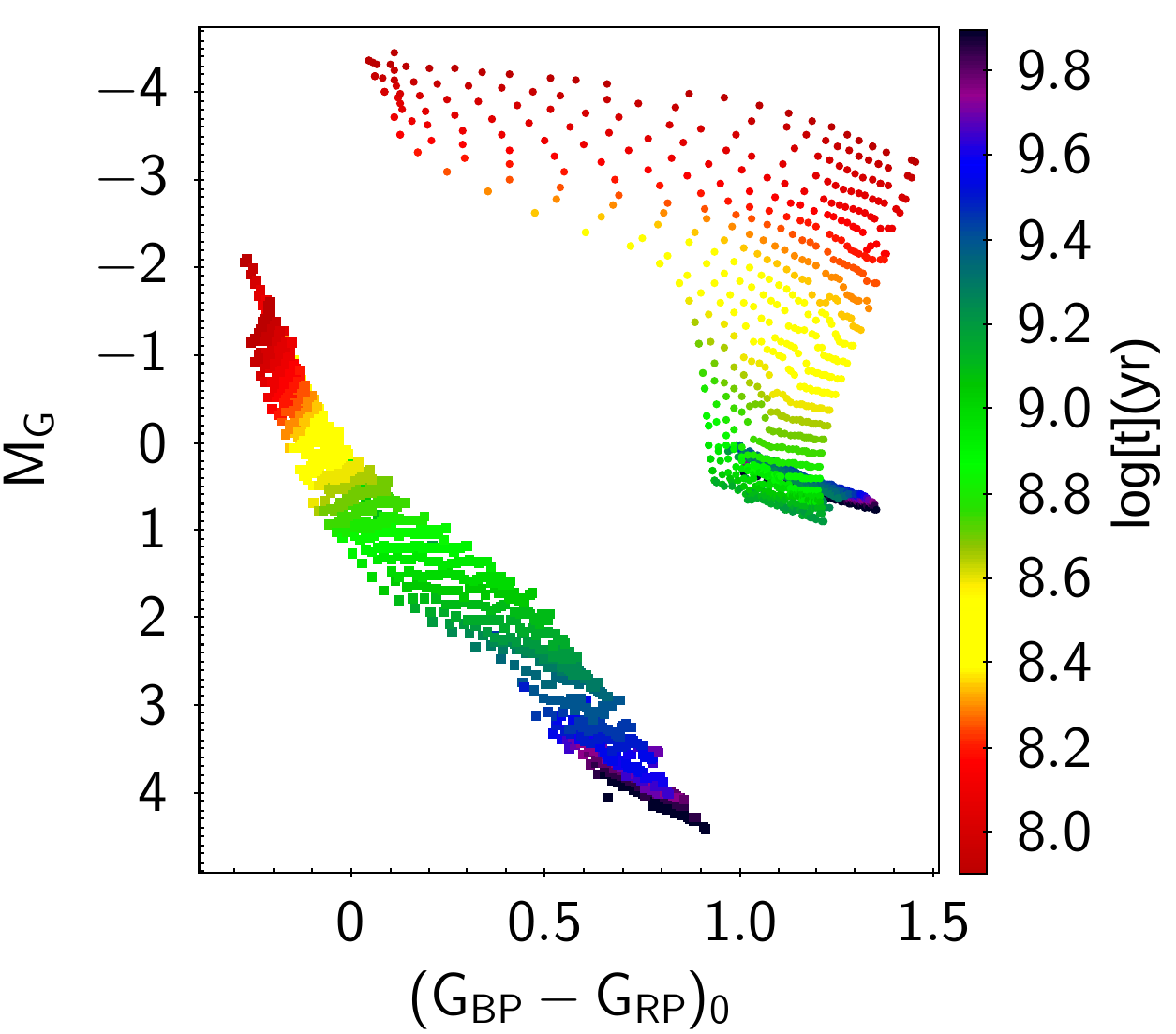}

\caption{Top: Examples of the method used to determine the colour and magnitude for RC and turnoff for solar metallicity PARSEC isochrones of age $\log[t({\rm yr})]=8.3$ (left), 9.1 (middle) and 9.5 (right). Bottom left panel: CMD of all isochrones used, where the red region corresponds to MS stars (\textit{label}=1) and the blue region corresponds to the beginning of the RC phase (\textit{label}=4,5,6). Bottom right panel: RC and turnoff positions identified from the models, where the colourbar indicates the ages.}
\label{fig:rc_demarca_modelos}
\end{figure}

The RC positions according to the PARSEC isochrones correspond to the point of maximum effective temperature (similar to those displayed in Figure 8 of \citealp{2018A&A...609A.116R}). Fig. \ref{fig:rc_demarca_modelos} shows the procedure applied for different ages and the result for the entire range of ages and metallicities. The calculated RC colours and magnitudes, morphological age indices (see next section) and their uncertainties are presented in Tabs. \ref{Tab:ocs_mags} and \ref{Tab:ocs_deltas}, respectively. Detailed tables with those parameters and all other useful information concerning our OC sample and also the calculations performed on the models are available electronically through
Vizier\footnote{http://cdsarc.u-strasbg.fr/vizier/cat/J/MNRAS/{\bf vol/page}}.

\section{Results and discussion}
  \label{sect:result}
 
 \subsection{The RC position}
\label{subsection:RC}

From the mean values of colour and magnitude for RC stars identified in Sect \ref{sect:CMD}, we obtained their absolute values. We took distances and colour excesses from D2021 to obtain the distance modulus and the interstellar extinction for each passband. To convert colour excesses values into interstellar extinction, we used values from the relationship $A_{V}/A_{\lambda}$ based on an established extinction law \citep{Cardelli:1989}, with $R_{V}=3.1$. Using these values, we obtained the absolute RC magnitudes $M_{G}$ and $M_{K}$ and intrinsic colour indices $(G_{BP}-G_{RP})_0$ and $(J-K)_{0}$.

\begin{figure*}
\centering
\includegraphics[width=0.31\linewidth,trim={0.7cm 0.2cm 0.5cm 0.5 cm}]{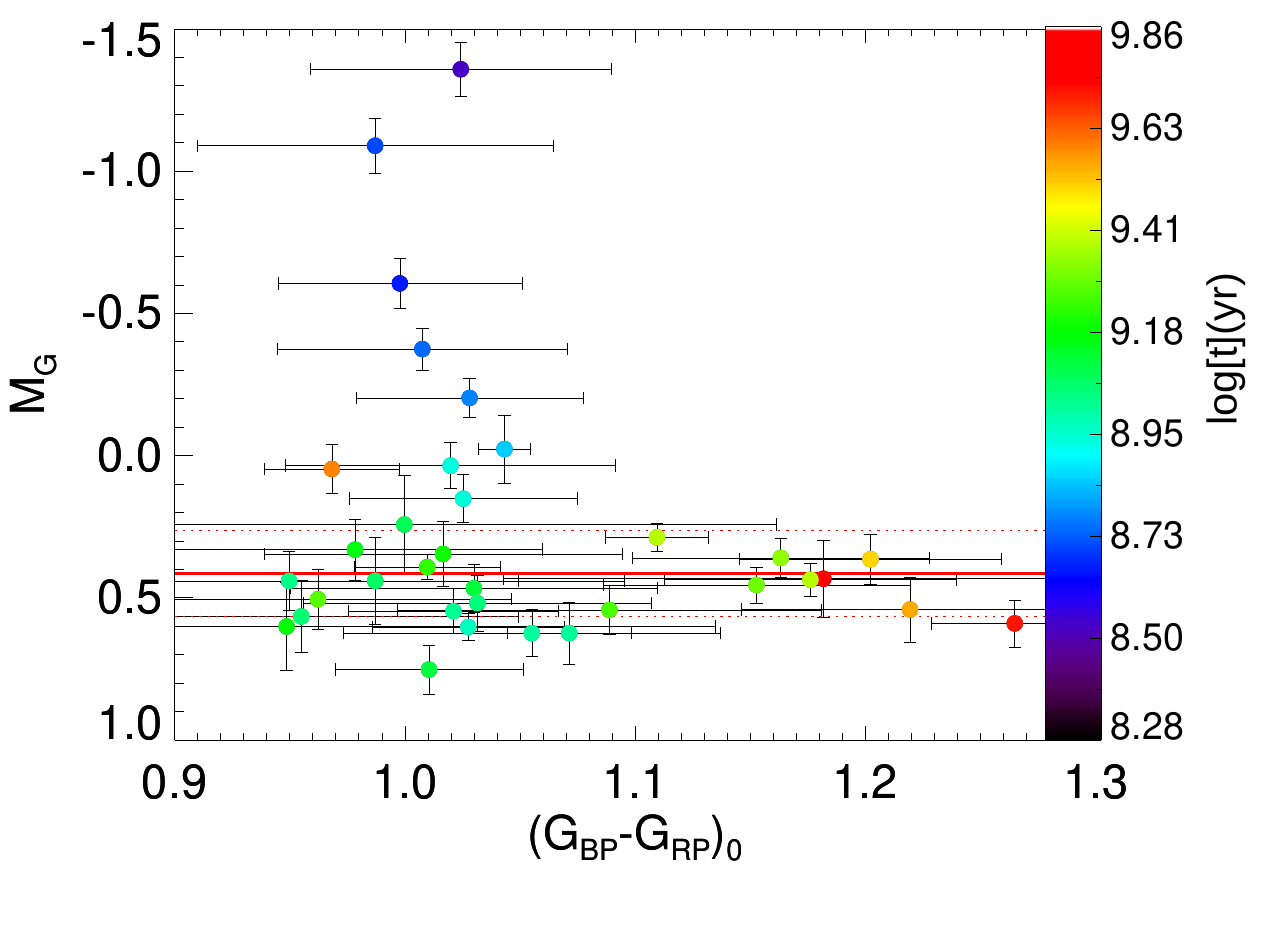}
\includegraphics[width=0.33\linewidth,trim={0.3cm 0.2cm 0.5cm 0.5 cm}]{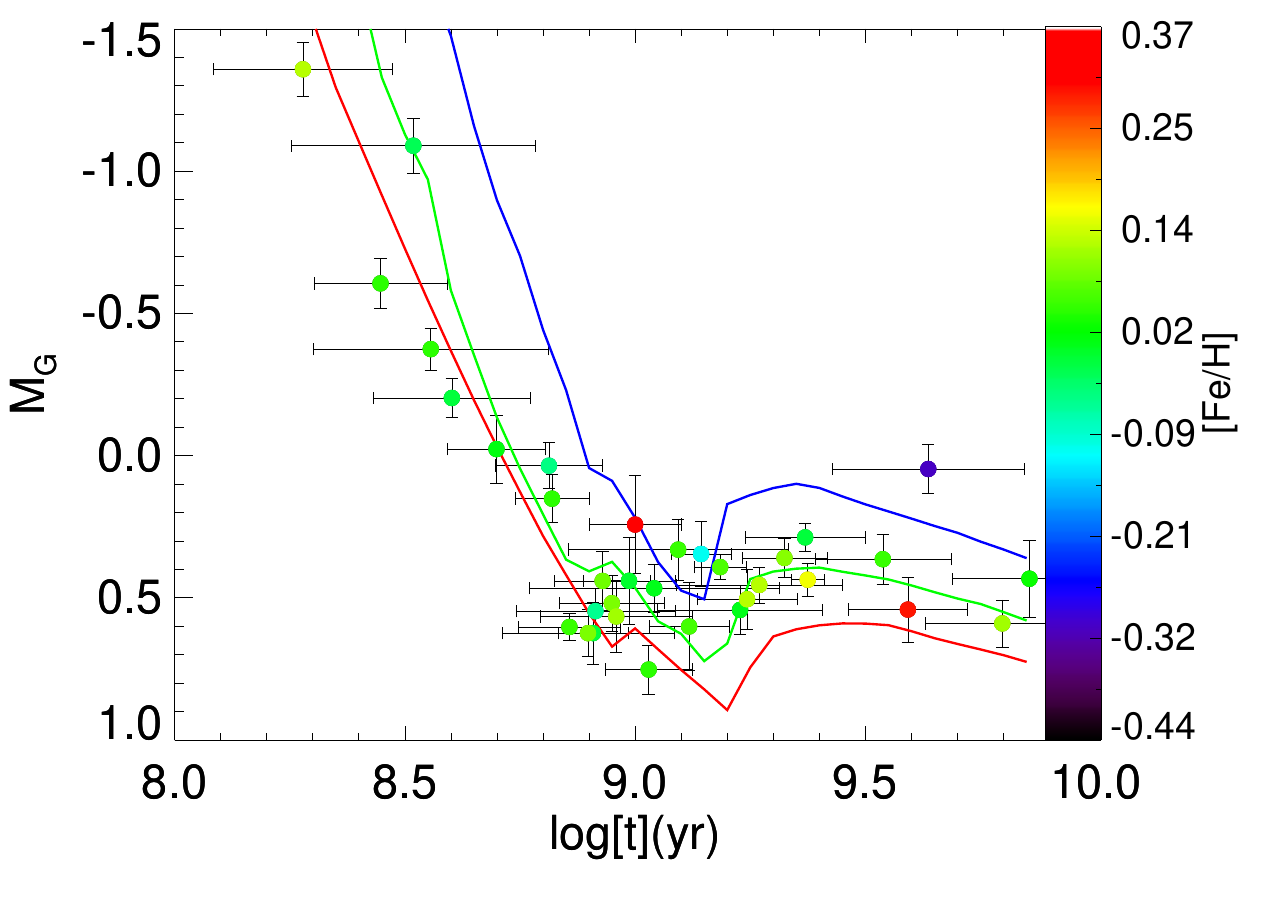}
\includegraphics[width=0.30\linewidth,trim={0.3cm 0.2cm 1.9cm 0.5 cm}]{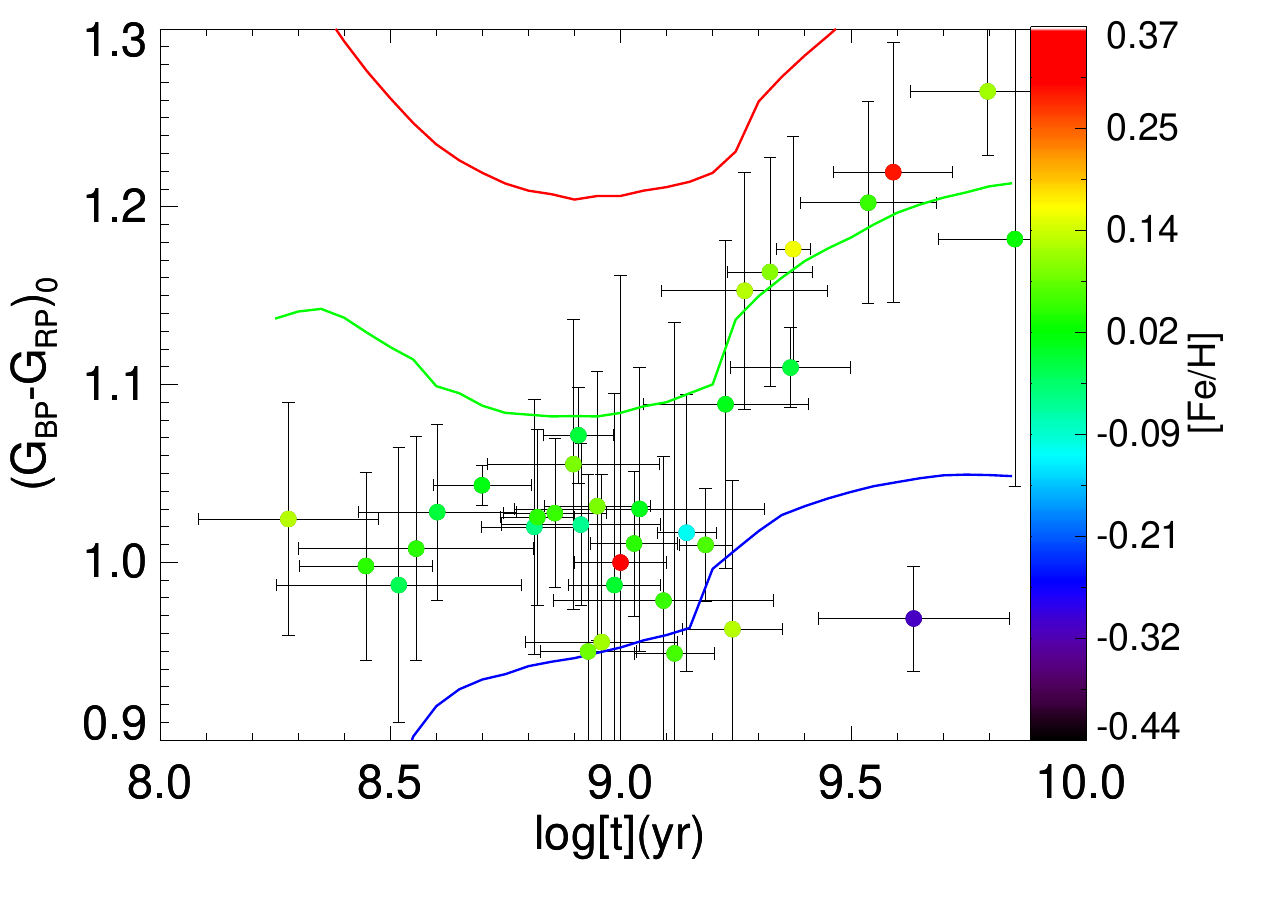}
\includegraphics[width=0.32\linewidth,trim={0.7cm 0.2cm 0.5cm 0.5 cm}]{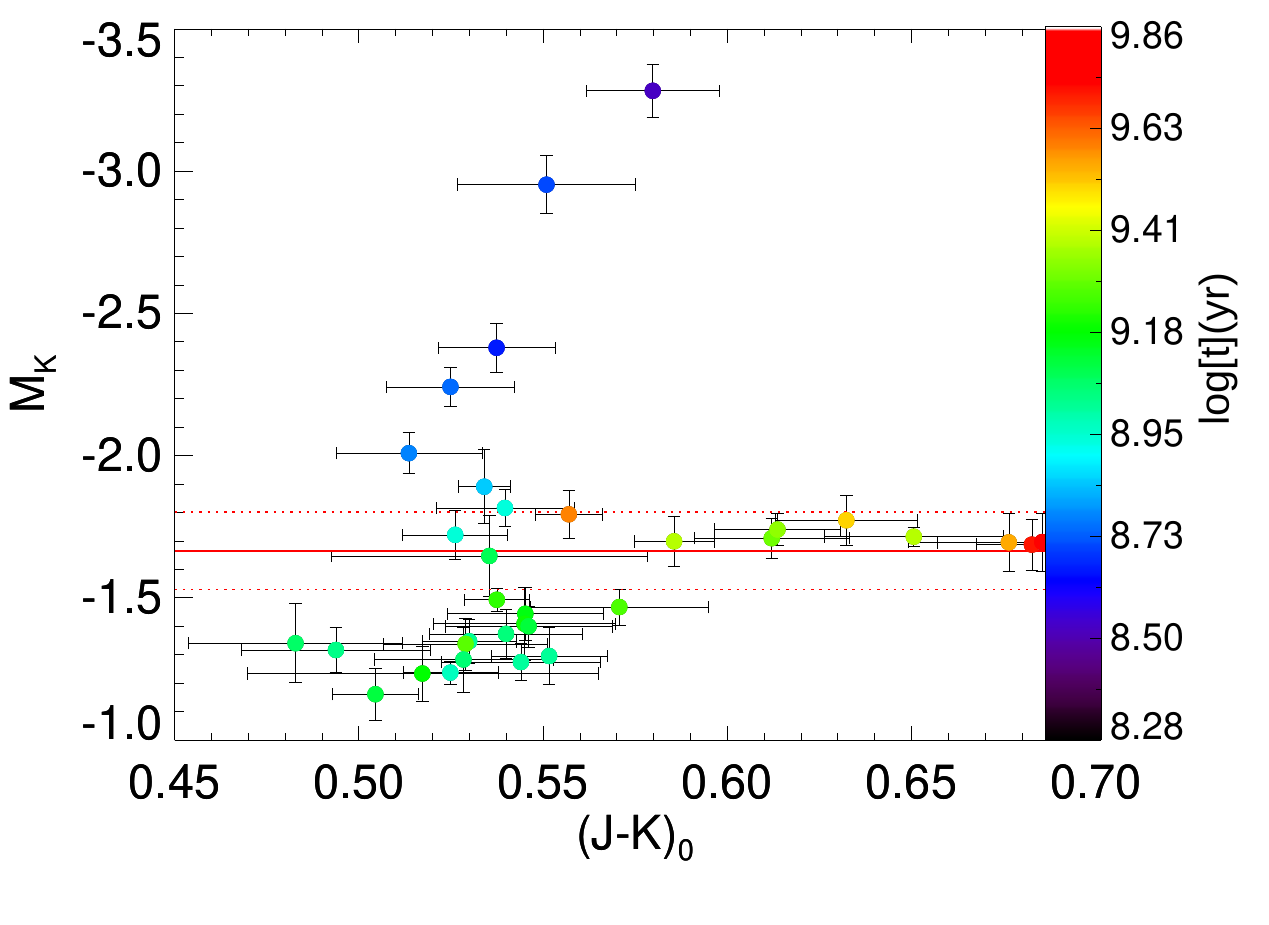}
\includegraphics[width=0.32\linewidth,trim={0.3cm 0.2cm 0.5cm 0.5 cm}]{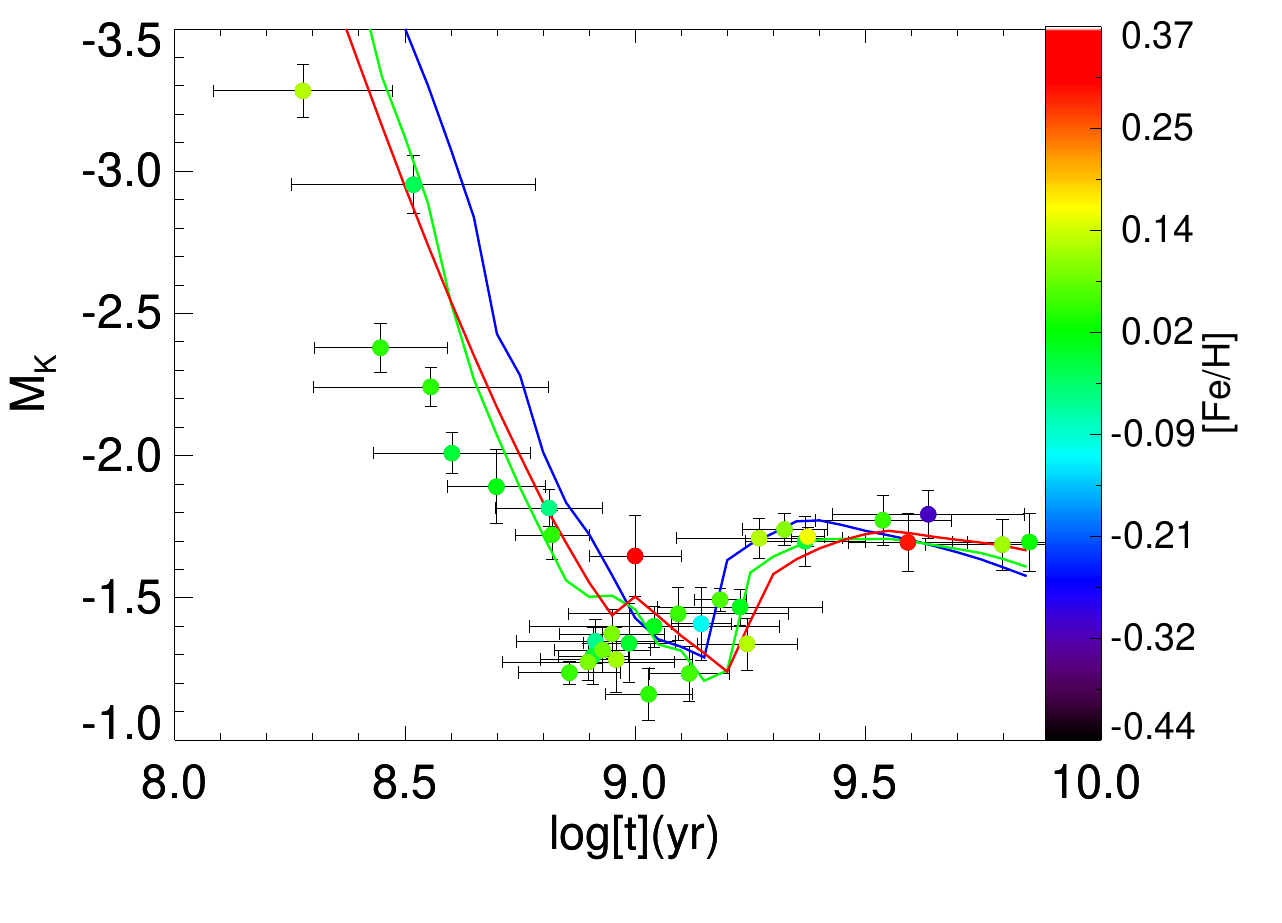}
\includegraphics[width=0.32\linewidth,trim={0.3cm 1.2cm 1.9cm 0.5 cm}]{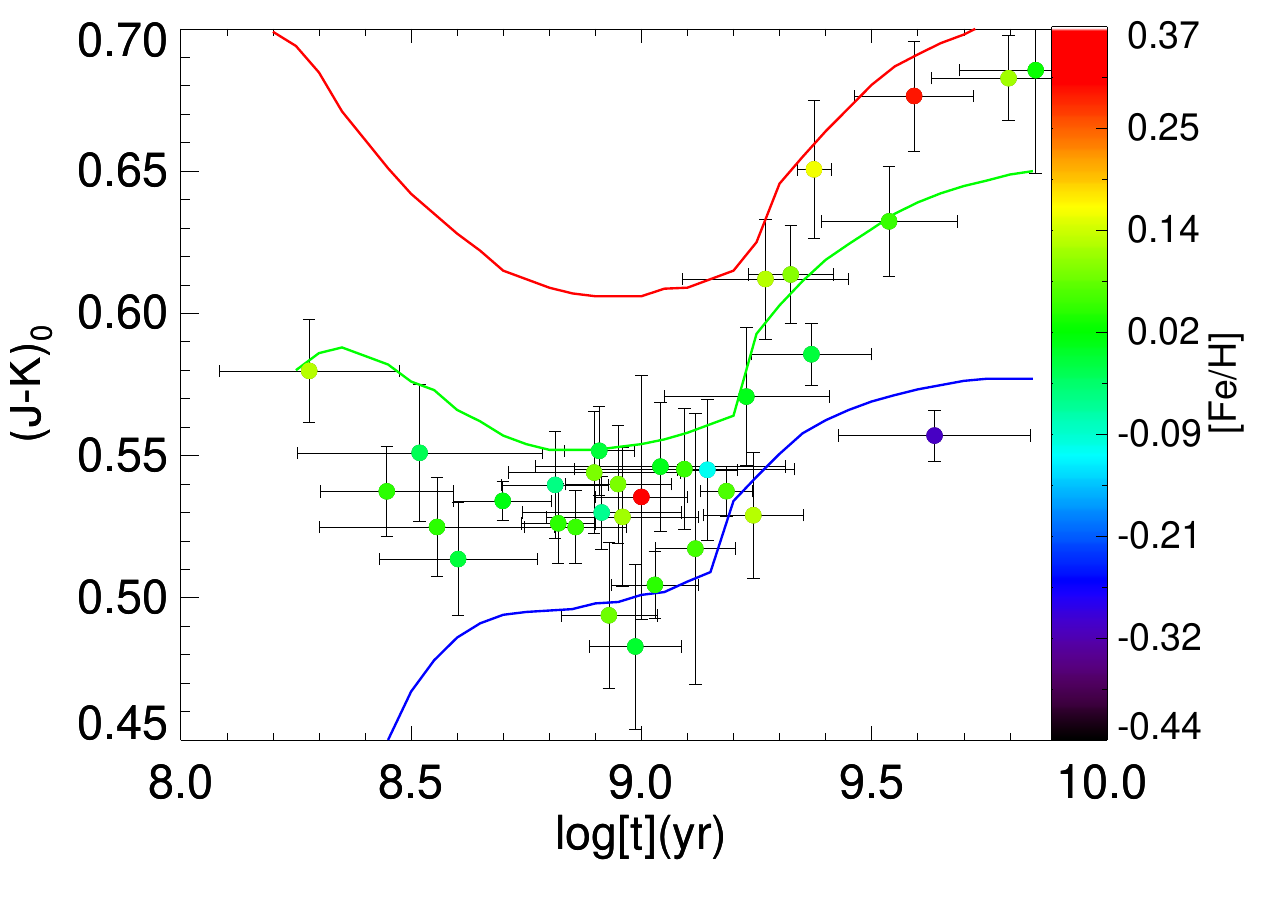}
\caption{RC colours and average magnitudes from our OC sample. Top panels: \textit{Gaia} DR3 bands. Bottom panels: 2MASS bands. The first column presents colour properties and mean magnitude in addition to the mean value (solid red line) and standard deviation (dotted red lines) of the clump magnitude for $\log[t({\rm yr})]>9.2$ and the colour bar represents the cluster age. The second column exhibits the relationship between age and average RC magnitude, while the third column of panels shows the relationship between age and average RC colour. For the second and third column of panels, the colour bars represent the metallicity and the lines represent the RC predictions according to PARSEC models for three different metallicity values: maximum metallicity of the cluster sample (red), minimum metallicity (blue) and solar metallicity (green). The error bars in absolute magnitudes and intrinsic colors were determined by error propagation.}
\label{fig:clumps_gigantes}
\end{figure*}

We verified the variation of the RC absolute magnitude and intrinsic colour as a function of age, as can be seen in Fig. \ref{fig:clumps_gigantes}. We noted that for both visible (\textit{Gaia}, upper panels) and infrared filters (2MASS, lower panels), RC stars tend to get significantly dimmer with age, within $8.3<\log[t({\rm yr})]<9.0$. Through this age interval, according to \textit{Gaia} filters, the RC colour index seems not to change significantly, on the other hand, according to the infrared filters, the RC tends to get hotter. From $\log[t({\rm yr})] \sim 9.0$ onwards, for both visible and infrared filters, the RC tend to become cooler and a slightly brighter until the age of $\log[t({\rm yr})] \sim 9.2$. For older OCs, the values of $M_{G}$ and $M_{K}$ tend to remain approximately constant, but $M_{K}$ is less affected by metallicity and exhibits a tighter distribution for older RC populations. On average the RC position tends to become cooler with age, in agreement with the PARSEC models and the analysis in \cite{2002AJ....123.1603G} for infrared bands.

%\cite{2019MNRAS.486.5600O}

 The main structures formed in observational HR diagrams of field stars have already been pointed out in the literature (see Fig. 10 from \cite{2018A&A...616A..10G}). For example, the secondary red clump (SRC) is a structure more extended in its bluest part towards fainter magnitudes than the RC. It tends to appear around $(G_{BP} - G_{RP})_{0} = 1.10$, $M_{G} = 0.60$ in observational \textit{Gaia} HR diagrams, which corresponds to younger more massive RC stars. Indeed, the RC magnitude and colour from our observed sample reflect such extended structure. In the left panels of Fig. \ref{fig:clumps_gigantes} we see this structure in \textit{Gaia} passbands around the expected position and in 2MASS passbands this structure is more remarkable and lies around $(J-K)_{0} = 0.53$, $M_{K} = -1.30$. According to our data, those RC stars are comprised in the age range of $8.80<\log[t({\rm yr})]<9.20$.

A bluer and vertical structure that is called the Vertical Red Clump (VRC) is also present, in which core-helium burning stars that are even more massive are more luminous than the RC and lie still on the blue part of it. This structure is present in the left panels of \ref{fig:clumps_gigantes}, for OCs younger than $\log[t({\rm yr})]=8.80$ where the absolute magnitude decreases strongly with age.

%trim={2.5 cm 2.5cm 2.5cm 2.5 cm}  trim={left bottom right top}

%\includegraphics[width=0.48\linewidth,trim={0.5cm 1.2cm 1.9cm 0.5 cm}]{fig/membership

%The characteristic of the second clump of SRC giants (Fig. 7 from \cite{2019MNRAS.486.5600O}), where the stars, in the infrared, present a luminosity disparity in $M_ {K}$ greater with respect to the average magnitude of the giant clump than visible with \textit{Gaia} data in $M_{G}$. The same can be seen in our data in Figure \ref{fig:clumps_gigantes} for objects between $8.8 \lesssim logt[t] \lesssim 9.2$, where the average position of the clumps is mainly in $M_{K} \sim - 1.2$ and , $M_{G} \sim 0.4$ which is probably the typical range of ages of stars observed in the SRC. On the other hand, the vertical giant clump or VRC (Figure \ref{fig:hr_observacional_gigantes}, Chapter \ref{cap:cap1}) is made up of objects below this age limit, presenting the position of the clump brighter than average , represented by clusters younger than $log[t] \sim 8.8$ in our sample.

In general, for both visible and infrared wavelengths, the values of $M_{G}$ and $M_{K}$ as a function of age are well represented by PARSEC isochrones (second column of panels in Fig. \ref{fig:clumps_gigantes}), especially for the older RC population \citep{2019MNRAS.486.5600O,2018A&A...609A.116R}. We note that the infrared models tend to present values of $M_{K}$ systematically brighter for objects younger than $\log[t({\rm yr})]\sim 9.0$, as seen in the Figs. 6 and 7 from \cite{2007A&A...463..559V}.

When comparing the RC colour indices (third column of panels in Fig. \ref{fig:clumps_gigantes}), we noticed that the nearly solar metallicity models represent well the population older than $\log[t({\rm yr})]\sim 9.0$. On the other hand, they are systematically redder than the observed younger OCs. The RC colours, according to PARSEC isochrones, also exhibit a minimum at $\log[t({\rm yr})]\sim 9.0$ for metallicities $[Fe/H] \gtrsim -0.2$, however this trend is better followed by infrared data compared with visible data (right panels of Fig. \ref{fig:clumps_gigantes}). This discrepancy with the models was already noted in \cite{1998ApJS..116..263P} and can also be observed in recent works with objects of age $\log[t({\rm yr})] \lesssim 8.8$ \citep{10.1111/j.1365-2966.2011.18627.x,10.1093/mnras/stac2496,10.1093/mnras/staa647,2019A&A...623A.108B}, in which the theoretical isochrones, when well fitted to the MS, do not pass through the RC stars, predicting a RC \textit{locus} systematically redder. Such discrepancies may be caused by a combination of effects such as: unresolved binary stars, mass distribution of giant parent clump stars (including mass loss), differential reddening or the presence of SRC stellar populations simultaneously with the RC for some objects, making the observed average RC value bluer. Alternatively, this may indicate that the models still have deficiencies in bolometric corrections, colour transformations and effective temperatures in this age range \citep{1999MNRAS.308..818G,2019ApJ...879...81A,2020AJ....159...96S}.

%Some differences cannot be explained by the error bars, for example, Trumpler 5 in our sample. Trumpler 5 is the object with the greatest heliocentric distance in our sample, whereas Trumpler 5 has smaller luminosity discrepancies in the infrared, which could mean an overestimated extinction value. 

Our sample of clusters has approximately solar metallicity, with an average value of $[Fe/H]=0.02$ and $\sigma_{[Fe/H]}=0.12$ dex, with few clusters outside this range. We note that OCs with metallicities that are more discrepant from the average tend to present large variations in RC colour, but more moderate variations in magnitude, mainly for the infrared, in agreement with the expectation that the average RC magnitude is a good indicator of distance even with variations in metallicity.
%There is a distribution of colour and magnitudes in the range $9<log[t]<9.2$, which, according to the models, is exactly the point at which RC magnitude reaches a minimum and varies very sensitively with age.

We also determined the average RC magnitude for our sample by adopting an interval of ages less affected by population effects. For this purpose, we used objects older than $\log[t({\rm yr})]= 9.2$, a similar value adopted in \cite{2002AJ....123.1603G} to estimate the mean RC value for $M_{K}$. We found $M_{G}=0.42 \pm 0.05$ and $ M_{K}=-1.66 \pm 0.04$, which is in good agreement with the literature (Table \ref{Tab:mag_clump_literatura}).

 % \begin{table}
%  \small
%\begin{tabular}{|l|r|r|r|}
%\hline
 %  \multicolumn{1}{|c|}{$source$} &
 %  \multicolumn{1}{c|}{$M_{G}$} &
 % \multicolumn{1}{c|}{$M_{K}$} \\
%\hline
%This work & $0.42 \pm 0.05$ & $-1.66 \pm 0.04$ \\
%  \hline
%\cite{2018A&A...609A.116R} & $0.495 \pm 0.009 $ & - \\
 % \hline
 % \cite{10.1093/mnras/stx1655} & $0.44 \pm 0.01 $ & $-1.61 \pm 0.01$ \\
 %  \hline
 % \cite{2007A&A...463..559V} & $- $ & $-1.57 \pm 0.05$ \\
%   \hline
%  \cite{2002AJ....123.1603G} & $ - $ & $-1.61 \pm 0.04 $ \\
%\hline\end{tabular}
%\caption{Average values calculated for $M_{G}$ and $M_{K}$.}
 %  \label{tab:mag_clump_literatura}
%\end{table}

  \begin{table}
\caption{RC average values calculated for $M_{G}$ and $M_{K}$ from this work and from the literature: Ruiz-Dern et al. (2018) (RD2018), Hawkins et al. (2017) (H2017), van Helshoecht \& Groenewegen (2007) (vH2007) and Grocholski \& Sarajedini (2002) (Gro2002).}
  \small
\begin{tabular}{|l|r|r|r|r|}
\hline
   \multicolumn{1}{|c|}{Source} &
   \multicolumn{1}{c|}{$M_{G}$} &
  \multicolumn{1}{c|}{$M_{K}$} \\
   %\multicolumn{1}{c|}{$Method$} \\
\hline
  This work &$ 0.42 \pm 0.05 $&$ -1.66 \pm 0.04$  \\
  \hline
  RD2018 & $0.495 \pm 0.009$ & -   \\
  \hline
  H2017  & $0.44 \pm 0.01 $ & $-1.61 \pm 0.01$  \\
  \hline
  vH2007  & -& $-1.57 \pm 0.05$  \\
  \hline
   Gro2002 & - & $-1.61 \pm 0.04$  \\
  \hline

\end{tabular}
   \label{Tab:mag_clump_literatura}
\end{table}

 \subsection{Morphological age indices} % \textcolor{red}{and evolution with age}}
 \label{subsection:deltas}
 
 In this section, we investigated how turnoff-RC differences relate to age and metallicity. Using the quantities determined in Section \ref{sect:CMD}, we determined the indices $\Delta G$ and $\Delta K$ as the difference in magnitude between the RC and the turnoff, as well as the colour difference $\Delta BR$ and $\Delta JK$ for both \textit{Gaia} and 2MASS passbands, respectively.
 
 % The errors established for ∆G, ∆(G BP −G RP ) 0 , ∆K, ∆(J − K) 0 and ∆V were calculated by summing in quadrature from the values of individual errors in magnitudes and colors of the turnoff and clump of giants.

\begin{figure*}
\centering
\includegraphics[width=0.31\linewidth,trim={0.7cm 0.2cm 0.5cm 0.5 cm}]{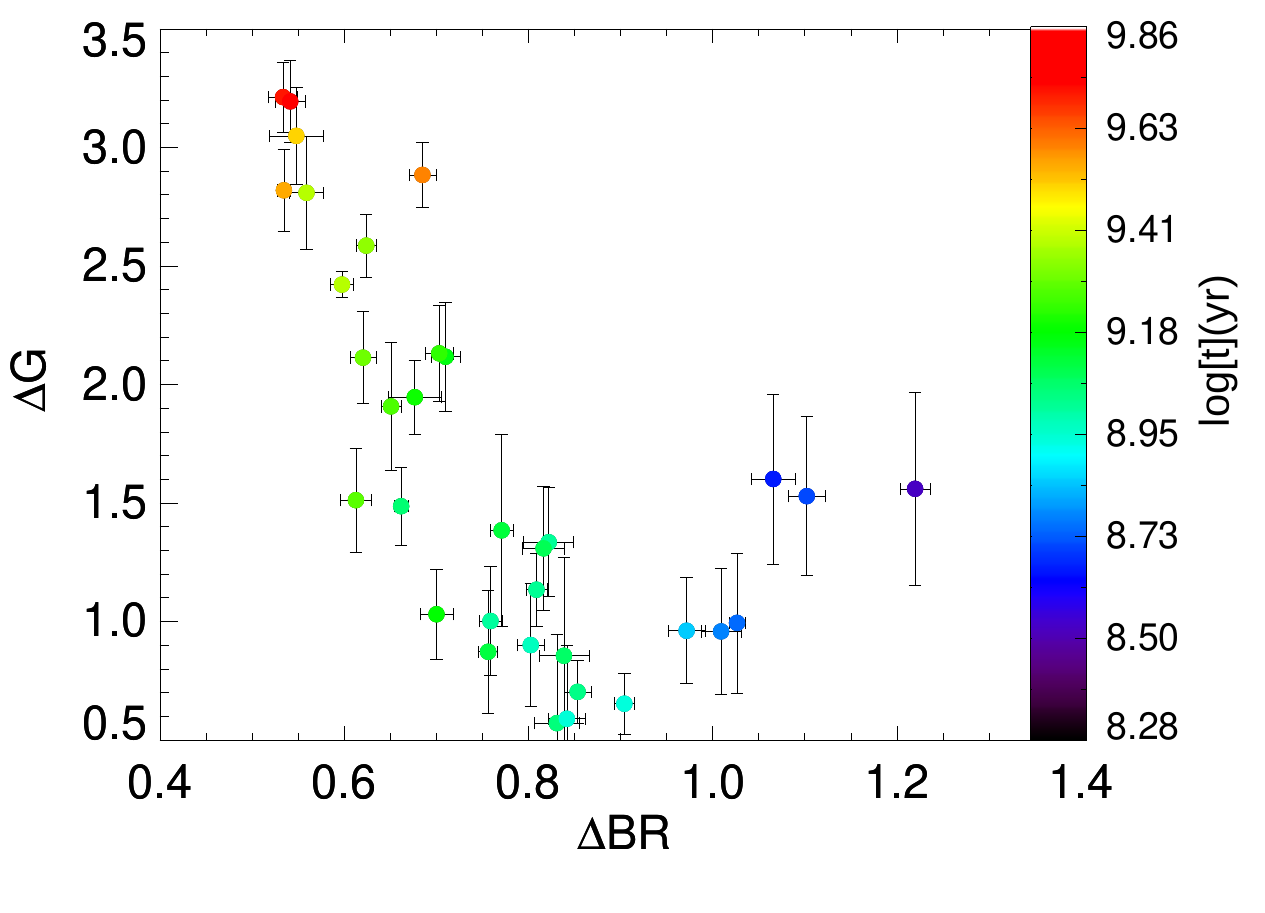}
\includegraphics[width=0.33\linewidth,trim={0.3cm 0.2cm 0.5cm 0.5 cm}]{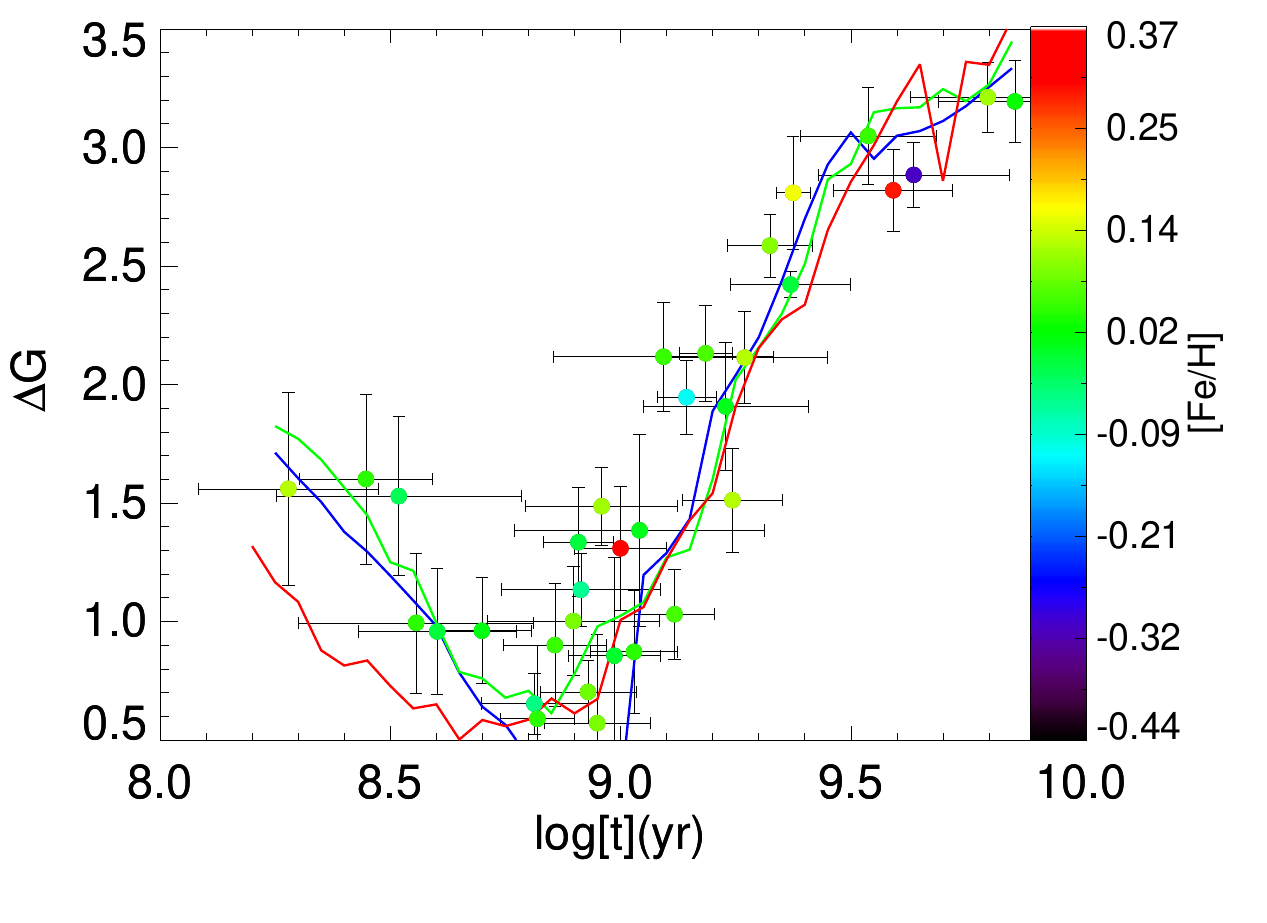}
\includegraphics[width=0.30\linewidth,trim={0.3cm 0.2cm 1.9cm 0.5 cm}]{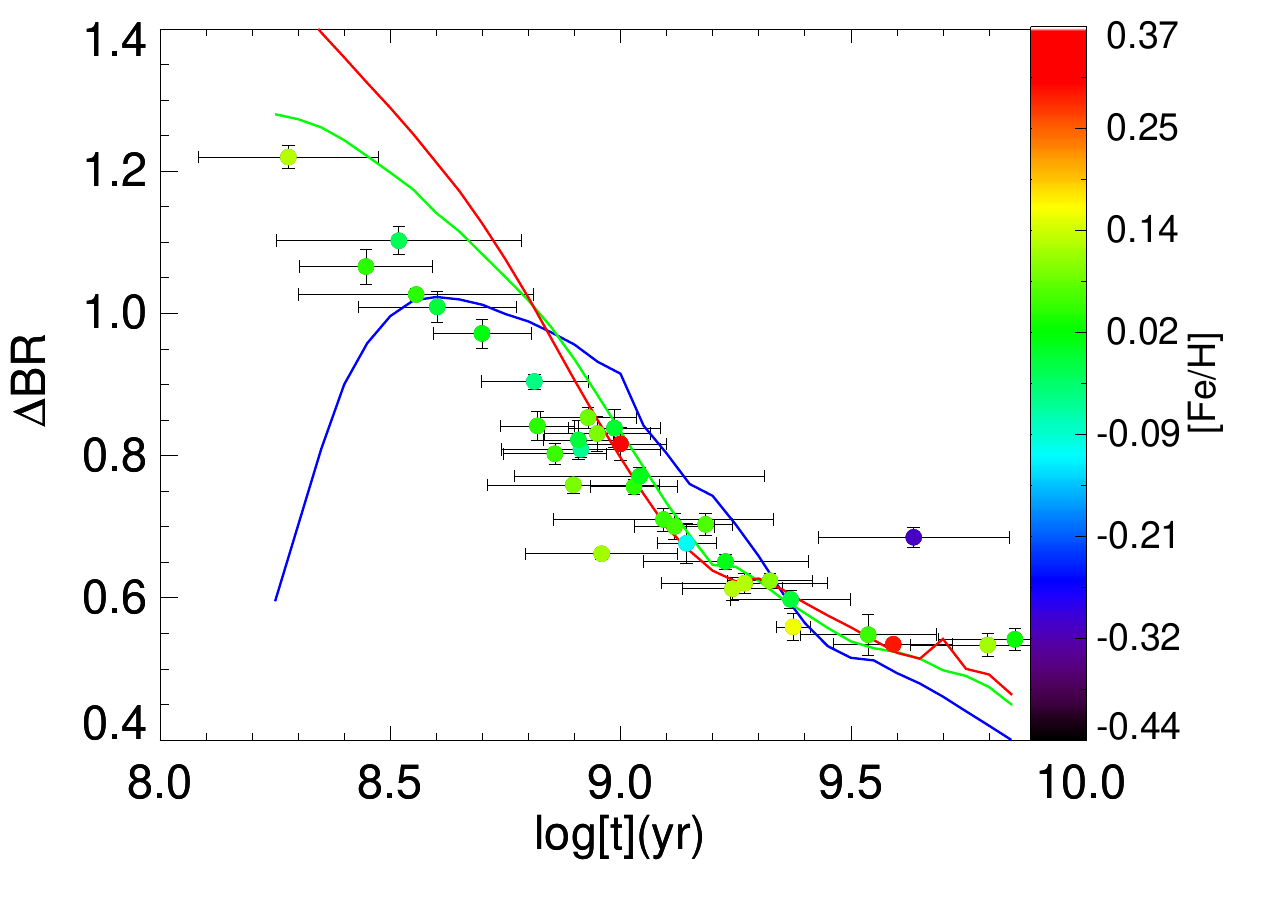}

\includegraphics[width=0.32\linewidth,trim={0.7cm 0.2cm 0.5cm 0.5 cm}]{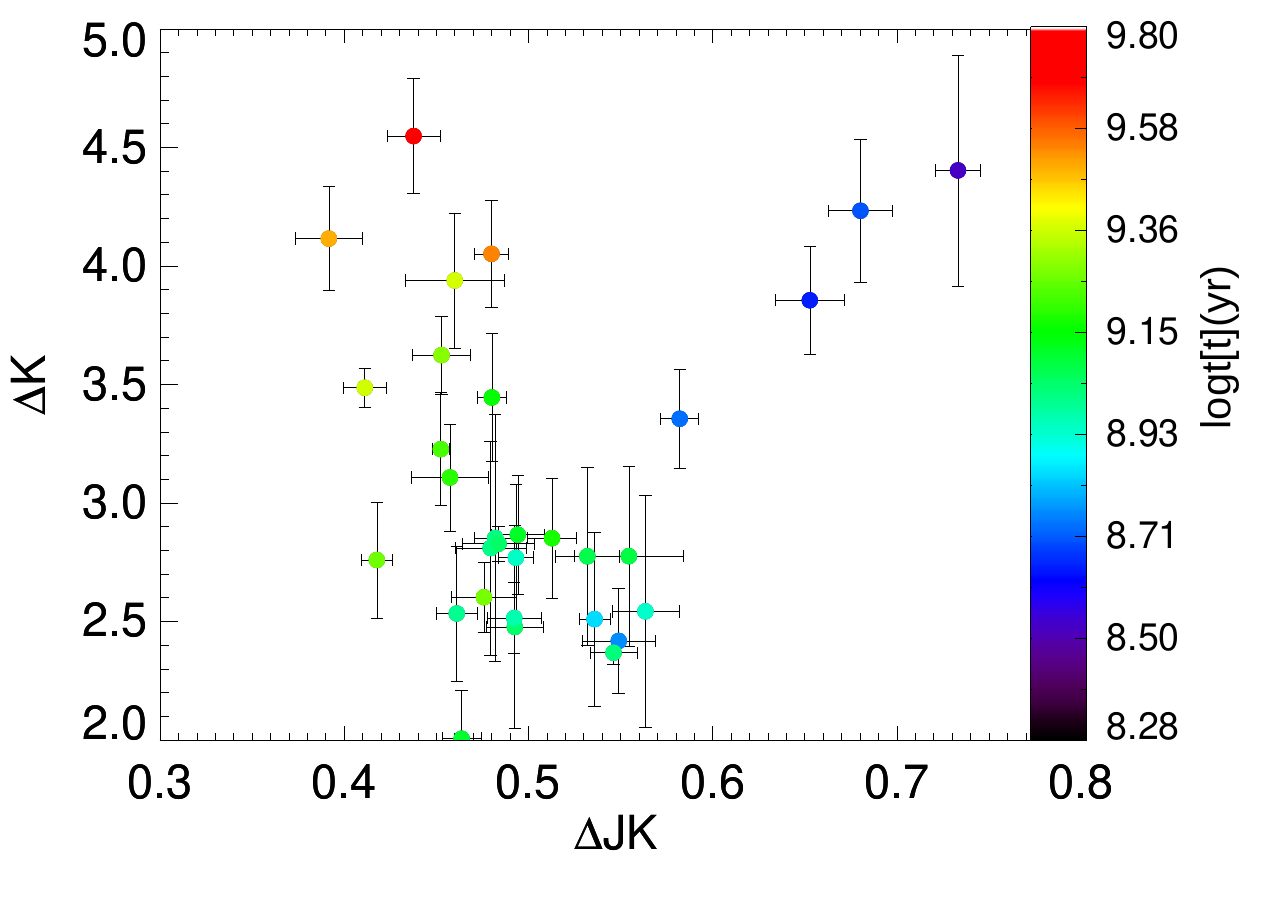}
\includegraphics[width=0.32\linewidth,trim={0.3cm 0.2cm 0.5cm 0.5 cm}]{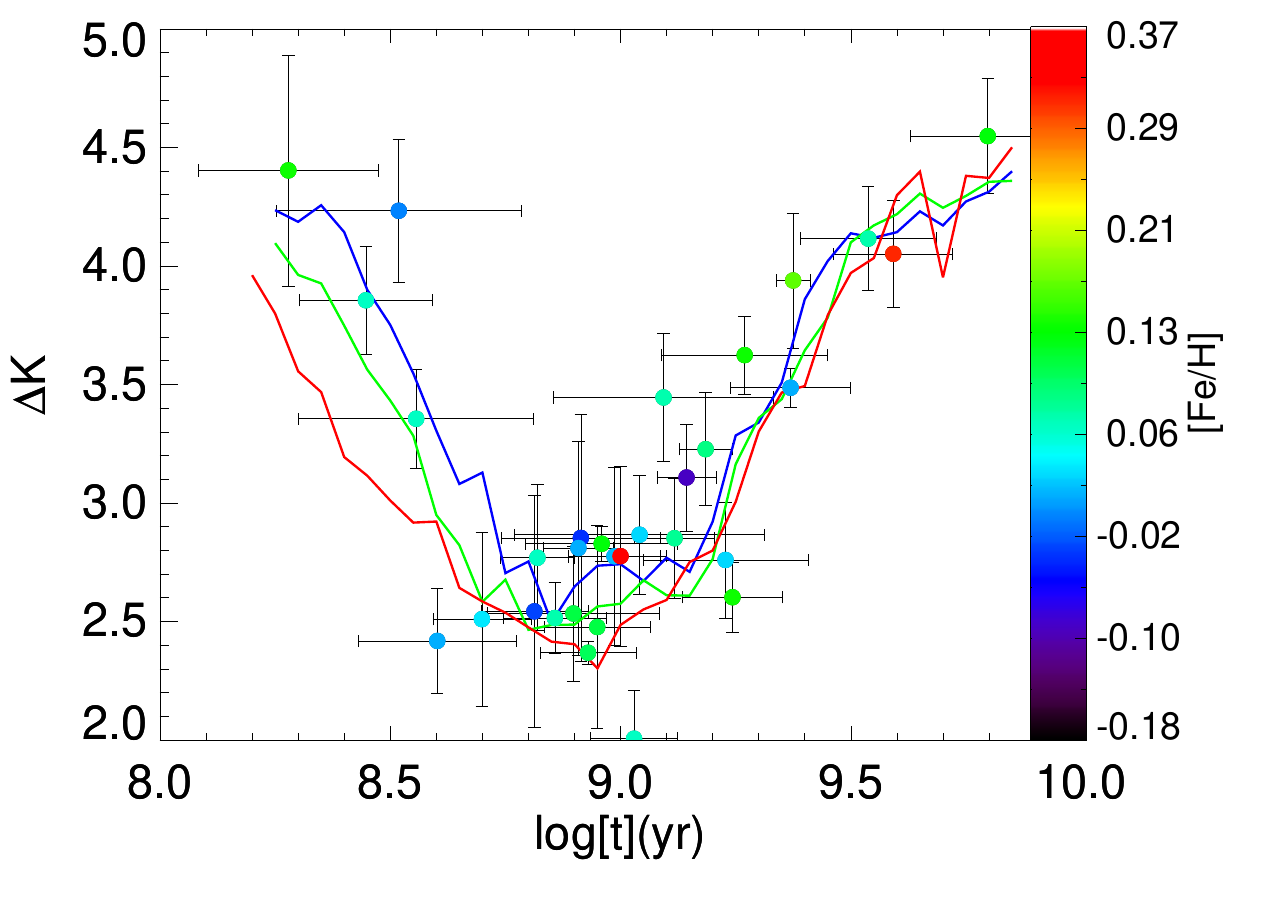}
\includegraphics[width=0.32\linewidth,trim={0.3cm 1.2cm 1.9cm 0.5 cm}]{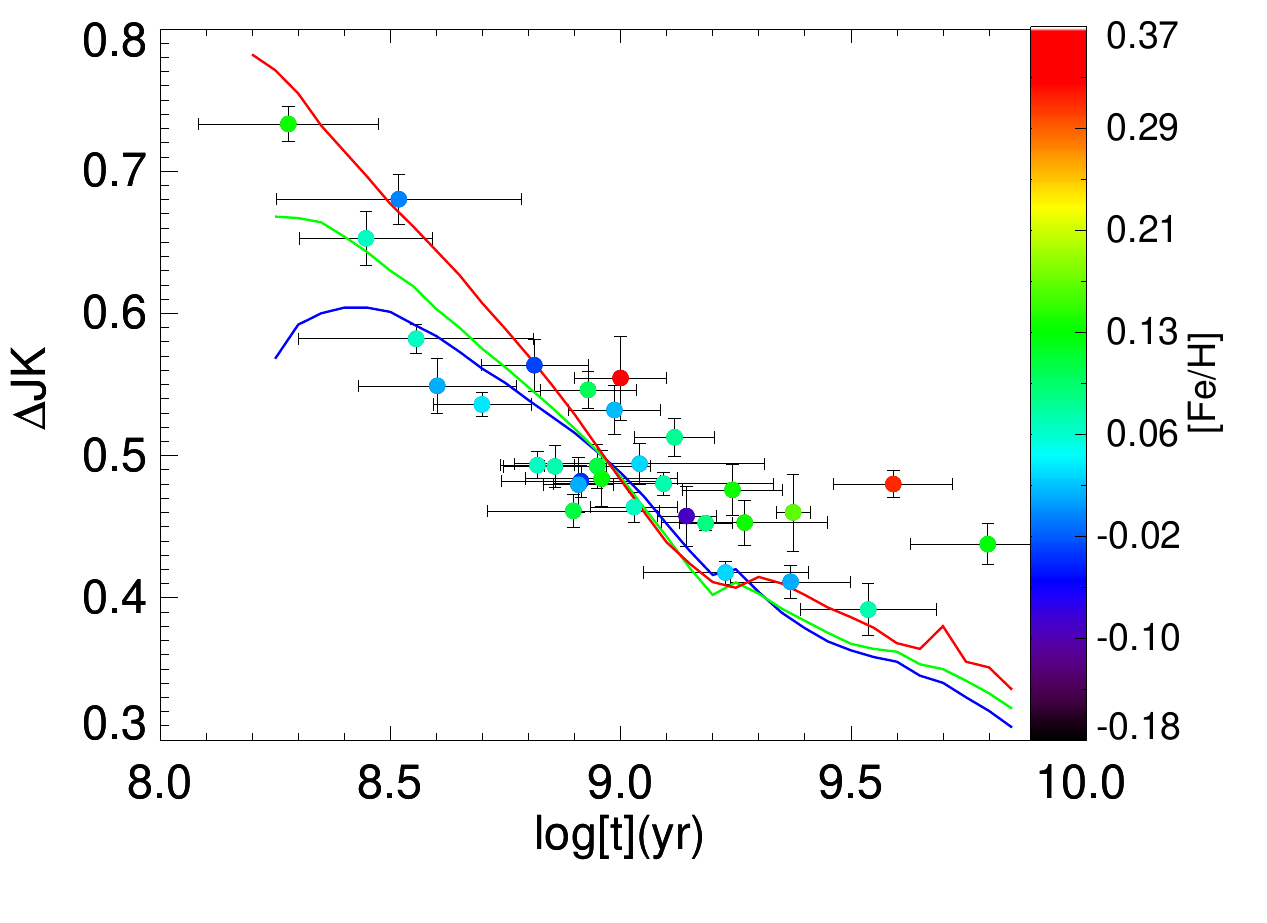}

\caption{Age evolution of morphological colour indices. Top panels: \textit{Gaia} DR3 indices. Bottom panels: Same, but for 2MASS indices. The first column presents the differences in colour and magnitude from the RC and turnoff positions, the colour bar represents age. The second column represents the relationship between the indices $\Delta G$ versus age (top) and  $\Delta K$ versus age (bottom), while the third column represents the relationship between the indices $\Delta BR$ versus age (top) and $\Delta JK$ versus age (bottom). For the second and third column of panels the colour bars represent the metallicity and the lines indicate the indices predictions according to PARSEC models for three different metallicities: maximum metallicity of the cluster sample (blue), minimum metallicity (red) and solar metallicity (green). The error bars are also represented.}
\label{fig:indice_morf}
\end{figure*}

The top panels of Fig. \ref{fig:indice_morf} show how the indices $\Delta G$ and $\Delta BR$ are related with age and metallicity and the bottom ones illustrate the same relations for $\Delta K$  and  $\Delta JK$ indices. It is evident that there appear to be two approximate linear relationships of $\Delta G$ with $\log[t({\rm yr})]$: one for objects with ages younger than $\log[t({\rm yr})] \sim 8.8$ and another for objects older than this limit. Our older OCs present a similar linear trend to that determined by \cite{1994AJ....107.1079P} for the indices in the planes $\delta_{1}(BV)$ vs $\delta V$ and $\delta_{1}(VI)$ vs $\delta V$. 

At least in the range spanned by OCs in our sample, metallicity does not appear to affect strongly the $\Delta G$ values for $\log[t({\rm yr})]>9.0$, which has already been verified by other authors \citep{1994AJ....107.1079P,1994A&A...287..761C,2004A&A...414..163S,2009A&A...508.1279B}. This effect is confirmed by the overplotted PARSEC models.%, especially for older OCs. %Despite the models presenting a stronger trend in magniude with metalicity, our younger OCs are present with approximately solar metallicity, thus we are not able to point out this trend in our data.  %(Phelps et al 1994, Carraro et al. 2004, Salaris et al. 2004, Beletsky et al. 2009).

According to our data and the PARSEC models, the index $\Delta BR$ also shows little dependence on metallicity for objects older than $\log[t({\rm yr})] \sim 8.8$. The models indicate a more significant dependence on metallicity for younger objects (see Fig. \ref{fig:indice_morf}, top-right panel).

In general, the models show good agreement with the data on the relation $\Delta G$ versus $\log[t({\rm yr})]$. Regarding the values of $\Delta BR$, apparently for younger objects ($\log[t({\rm yr})]< 8.8$), the isochrones tend to overestimate this index, which is related to the prediction of redder RC in this age interval (see Sect. \ref{subsection:RC}). %The models also appear to fail to predict the values of $\Delta BR$ for objects older than $\log[t({\rm yr})]=9.5$.

%Regarding the infrared indices $\Delta K$ and $\Delta JK$ (bottom panels in Fig. \ref{fig:indice_morf}), it is possible to note trends analogous to those for \textit{Gaia} indices, but both indices present a wider distribution when compared with \textit{Gaia} ones due a noiser photometry. In general, PARSEC models show good agreement with the data for the relation $\Delta K$ versus $\log[t({\rm yr})]$. However, the index $\Delta JK$ seems to be much more affected by the 2MASS colour indices errors, especially for older OCs. The difference $\Delta JK$ of our OCs exhibits a range of $\sim 0.35$ mag (considering the overall age interval), although the average error of the colour index $(J-K)$ for our clusters members are about $\sim 0.05$ mag and the same errors near the 2MASS observational limit exceed $0.10$ mag, which leads to the turnoff being bluer than expected. Therefore, we observed a high dispersion on that index, especially for older objects ($\log[t({\rm yr})]\gtrsim 9.0$),
%which present a distribution within $0.4 <\Delta JK< 0.5$.  We did not see the same problem with \textit{Gaia} photometry with the index $\Delta BR$, because this index from our OCs star members spreads over a wider range ($\sim 1$ mag), where the mean colour index $(G_{BP}-G_{RP})$ error is about $\sim 0.008$ mag and only $4\%$ of the stars present errors as high as $0.04$ mag, in other words this index is not very affected by photometric errors.

Regarding the infrared indices $\Delta K$ and $\Delta JK$ (bottom panels in Fig. \ref{fig:indice_morf}), it is possible to note trends analogous to those for \textit{Gaia} indices, but both indices present a wider distribution when compared with \textit{Gaia} ones due the larger photometric errors. In general, PARSEC models show good agreement with the data for the relation $\Delta K$ versus $\log[t({\rm yr})]$. However, the index $\Delta JK$ seems to be much more affected by the 2MASS colour indices errors, especially for older OCs ($\log[t({\rm yr})]\gtrsim 9.0$), which present a distribution within $0.4 <\Delta JK< 0.5$. The difference $\Delta JK$ of our OCs exhibits a range of $\sim 0.35$ mag (considering the overall age interval), although the average error of the colour index $(J-K)$ for our clusters members are about $\sim 0.05$ mag. We did not see the same problem with \textit{Gaia} photometry with the index $\Delta BR$, because this index from our OCs star members spreads over a wider range ($\sim 1$ mag), where the mean colour index $(G_{BP}-G_{RP})$ error is about $\sim 0.008$ mag, in other words this index is not very affected by photometric errors.

%Pode-se notar que todos os CMDs presentes em [Beletsky, Carraro & Ivanov 2009] possuem amostras consideráveis de estrelas mais azuis do que o ponto de turnoff calculado, porém, diferente do nosso método, os pontos de turnoff e valores médios de M K foram visualmente apontados pelos autores. Dessa forma, uma análise da diferença de cor entre o turnoff e o clump de gigantes, em regimes de altos erros fotométricos relativos, torna-se imprecisa. Porém, como veremos na próxima seção, a diferença ∆K ainda pôde ser avaliada com precisão. 

 % \begin{figure}
%  \includegraphics[width=0.31\linewidth]{fig/output_gaia_Delta_mag_cor_idade.eps}

 %\includegraphics[width=0.48\linewidth]{fig/FilipeFerreira2Fig17.eps}

 %\includegraphics[width=0.48\linewidth]{fig/FilipeFerreira2Fig18.eps}

 %\caption{Left: Relation between $\DeltaBR$ and $\log[t({\rm yr})]$. The colour bar represent the metalicity [Fe/H]. The lines represent the values from PARSEC isochrones, where the colours represent the maximum (red), minimum (blue) and solar (green) values of $[Fe/H]$. Right: The same, but for the $\Delta G$ versus $\log[t({\rm yr})]$ relation.}

 %\label{fig:indice_morf}

 %\end{figure}

\begin{figure}
\centering
\includegraphics[width=0.90\linewidth]{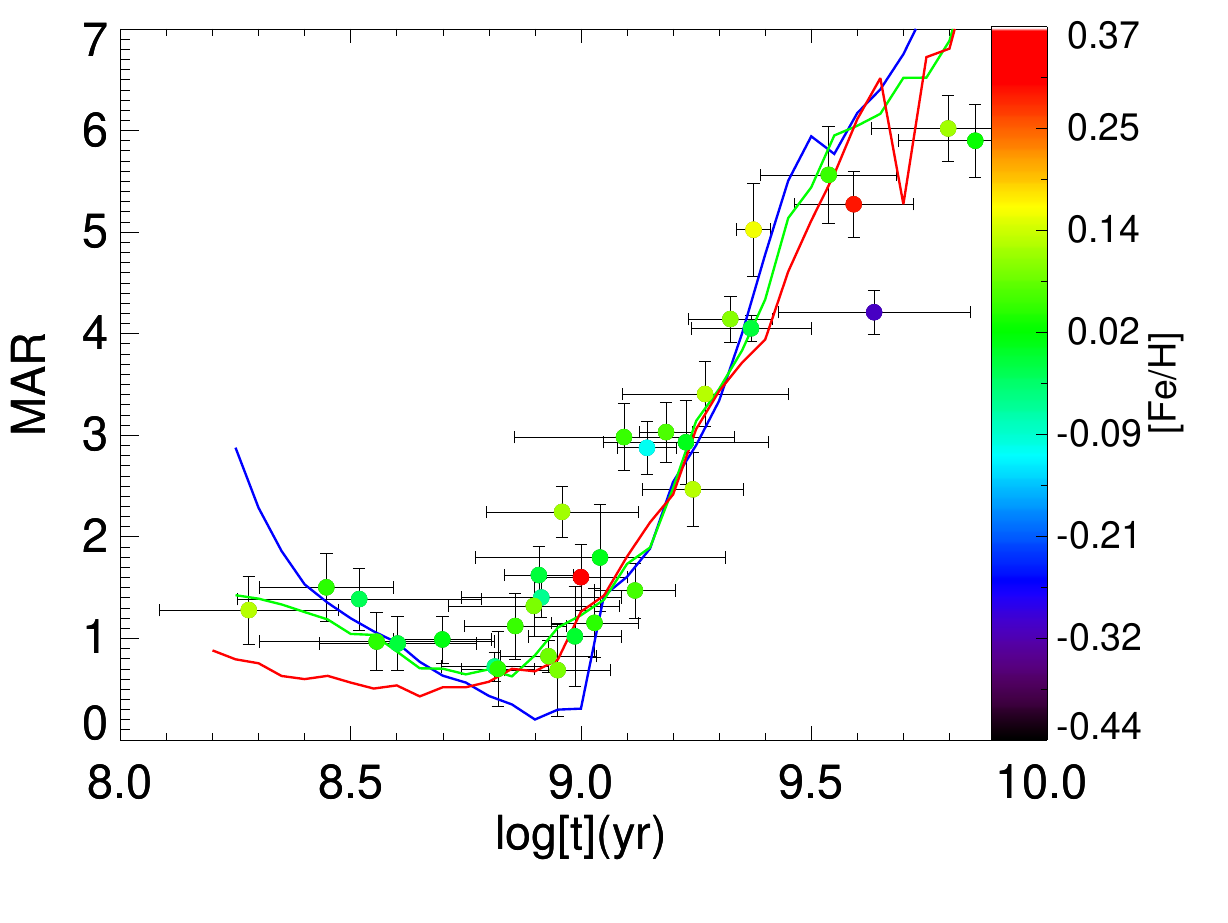}

\caption{The evolution of the morphological age index MAR. The colour bar represents the metallicity. The lines represent the same index predicted by  PARSEC models for three different metallicity values: maximum metallicity of the cluster sample (red), minimum metallicity (blue) and solar metallicity (green). The error bars were determined by propagating the errors from the ratio $\Delta G/ \Delta BR$.}
\label{fig:MAR}
\end{figure}

We also calculated a similar age index to MAR, defined in \cite{1985ApJ...291..595A}, taking the ratio $\Delta G$/$\Delta BR$. We do not define the same index for 2MASS bands due to the scatter observed in index $\Delta JK$. The same ratio was determined for PARSEC isochrones and a comparison with data is shown in Fig. \ref{fig:MAR}. Despite the discrepancies of the index $\Delta BR$ for younger OCs, the index MAR is very well reproduced by PARSEC isochrones, with the exception of OCs older than $\log[t({\rm yr})]>9.5$. In \cite{1985ApJ...291..595A} a linear correlation with age is obtained from MAR (ratio $\Delta V$/$\Delta (B-V)$) for objects older than 2 Gyr ($\log[t({\rm yr})]=9.3$), however, our work shows that this relation is clearly non-linear, especially for objects younger than 1 Gyr and older than 3 Gyr.

 \subsection{Age calibrations}
 
\begin{figure}
\includegraphics[width=0.85\linewidth]{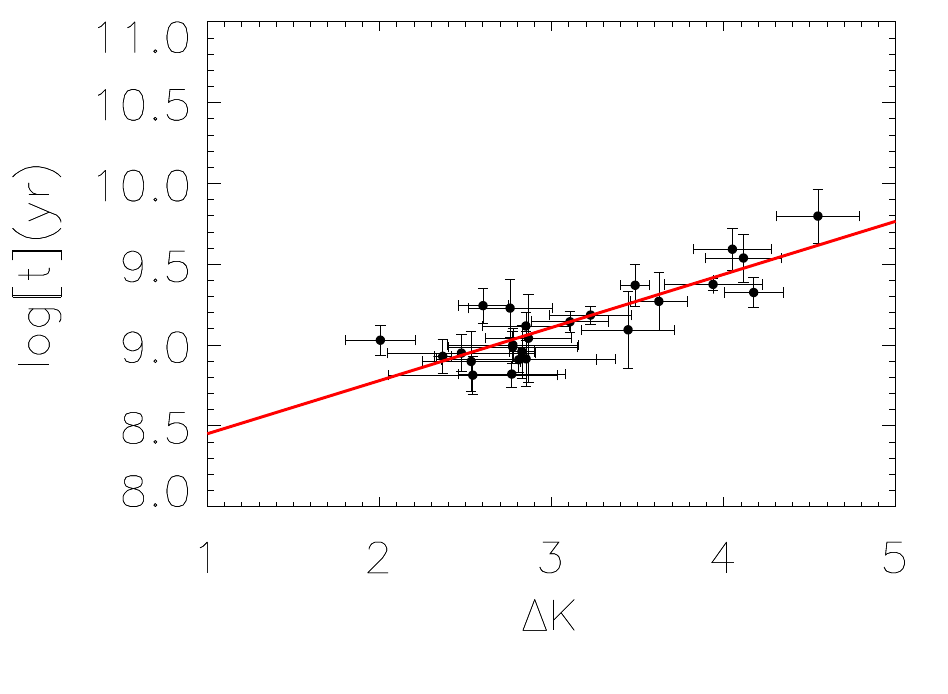}
\includegraphics[width=0.85\linewidth]{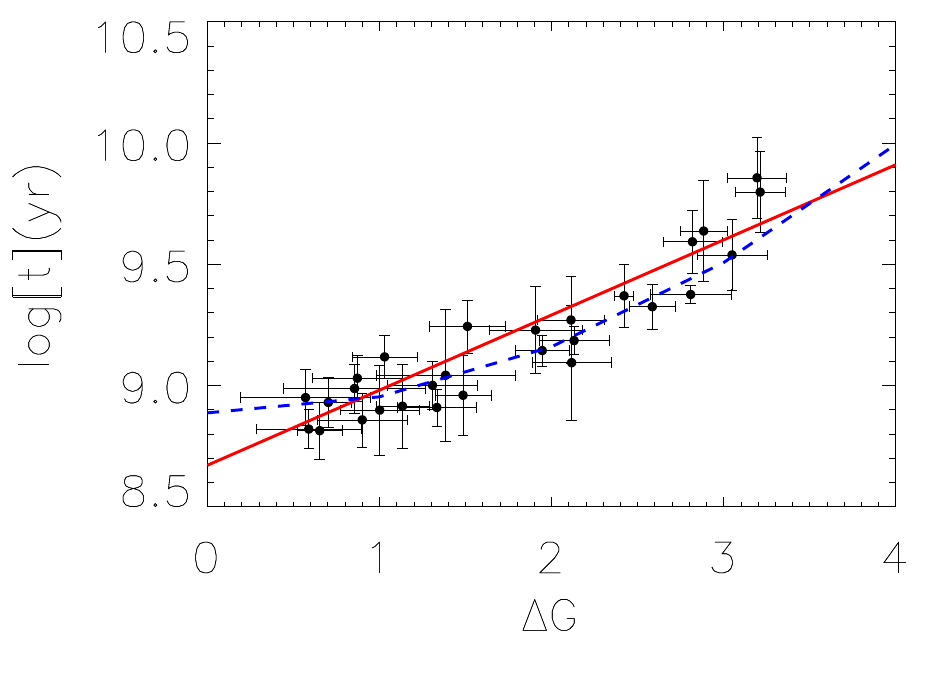}
\caption{Top: Linear (red line) relation of $\log[t({\rm yr})]$ versus $\Delta K$. Bottom: Linear (red line) and quadratic (blue dotted line) of the relation $\log[t({\rm yr})]$ versus $\Delta G$.}
\label{fig:fit_delta_k_g}
\end{figure}

As far as we are aware, age calibrations for \textit{Gaia} data ($\Delta G$ and $\Delta BR$ indices) are not yet available in the literature, so here we will provide them for both \textit{Gaia} morphological indices for the first time. We also performed calibrations for the $\Delta K$ index as a benchmark. In Fig. \ref{fig:indice_morf} it is possible to note that both indices $\Delta G$ and $\Delta K$ increase with age for OCs older than $ \log[t({\rm yr})] = 8.8$, for which we determined calibrations (Eqs. \ref{eq:eq_delta_G} and \ref{eq:eq_delta_K}) similar to those present in \cite{2009A&A...508.1279B}. In addition, we determined two more calibrations for $\Delta G$: quadratic in $\Delta G$ (Eq. \ref{eq:eq_delta_G_quadra}) and a relation that takes into account the metallicity (Eq. \ref{eq:eq_delta_G_metal}). Fig. \ref{fig:fit_delta_k_g} summarizes all these calibrations. Mean residuals, correlation coeficients and the age range of the calibration equations established in this work are present in Table \ref{tab:coef_quali_eqs}.

% \begin{table*}[h] 
% \centering
% \small
%\begin{tabular}{|l|r|r|r|r|r|r|}
%\hline
%\multicolumn{1}{|c|}{} &
%\multicolumn{1}{c|}{$a$} &
%%\\multicolumn{1}{|c|}{$\delta_{a}$} &
%\multicolumn{1}{c|}{$b$} &
%\\multicolumn{1}{|c|}{$\delta_{b}$} &
%\multicolumn{1}{c|}{$c$} &
%\\multicolumn{1}{|c|}{$\delta_{c}$} &
%\multicolumn{1}{c|}{$d$} &
%\\multicolumn{1}{|c|}{$\delta_{d}$} &
%\multicolumn{1}{c|}{$e$} &
%\\multicolumn{1}{|c|}{$\delta_{e}$} &
%\multicolumn{1}{c|}{$f$} \\
%\multicolumn{1}{|c|}{$\delta_{f}$}   \\
%\hline
%Variable &$log[t]_{0}$& $\Delta G $&$(\Delta G)^{2}$&$  \Delta BR$&$ (\Delta BR)^{2}$&$ [Fe/H]$ \\
%\hline
%Coef &$9.22 \pm  0.33 $    & $-0.15  \pm  0.07$ &    $  0.09 \pm 0.02$ & $  0.77  \pm 1.46$ &$-1.28  \pm 0.80 $&$ -0.13\pm 0.13 $\\
%\hline
%\end{tabular}
%\caption{Coefficients of the relationship established in Equation \ref{eq:eq_dela_G_delta_bprp}.}
%  \label{tab:coe_eq_quadratica}
%\end{table*}

%  \begin{figure}
%\includegraphics[width=0.85\linewidth]{fig/ajustes/Gaia_delta_mag_quadratico.pdf}
%\caption{Top: Linear (red line) and quadratic (blue dotted line) of the relation $\Delta G$ versus $\log[t({\rm yr})]$.}
%\label{fig:fit_delta_g}
%\end{figure}

 \small

\begin{equation}
\label{eq:eq_delta_G}
 \log[t(yr)]=8.63(\pm 0.05)+0.31(\pm0.03)\times \Delta G
 \end{equation}

 \begin{equation}
\label{eq:eq_delta_K}
 \log[t(yr)]=8.12(\pm 0.13)+0.33(\pm0.04)\times \Delta K
 \end{equation}

 \begin{equation}
\label{eq:eq_delta_G_quadra}
 \log[t(yr)]=8.89 (\pm 0.10)-0.01(\pm 0.13)\times \Delta G + 0.07(\pm0.04)\times (\Delta G)^{2}
 \end{equation}
 
\begin{equation}
\label{eq:eq_delta_G_metal}
 \log[t(yr)]=8.63(\pm 0.05)+0.31(\pm0.03)\times \Delta G -0.02 (\pm0.15) \times [Fe/H]
 \end{equation}
\normalsize

 \begin{table}
 \small
 \caption{Mean residuals and correlation coeficients of the calibration equations established in this work. The age range in which each equation are calculated, and therefore applicable are also indicated.}
\begin{tabular}{|l|r|r|r|}
\hline
\multicolumn{1}{|c|}{Eq.} &
\multicolumn{1}{c|}{Residuals} &
\multicolumn{1}{c|}{Correlation} &
\multicolumn{1}{c|}{Validity range} \\
\hline

\ref{eq:eq_delta_G} & 0.09  & 0.92 & $ 8.8<\log[({\rm yr})] <9.9$  \\
\hline
\ref{eq:eq_delta_K} &  0.10  & 0.85  & $ 8.8<\log[({\rm yr})] <9.9$ \\
\hline
\ref{eq:eq_delta_G_quadra} & 0.08     & 0.95 & $8.8<\log[({\rm yr})] <9.9$  \\
\hline
\ref{eq:eq_delta_G_metal} & 0.09  & 0.92 & $8.8<\log[({\rm yr})] <9.9$  \\
\hline
\ref{eq:eq_delta_G_delta_bprp} & 0.06   & 0.98 & $8.3<\log[({\rm yr})]<9.9 $\\
\hline
\end{tabular}
  \label{tab:coef_quali_eqs}
\end{table}
 
 Regarding the age calibration for the $\Delta K$ index (Eq. \ref{eq:eq_delta_K}), we see a very good agreement with \cite{2009A&A...508.1279B}, showing that our method provides similar results in comparison with those obtained by visual inspection of the CMDs, which validates its application on \textit{Gaia} data. As seen in Sect. \ref{subsection:deltas} and already evidenced by other authors for visible and infrared morphologycal age indices $\Delta V$ and $\Delta K$ \citep{1994AJ....107.1079P,1994A&A...287..761C,2004A&A...414..163S,2009A&A...508.1279B}, the dependence of the index $\Delta G$ with metallicity is small (see Fig. \ref{fig:indice_morf}), so that in many cases the metallicity term can be neglected. As accurate metallicity determinations are scarce in the literature, equations \ref{eq:eq_delta_G} and \ref{eq:eq_delta_G_quadra} become useful tools for determining ages of Galactic OCs whenever their turnoffs and RC positions are measured with good accuracy on their CMDs.

 \subsection{General formula}
  
\begin{figure*}
\includegraphics[width=0.85\linewidth]{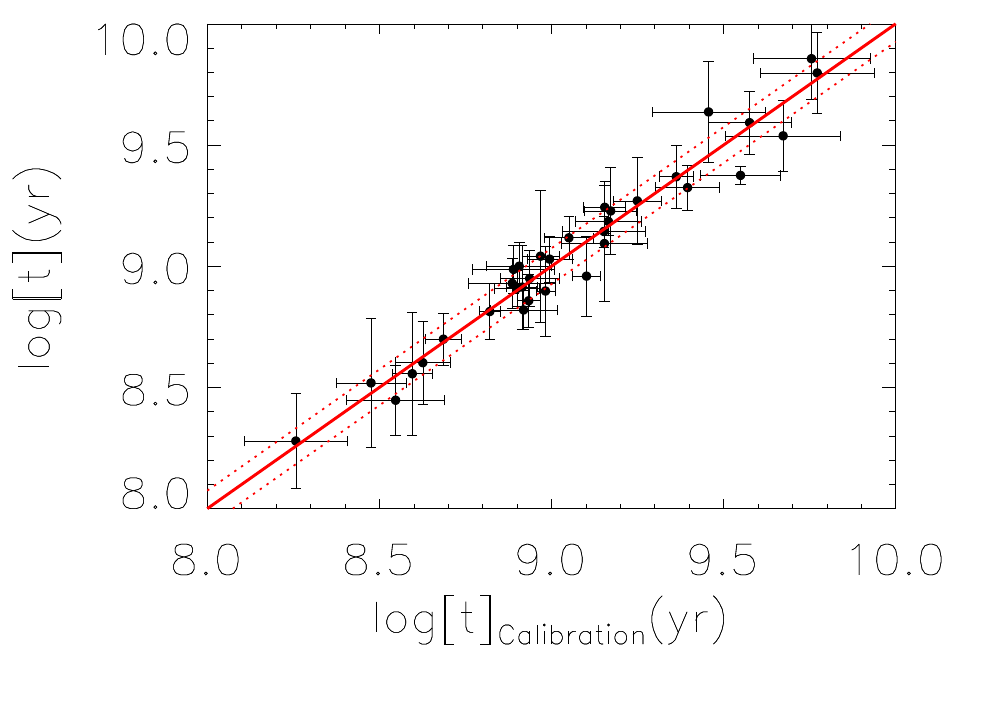}
\caption{Relation between ages of OCs from the literature as function of their ages determined from Eq. \ref{eq:eq_delta_G_delta_bprp}. The continuous line gives a linear fit to the relation and the dashed lines correspond to its $1-\sigma$ uncertainties. The relation is given by $\log[t({\rm yr})]=A\times logt_{fit}+B$, where A=1.00 and B=0.00. The error bars in the X axis were determined by error propagating from Eq. \ref{eq:eq_delta_G_delta_bprp} and the Y axis are erros in $\log[t({\rm yr})]$ from the literature.}
\label{fig:fig_fit_delta_G_delta_bprp}
\end{figure*}

 As seen in Fig. \ref{fig:indice_morf}, when we took into account the relation $\log[t(yr)]$ \textit{versus} $\Delta G$, we realize that objects with ages in the range $8.3 < \log[t(yr)] < 8.8$ may have the same $\Delta G$ values as those of $8.8< \log[t(yr)] < 9.5$. However, the parameter $\Delta BR$ has an almost linear dependence on age, i.e., younger objects exhibit greater $\Delta BR$ than the older ones. In this case, using the index $\Delta BR$ breaks the degeneracy of the $\Delta G$ values as an age indicator.  
  Thus, by extending the age calibration to the interval $\sim 8.3 < \log[t(yr)] < 9.0$, we were able to establish a fitting function with quadratic and linear terms in both indices $\Delta G$ and $\Delta BR$ (Eq. \ref{eq:eq_delta_G_delta_bprp} and Table \ref{Tab:coe_eq_quadratica}). A comparison of OCs ages from our sample with that determined using Eq. \ref{eq:eq_delta_G_delta_bprp} is shown in Fig.\ref{fig:fig_fit_delta_G_delta_bprp}, where we note small residuals. Therefore, Eq. \ref{eq:eq_delta_G_delta_bprp} represents an important tool to determine OCs ages within a wide range using \textit{Gaia} data.

%Table 9 shows the average residual values and correlation coefficients of the adjustments
%settled down.
%See bottom panel of Fig. \ref{fig:lin_fit}
\small 
\begin{equation}
\begin{aligned}
 & \log[t(yr)]=a+b\ \Delta G + c\ (\Delta G)^{2}+ d\ (\Delta BR)^{2}
 \end{aligned}
\label{eq:eq_delta_G_delta_bprp}
 \end{equation}
\normalsize

% 10.1-0.71*delta_G+0.17*delta_G*delta_G-0.45*delta_br-0.85*delta_br*delta_br+0.35*delta_br*delta_g-0.1*fe_h_1
%eixo x do topcat

\begin{table}
 \centering
 \caption{Coefficients of Eq. \ref{eq:eq_delta_G_delta_bprp}.}
 \small
\begin{tabular}{|l|r|}
\hline
Coef & Value \\
\hline
a&$9.64 \pm  0.34 $   \\
b&$-0.32  \pm  0.09$ \\
c&$  0.14 \pm 0.03$ \\
d&$  -0.81  \pm 0.06$ \\
\hline
\end{tabular}
  \label{Tab:coe_eq_quadratica}
\end{table}

\section{Conclusions}
\label{sect:concl}
%Neste trabalho fomos capazes de produzir uma metodologia uniforme para a caracterização de aglomerados abertos de estrelas a partir de dados precisos de astrometria e fotometria do Gaia DR2. Os objetos estão compreendidos num intervalo de distâncias (300pc . d . 3kpc), idades (7.5 . log[t] . 9.8), excessos de cor (0.0 . E(B−V ) . 0.8) e número de membros (60 . N . 2000), o que impõe uma grande variedade de cenários diferentes no que diz respeito ao contraste desses objetos sobre as populações estelares de campo. 
In this work we were able to obtain member star lists for a set of 34 star clusters from accurate astrometry and photometry from \textit{Gaia} DR3 data using a uniform methodology. CMDs were constructed and morphological age indices were measured.

We presented an observational view of the RC population through the interval $8.3< \log[t({\rm yr})]<9.9$. In general, for both visible and infrared, the values of the absolute magnitudes of the RC, $M_{G}$ and $M_{K}$ as a function of age are well represented by PARSEC isochrones. We note that for the infrared the models tend to present values of $M_{K}$ systematically brighter for younger objects. % especially through the SRC region.

When comparing the RC colour indices, we noticed that the models represent well the older population ($\log[t({\rm yr})]> 9.0$). However, they are systematically redder than observed younger OCs. The same conclusions were drawn by other authors for objects with age $\log[t({\rm yr})] \lesssim 8.8$ \citep{10.1111/j.1365-2966.2011.18627.x,10.1093/mnras/stac2496,10.1093/mnras/staa647,2019A&A...623A.108B}. Such discrepancies may be caused by a combination of different effects. For example, the presence of unresolved binary stars, the mass distribution of the RC progenitors stars (including mass loss), presence of differential reddening, or the presence of SRC stellar populations simultaneously with the RC for some objects, making the observed average RC colour bluer than expected. Alternatively, this may indicate that the models still present limitations such as bolometric corrections, colour transformations and effective temperatures for this age range. Uncertainties of transformations from the theoretical to the observational plane also play a role.

Comparisons of the established morphological indices showed that the models tend to satisfactorily predict the $\Delta G$ index for the entire age range  explored here. But for ages $\log[t({\rm yr})]< 8.8$, the models tend to present the index
$\Delta BR$ systematically greater than observed (probably an effect caused by the inefficiency of predicting the RC colours for younger ages.  According to our data, the models also face challenges in accurately predicting the index $\Delta BR$ for older objects ($\log[t(yr)] > 9.6$), although few OCs are that old to provide a useful comparison.

Finally, we provided a set of age calibration functions based on \textit{Gaia} morphological indices for the first time, allowing an estimate of star cluster ages based on such indices. In particular, a non-linear fitting function was obtained by using both indices $\Delta BR$ and $\Delta G$, extending the age determination using such method to younger objects. We have demonstrated its accuracy for the range $8.3<\log[t({\rm yr})]< 9.8$ by a direct comparison with ages from the literature, obtaining a mean residuals of 0.06 dex in $\log[t(yr)]$.

Although out the scope of the present paper, asteroseismology is an alternative approach to obtain precise ages of OCs, from which new color calibrations can be derived. However, due to the lack of OCs with ages measured via asteroseismology (\citealp[e.g.][]{2023A&A...679A..23B, 2025ApJ...979..135A}), we still face difficulties incorporating a significant number of objects with isochrone-independent age measurements. In the future, as a sizeable sample of stars in several clusters is measured using asteroseismology, the results from this age-dating method may then be used to establish new color calibrations.

%RETIRE A LINHA ABAIXO SE NÃO QUISER AGRADECER A ALGUÉM
%\begin{acknowledgements} 

%The authors wish to thank the Brazilian financial agencies FAPEMIG, CNPq and CAPES (finance code 001).
%W. Corradi wishes to thank the LNA staff. This research has made use of the VizieR catalogue access tool, CDS, Strasbourg, France and has made use of data from the European Space Agency (ESA) mission \textit{Gaia} (\url{https://www.cosmos.esa.int/gaia}), processed by the \textit{Gaia} Data Processing and Analysis Consortium (DPAC, \url{https://www.cosmos.esa.int/web/gaia/dpac/consortium}). Funding for the DPAC has been provided by national institutions, in particular the institutions participating in the \textit{Gaia} Multilateral Agreement.
%\end{acknowledgements} 
%

%%%%%%%%%%%%%%%%%%%%%%%%%%%%%%%%%%%%%%%%%%%%%%%%%%
\section*{Acknowledgements}

We thank the referee for a critical, positive and motivating report. The authors wish to thank the Brazilian financial agencies FAPEMIG, CNPq and CAPES (finance code 001). W.Corradi acknowledges the support from CNPq - BRICS 440142/2022-9, FAPEMIG APQ 02493-22  and FNDCT/FINEP/REF 0180/22. 
F.F.S.M. acknowledges financial support from Conselho Nacional de Desenvolvimento Cient\'ifico e Tecnol\'ogico – CNPq (proc. 404482/2021-0) and from FAPERJ (proc. E-26/201.386/2022 and E-26/211.475/2021).
%{\textbf{We thank the referee for the comments that helped us to improve the paper}}.
This research has made use of the VizieR catalogue access tool, CDS, Strasbourg, France. This work has made use of data from the European Space Agency (ESA) mission \textit{Gaia} (\url{https://www.cosmos.esa.int/gaia}), processed by the \textit{Gaia} Data Processing and Analysis Consortium (DPAC, \url{https://www.cosmos.esa.int/web/gaia/dpac/consortium}). Funding for the DPAC has been provided by national institutions, in particular the institutions participating in the \textit{Gaia} Multilateral Agreement. This  research  has  made  use  of  TOPCAT \citep{Taylor:2005}.

\section*{Data availability}
The data underlying this article is publicly available (\textit{Gaia DR3} and 2MASS) or is available in the article.

%%%%%%%%%%%%%%%%%%%% REFERENCES %%%%%%%%%%%%%%%%%%

% The best way to enter references is to use BibTeX:

\bibliographystyle{mnras}
\bibliography{references.bib} % if your bibtex file is called example.bib

\begin{thebibliography}{}
\makeatletter
\relax
\def\mn@urlcharsother{\let\do\@makeother \do\$\do\&\do\#\do\^\do\_\do\%\do\~}
\def\mn@doi{\begingroup\mn@urlcharsother \@ifnextchar [ {\mn@doi@}
  {\mn@doi@[]}}
\def\mn@doi@[#1]#2{\def\@tempa{#1}\ifx\@tempa\@empty \href
  {http://dx.doi.org/#2} {doi:#2}\else \href {http://dx.doi.org/#2} {#1}\fi
  \endgroup}
\def\mn@eprint#1#2{\mn@eprint@#1:#2::\@nil}
\def\mn@eprint@arXiv#1{\href {http://arxiv.org/abs/#1} {{\tt arXiv:#1}}}
\def\mn@eprint@dblp#1{\href {http://dblp.uni-trier.de/rec/bibtex/#1.xml}
  {dblp:#1}}
\def\mn@eprint@#1:#2:#3:#4\@nil{\def\@tempa {#1}\def\@tempb {#2}\def\@tempc
  {#3}\ifx \@tempc \@empty \let \@tempc \@tempb \let \@tempb \@tempa \fi \ifx
  \@tempb \@empty \def\@tempb {arXiv}\fi \@ifundefined
  {mn@eprint@\@tempb}{\@tempb:\@tempc}{\expandafter \expandafter \csname
  mn@eprint@\@tempb\endcsname \expandafter{\@tempc}}}

\bibitem[\protect\citeauthoryear{{Alfonso}, {Garc{\'\i}a-Varela}  \&
  {Vieira}}{{Alfonso} et~al.}{2024}]{2024A&A...689A..18A}
{Alfonso} J.,  {Garc{\'\i}a-Varela} A.,   {Vieira} K.,  2024, \mn@doi [\aap]
  {10.1051/0004-6361/202450901}, \href
  {https://ui.adsabs.harvard.edu/abs/2024A&A...689A..18A} {689, A18}

\bibitem[\protect\citeauthoryear{{An}, {Pinsonneault}, {Terndrup}  \&
  {Chung}}{{An} et~al.}{2019}]{2019ApJ...879...81A}
{An} D.,  {Pinsonneault} M.~H.,  {Terndrup} D.~M.,   {Chung} C.,  2019, \mn@doi
  [\apj] {10.3847/1538-4357/ab23ed}, \href
  {https://ui.adsabs.harvard.edu/abs/2019ApJ...879...81A} {879, 81}

\bibitem[\protect\citeauthoryear{{Angelo}, {Santos}, {Corradi}  \&
  {Maia}}{{Angelo} et~al.}{2019}]{2019A&A...624A...8A}
{Angelo} M.~S.,  {Santos} J.~F.~C.,  {Corradi} W.~J.~B.,   {Maia} F.~F.~S.,
  2019, \mn@doi [\aap] {10.1051/0004-6361/201832702}, \href
  {https://ui.adsabs.harvard.edu/abs/2019A&A...624A...8A} {624, A8}

\bibitem[\protect\citeauthoryear{{Angelo}, {Corradi}, {Santos}, {Maia}  \&
  {Ferreira}}{{Angelo} et~al.}{2021}]{2021MNRAS.500.4338A}
{Angelo} M.~S.,  {Corradi} W.~J.~B.,  {Santos} J.~F.~C. J.,  {Maia} F.~F.~S.,
  {Ferreira} F.~A.,  2021, \mn@doi [\mnras] {10.1093/mnras/staa3192}, \href
  {https://ui.adsabs.harvard.edu/abs/2021MNRAS.500.4338A} {500, 4338}

\bibitem[\protect\citeauthoryear{{Anthony-Twarog} \& {Twarog}}{{Anthony-Twarog}
  \& {Twarog}}{1985}]{1985ApJ...291..595A}
{Anthony-Twarog} B.~J.,  {Twarog} B.~A.,  1985, \mn@doi [\apj]
  {10.1086/163100}, \href
  {https://ui.adsabs.harvard.edu/abs/1985ApJ...291..595A} {291, 595}

\bibitem[\protect\citeauthoryear{{Ash}, {Pinsonneault}, {Vrard}  \&
  {Zinn}}{{Ash} et~al.}{2025}]{2025ApJ...979..135A}
{Ash} A.~L.,  {Pinsonneault} M.~H.,  {Vrard} M.,   {Zinn} J.~C.,  2025, \mn@doi
  [\apj] {10.3847/1538-4357/ad9b18}, \href
  {https://ui.adsabs.harvard.edu/abs/2025ApJ...979..135A} {979, 135}

\bibitem[\protect\citeauthoryear{{Battinelli}, {Brandimarti}  \&
  {Capuzzo-Dolcetta}}{{Battinelli} et~al.}{1994}]{1994A&AS..104..379B}
{Battinelli} P.,  {Brandimarti} A.,   {Capuzzo-Dolcetta} R.,  1994, \aaps,
  \href {https://ui.adsabs.harvard.edu/abs/1994A&AS..104..379B} {104, 379}

\bibitem[\protect\citeauthoryear{{Beletsky}, {Carraro}  \& {Ivanov}}{{Beletsky}
  et~al.}{2009}]{2009A&A...508.1279B}
{Beletsky} Y.,  {Carraro} G.,   {Ivanov} V.~D.,  2009, \mn@doi [\aap]
  {10.1051/0004-6361/200912697}, \href
  {https://ui.adsabs.harvard.edu/abs/2009A&A...508.1279B} {508, 1279}

\bibitem[\protect\citeauthoryear{{Bilir}, {{\"O}nal}, {Karaali},
  {Cabrera-Lavers}  \& {{\c{C}}akmak}}{{Bilir}
  et~al.}{2013}]{2013Ap&SS.344..417B}
{Bilir} S.,  {{\"O}nal} {\"O}.,  {Karaali} S.,  {Cabrera-Lavers} A.,
  {{\c{C}}akmak} H.,  2013, \mn@doi [\apss] {10.1007/s10509-012-1342-9}, \href
  {https://ui.adsabs.harvard.edu/abs/2013Ap&SS.344..417B} {344, 417}

\bibitem[\protect\citeauthoryear{{Bossini} et~al.,}{{Bossini}
  et~al.}{2019}]{2019A&A...623A.108B}
{Bossini} D.,  et~al., 2019, \mn@doi [\aap] {10.1051/0004-6361/201834693},
  \href {https://ui.adsabs.harvard.edu/abs/2019A%26A...623A.108B} {623, A108}

\bibitem[\protect\citeauthoryear{{Bressan}, {Marigo}, {Girardi}, {Salasnich},
  {Dal Cero}, {Rubele}  \& {Nanni}}{{Bressan} et~al.}{2012}]{bressan2012}
{Bressan} A.,  {Marigo} P.,  {Girardi} L.,  {Salasnich} B.,  {Dal Cero} C.,
  {Rubele} S.,   {Nanni} A.,  2012, \mn@doi [\mnras]
  {10.1111/j.1365-2966.2012.21948.x}, \href
  {http://adsabs.harvard.edu/abs/2012MNRAS.427..127B} {427, 127}

\bibitem[\protect\citeauthoryear{{Brogaard} et~al.,}{{Brogaard}
  et~al.}{2023}]{2023A&A...679A..23B}
{Brogaard} K.,  et~al., 2023, \mn@doi [\aap] {10.1051/0004-6361/202347330},
  \href {https://ui.adsabs.harvard.edu/abs/2023A&A...679A..23B} {679, A23}

\bibitem[\protect\citeauthoryear{{Cantat-Gaudin} et~al.,}{{Cantat-Gaudin}
  et~al.}{2018}]{cjv18}
{Cantat-Gaudin} T.,  et~al., 2018, \mn@doi [\aap]
  {10.1051/0004-6361/201833476}, \href
  {https://ui.adsabs.harvard.edu/abs/2018A&A...618A..93C} {618, A93}

\bibitem[\protect\citeauthoryear{{Cantat-Gaudin} et~al.,}{{Cantat-Gaudin}
  et~al.}{2020}]{2020A&A...640A...1C}
{Cantat-Gaudin} T.,  et~al., 2020, \mn@doi [\aap]
  {10.1051/0004-6361/202038192}, \href
  {https://ui.adsabs.harvard.edu/abs/2020A&A...640A...1C} {640, A1}

\bibitem[\protect\citeauthoryear{{Cardelli}, {Clayton}  \& {Mathis}}{{Cardelli}
  et~al.}{1989}]{Cardelli:1989}
{Cardelli} J.~A.,  {Clayton} G.~C.,   {Mathis} J.~S.,  1989, \mn@doi [\apj]
  {10.1086/167900}, \href {http://adsabs.harvard.edu/abs/1989ApJ...345..245C}
  {345, 245}

\bibitem[\protect\citeauthoryear{{Carraro} \& {Chiosi}}{{Carraro} \&
  {Chiosi}}{1994}]{1994A&A...287..761C}
{Carraro} G.,  {Chiosi} C.,  1994, \aap, \href
  {https://ui.adsabs.harvard.edu/abs/1994A&A...287..761C} {287, 761}

\bibitem[\protect\citeauthoryear{{Carraro}, {Geisler}, {Villanova},
  {Frinchaboy}  \& {Majewski}}{{Carraro} et~al.}{2007}]{2007A&A...476..217C}
{Carraro} G.,  {Geisler} D.,  {Villanova} S.,  {Frinchaboy} P.~M.,   {Majewski}
  S.~R.,  2007, \mn@doi [\aap] {10.1051/0004-6361:20078113}, \href
  {https://ui.adsabs.harvard.edu/abs/2007A%26A...476..217C} {476, 217}

\bibitem[\protect\citeauthoryear{{Castro-Ginard} et~al.,}{{Castro-Ginard}
  et~al.}{2020}]{cjl20}
{Castro-Ginard} A.,  et~al., 2020, \mn@doi [\aap]
  {10.1051/0004-6361/201937386}, \href
  {https://ui.adsabs.harvard.edu/abs/2020A&A...635A..45C} {635, A45}

\bibitem[\protect\citeauthoryear{{Chen} \& {Zhao}}{{Chen} \&
  {Zhao}}{2020}]{2020MNRAS.495.2673C}
{Chen} Y.~Q.,  {Zhao} G.,  2020, \mn@doi [\mnras] {10.1093/mnras/staa1079},
  \href {https://ui.adsabs.harvard.edu/abs/2020MNRAS.495.2673C} {495, 2673}

\bibitem[\protect\citeauthoryear{Dias, Monteiro, Moitinho, Lépine, Carraro,
  Paunzen, Alessi  \& Villela}{Dias et~al.}{2021}]{10.1093/mnras/stab770}
Dias W.~S.,  Monteiro H.,  Moitinho A.,  Lépine J. R.~D.,  Carraro G.,
  Paunzen E.,  Alessi B.,   Villela L.,  2021, \mn@doi [Monthly Notices of the
  Royal Astronomical Society] {10.1093/mnras/stab770}, 504, 356

\bibitem[\protect\citeauthoryear{{D{\"u}rbeck}}{{D{\"u}rbeck}}{1960}]{1960ZA.....49..214D}
{D{\"u}rbeck} W.,  1960, \zap, \href
  {https://ui.adsabs.harvard.edu/abs/1960ZA.....49..214D} {49, 214}

\bibitem[\protect\citeauthoryear{{Evans} et~al.,}{{Evans}
  et~al.}{2018}]{Evans:2018}
{Evans} D.~W.,  et~al., 2018, \mn@doi [\aap] {10.1051/0004-6361/201832756},
  \href {https://ui.adsabs.harvard.edu/#abs/2018A&A...616A...4E} {616, A4}

\bibitem[\protect\citeauthoryear{{Ferreira}, {Santos}, {Corradi}, {Maia}  \&
  {Angelo}}{{Ferreira} et~al.}{2019}]{2019MNRAS.483.5508F}
{Ferreira} F.~A.,  {Santos} J.~F.~C.,  {Corradi} W.~J.~B.,  {Maia} F.~F.~S.,
  {Angelo} M.~S.,  2019, \mn@doi [\mnras] {10.1093/mnras/sty3511}, \href
  {https://ui.adsabs.harvard.edu/abs/2019MNRAS.483.5508F} {483, 5508}

\bibitem[\protect\citeauthoryear{Ferreira, Corradi, Maia, Angelo  \&
  Santos}{Ferreira et~al.}{2020}]{10.1093/mnras/staa1684}
Ferreira F.~A.,  Corradi W. J.~B.,  Maia F. F.~S.,  Angelo M.~S.,   Santos
  J~F~C J.,  2020, \mn@doi [\mnras] {10.1093/mnras/staa1684}, 496, 2021

\bibitem[\protect\citeauthoryear{{Ferreira}, {Corradi}, {Maia}, {Angelo}  \&
  {Santos}}{{Ferreira} et~al.}{2021}]{2021MNRAS.502L..90F}
{Ferreira} F.~A.,  {Corradi} W.~J.~B.,  {Maia} F.~F.~S.,  {Angelo} M.~S.,
  {Santos} J.~F.~C. J.,  2021, \mn@doi [\mnras] {10.1093/mnrasl/slab011}, \href
  {https://ui.adsabs.harvard.edu/abs/2021MNRAS.502L..90F} {502, L90}

\bibitem[\protect\citeauthoryear{{Friel}}{{Friel}}{1995}]{friel1995}
{Friel} E.~D.,  1995, \mn@doi [\araa] {10.1146/annurev.aa.33.090195.002121},
  \href {http://adsabs.harvard.edu/abs/1995ARA&A..33..381F} {33, 381}

\bibitem[\protect\citeauthoryear{{Gaia Collaboration} et~al.,}{{Gaia
  Collaboration} et~al.}{2018}]{2018A&A...616A..10G}
{Gaia Collaboration} et~al., 2018, \mn@doi [\aap]
  {10.1051/0004-6361/201832843}, \href
  {https://ui.adsabs.harvard.edu/abs/2018A&A...616A..10G} {616, A10}

\bibitem[\protect\citeauthoryear{{Gaia Collaboration} et~al.,}{{Gaia
  Collaboration} et~al.}{2021}]{2021A&A...649A...1G}
{Gaia Collaboration} et~al., 2021, \mn@doi [\aap]
  {10.1051/0004-6361/202039657}, \href
  {https://ui.adsabs.harvard.edu/abs/2021A&A...649A...1G} {649, A1}

\bibitem[\protect\citeauthoryear{{Gaia Collaboration} et~al.,}{{Gaia
  Collaboration} et~al.}{2023}]{2023A&A...674A...1G}
{Gaia Collaboration} et~al., 2023, \mn@doi [\aap]
  {10.1051/0004-6361/202243940}, \href
  {https://ui.adsabs.harvard.edu/abs/2023A&A...674A...1G} {674, A1}

\bibitem[\protect\citeauthoryear{{Geisler}, {Bica}, {Dottori}, {Claria},
  {Piatti}  \& {Santos}}{{Geisler} et~al.}{1997}]{1997AJ....114.1920G}
{Geisler} D.,  {Bica} E.,  {Dottori} H.,  {Claria} J.~J.,  {Piatti} A.~E.,
  {Santos} Joao F.~C. J.,  1997, \mn@doi [\aj] {10.1086/118614}, \href
  {https://ui.adsabs.harvard.edu/abs/1997AJ....114.1920G} {114, 1920}

\bibitem[\protect\citeauthoryear{{Girardi}}{{Girardi}}{1999}]{1999MNRAS.308..818G}
{Girardi} L.,  1999, \mn@doi [\mnras] {10.1046/j.1365-8711.1999.02746.x}, \href
  {https://ui.adsabs.harvard.edu/abs/1999MNRAS.308..818G} {308, 818}

\bibitem[\protect\citeauthoryear{{Girardi}}{{Girardi}}{2016}]{2016ARA&A..54...95G}
{Girardi} L.,  2016, \mn@doi [\araa] {10.1146/annurev-astro-081915-023354},
  \href {https://ui.adsabs.harvard.edu/abs/2016ARA&A..54...95G} {54, 95}

\bibitem[\protect\citeauthoryear{{Grocholski} \& {Sarajedini}}{{Grocholski} \&
  {Sarajedini}}{2002}]{2002AJ....123.1603G}
{Grocholski} A.~J.,  {Sarajedini} A.,  2002, \mn@doi [\aj] {10.1086/339027},
  \href {https://ui.adsabs.harvard.edu/abs/2002AJ....123.1603G} {123, 1603}

\bibitem[\protect\citeauthoryear{Hawkins, Leistedt, Bovy  \& Hogg}{Hawkins
  et~al.}{2017}]{10.1093/mnras/stx1655}
Hawkins K.,  Leistedt B.,  Bovy J.,   Hogg D.~W.,  2017, \mn@doi [Monthly
  Notices of the Royal Astronomical Society] {10.1093/mnras/stx1655}, 471, 722

\bibitem[\protect\citeauthoryear{{He}, {Liu}, {Luo}, {Wang}  \& {Jiang}}{{He}
  et~al.}{2022}]{2022arXiv220908504H}
{He} Z.,  {Liu} X.,  {Luo} Y.,  {Wang} K.,   {Jiang} Q.,  2022, arXiv e-prints,
  \href {https://ui.adsabs.harvard.edu/abs/2022arXiv220908504H} {p.
  arXiv:2209.08504}

\bibitem[\protect\citeauthoryear{Holanda, Ramos, Peña Suárez, Martinez  \&
  Pereira}{Holanda et~al.}{2022}]{10.1093/mnras/stac2496}
Holanda N.,  Ramos A.~A.,  Peña Suárez V.~J.,  Martinez C.~F.,   Pereira
  C.~B.,  2022, \mn@doi [Monthly Notices of the Royal Astronomical Society]
  {10.1093/mnras/stac2496}, 516, 4484

\bibitem[\protect\citeauthoryear{{Hunt} \& {Reffert}}{{Hunt} \&
  {Reffert}}{2023}]{2023A&A...673A.114H}
{Hunt} E.~L.,  {Reffert} S.,  2023, \mn@doi [\aap]
  {10.1051/0004-6361/202346285}, \href
  {https://ui.adsabs.harvard.edu/abs/2023A&A...673A.114H} {673, A114}

\bibitem[\protect\citeauthoryear{{Im}, {Kim}, {Kyeong}, {Park}  \& {Lee}}{{Im}
  et~al.}{2023}]{2023AJ....165...53I}
{Im} H.,  {Kim} S.~C.,  {Kyeong} J.,  {Park} H.~S.,   {Lee} J.~H.,  2023,
  \mn@doi [\aj] {10.3847/1538-3881/aca7fb}, \href
  {https://ui.adsabs.harvard.edu/abs/2023AJ....165...53I} {165, 53}

\bibitem[\protect\citeauthoryear{{Kharchenko}, {Piskunov}, {R{\"o}ser},
  {Schilbach}  \& {Scholz}}{{Kharchenko} et~al.}{2005}]{2005A&A...438.1163K}
{Kharchenko} N.~V.,  {Piskunov} A.~E.,  {R{\"o}ser} S.,  {Schilbach} E.,
  {Scholz} R.~D.,  2005, \mn@doi [\aap] {10.1051/0004-6361:20042523}, \href
  {https://ui.adsabs.harvard.edu/abs/2005A&A...438.1163K} {438, 1163}

\bibitem[\protect\citeauthoryear{{Kharchenko}, {Piskunov}, {Schilbach},
  {R{\"o}ser}  \& {Scholz}}{{Kharchenko} et~al.}{2013}]{kharchenko2013}
{Kharchenko} N.~V.,  {Piskunov} A.~E.,  {Schilbach} E.,  {R{\"o}ser} S.,
  {Scholz} R.-D.,  2013, \mn@doi [\aap] {10.1051/0004-6361/201322302}, \href
  {http://adsabs.harvard.edu/abs/2013A&A...558A..53K} {558, A53}

\bibitem[\protect\citeauthoryear{{Lada} \& {Lada}}{{Lada} \&
  {Lada}}{2003}]{Lada:2003}
{Lada} C.~J.,  {Lada} E.~A.,  2003, \mn@doi [\araa]
  {10.1146/annurev.astro.41.011802.094844}, \href
  {http://adsabs.harvard.edu/abs/2003ARA%26A..41...57L} {41, 57}

\bibitem[\protect\citeauthoryear{{Lindegren} et~al.,}{{Lindegren}
  et~al.}{2018}]{Lindegren:2018}
{Lindegren} L.,  et~al., 2018, \mn@doi [\aap] {10.1051/0004-6361/201832727},
  \href {http://adsabs.harvard.edu/abs/2018A%26A...616A...2L} {616, A2}

\bibitem[\protect\citeauthoryear{{Lindegren} et~al.,}{{Lindegren}
  et~al.}{2021}]{2021A&A...649A...2L}
{Lindegren} L.,  et~al., 2021, \mn@doi [\aap] {10.1051/0004-6361/202039709},
  \href {https://ui.adsabs.harvard.edu/abs/2021A&A...649A...2L} {649, A2}

\bibitem[\protect\citeauthoryear{{Liu} \& {Pang}}{{Liu} \& {Pang}}{2019}]{lp19}
{Liu} L.,  {Pang} X.,  2019, \mn@doi [\apjs] {10.3847/1538-4365/ab530a}, \href
  {https://ui.adsabs.harvard.edu/abs/2019ApJS..245...32L} {245, 32}

\bibitem[\protect\citeauthoryear{{Loktin} \& {Popova}}{{Loktin} \&
  {Popova}}{2017}]{2017AstBu..72..257L}
{Loktin} A.~V.,  {Popova} M.~E.,  2017, \mn@doi [Astrophysical Bulletin]
  {10.1134/S1990341317030154}, \href
  {https://ui.adsabs.harvard.edu/abs/2017AstBu..72..257L} {72, 257}

\bibitem[\protect\citeauthoryear{Martinez, Holanda, Pereira  \& Drake}{Martinez
  et~al.}{2020}]{10.1093/mnras/staa647}
Martinez C.~F.,  Holanda N.,  Pereira C.~B.,   Drake N.~A.,  2020, \mn@doi
  [Monthly Notices of the Royal Astronomical Society] {10.1093/mnras/staa647},
  494, 1470

\bibitem[\protect\citeauthoryear{{Netopil}, {Paunzen}, {Heiter}  \&
  {Soubiran}}{{Netopil} et~al.}{2016}]{2016A&A...585A.150N}
{Netopil} M.,  {Paunzen} E.,  {Heiter} U.,   {Soubiran} C.,  2016, \mn@doi
  [\aap] {10.1051/0004-6361/201526370}, \href
  {https://ui.adsabs.harvard.edu/abs/2016A%26A...585A.150N} {585, A150}

\bibitem[\protect\citeauthoryear{{Onozato}, {Ita}, {Nakada}  \&
  {Nishiyama}}{{Onozato} et~al.}{2019}]{2019MNRAS.486.5600O}
{Onozato} H.,  {Ita} Y.,  {Nakada} Y.,   {Nishiyama} S.,  2019, \mn@doi
  [\mnras] {10.1093/mnras/stz1192}, \href
  {https://ui.adsabs.harvard.edu/abs/2019MNRAS.486.5600O} {486, 5600}

\bibitem[\protect\citeauthoryear{{Oralhan}, {Karata{\c{s}}}, {Schuster},
  {Michel}  \& {Chavarr{\'\i}a}}{{Oralhan} et~al.}{2015}]{2015NewA...34..195O}
{Oralhan} {\.I}.~A.,  {Karata{\c{s}}} Y.,  {Schuster} W.~J.,  {Michel} R.,
  {Chavarr{\'\i}a} C.,  2015, \mn@doi [\na] {10.1016/j.newast.2014.06.011},
  \href {https://ui.adsabs.harvard.edu/abs/2015NewA...34..195O} {34, 195}

\bibitem[\protect\citeauthoryear{{Parisi} et~al.,}{{Parisi}
  et~al.}{2014}]{2014AJ....147...71P}
{Parisi} M.~C.,  et~al., 2014, \mn@doi [\aj] {10.1088/0004-6256/147/4/71},
  \href {https://ui.adsabs.harvard.edu/abs/2014AJ....147...71P} {147, 71}

\bibitem[\protect\citeauthoryear{{Phelps}, {Janes}  \& {Montgomery}}{{Phelps}
  et~al.}{1994}]{1994AJ....107.1079P}
{Phelps} R.~L.,  {Janes} K.~A.,   {Montgomery} K.~A.,  1994, \mn@doi [\aj]
  {10.1086/116920}, \href
  {https://ui.adsabs.harvard.edu/abs/1994AJ....107.1079P} {107, 1079}

\bibitem[\protect\citeauthoryear{{Piatti}, {Clari{\'a}}  \& {Bica}}{{Piatti}
  et~al.}{1998}]{1998ApJS..116..263P}
{Piatti} A.~E.,  {Clari{\'a}} J.~J.,   {Bica} E.,  1998, \mn@doi [\apjs]
  {10.1086/313103}, \href {http://adsabs.harvard.edu/abs/1998ApJS..116..263P}
  {116, 263}

\bibitem[\protect\citeauthoryear{Piatti, Clariá  \& Ahumada}{Piatti
  et~al.}{2010}]{10.1111/j.1365-2966.2009.16106.x}
Piatti A.~E.,  Clariá J.~J.,   Ahumada A.~V.,  2010, \mn@doi [Monthly Notices
  of the Royal Astronomical Society] {10.1111/j.1365-2966.2009.16106.x}, 402,
  2720

\bibitem[\protect\citeauthoryear{Piatti, Clariá, Bica, Geisler, Ahumada  \&
  Girardi}{Piatti et~al.}{2011}]{10.1111/j.1365-2966.2011.18627.x}
Piatti A.~E.,  Clariá J.~J.,  Bica E.,  Geisler D.,  Ahumada A.~V.,   Girardi
  L.,  2011, \mn@doi [Monthly Notices of the Royal Astronomical Society]
  {10.1111/j.1365-2966.2011.18627.x}, 417, 1559

\bibitem[\protect\citeauthoryear{{Riello} et~al.,}{{Riello}
  et~al.}{2021}]{2021A&A...649A...3R}
{Riello} M.,  et~al., 2021, \mn@doi [\aap] {10.1051/0004-6361/202039587}, \href
  {https://ui.adsabs.harvard.edu/abs/2021A&A...649A...3R} {649, A3}

\bibitem[\protect\citeauthoryear{{Ruiz-Dern}, {Babusiaux}, {Arenou}, {Turon}
  \& {Lallement}}{{Ruiz-Dern} et~al.}{2018}]{2018A&A...609A.116R}
{Ruiz-Dern} L.,  {Babusiaux} C.,  {Arenou} F.,  {Turon} C.,   {Lallement} R.,
  2018, \mn@doi [\aap] {10.1051/0004-6361/201731572}, \href
  {https://ui.adsabs.harvard.edu/abs/2018A&A...609A.116R} {609, A116}

\bibitem[\protect\citeauthoryear{{Salaris}, {Weiss}  \& {Percival}}{{Salaris}
  et~al.}{2004}]{2004A&A...414..163S}
{Salaris} M.,  {Weiss} A.,   {Percival} S.~M.,  2004, \mn@doi [\aap]
  {10.1051/0004-6361:20031578}, \href
  {https://ui.adsabs.harvard.edu/abs/2004A&A...414..163S} {414, 163}

\bibitem[\protect\citeauthoryear{{Sandquist} et~al.,}{{Sandquist}
  et~al.}{2020}]{2020AJ....159...96S}
{Sandquist} E.~L.,  et~al., 2020, \mn@doi [\aj] {10.3847/1538-3881/ab68df},
  \href {https://ui.adsabs.harvard.edu/abs/2020AJ....159...96S} {159, 96}

\bibitem[\protect\citeauthoryear{{Sim}, {Lee}, {Ann}  \& {Kim}}{{Sim}
  et~al.}{2019}]{sla19}
{Sim} G.,  {Lee} S.~H.,  {Ann} H.~B.,   {Kim} S.,  2019, \mn@doi [JKAS]
  {10.5303/JKAS.2019.52.5.145}, \href
  {https://ui.adsabs.harvard.edu/abs/2019JKAS...52..145S} {52, 145}

\bibitem[\protect\citeauthoryear{{Skrutskie} et~al.,}{{Skrutskie}
  et~al.}{2006}]{Skrutskie:2006}
{Skrutskie} M.~F.,  et~al., 2006, \mn@doi [\aj] {10.1086/498708}, \href
  {http://adsabs.harvard.edu/abs/2006AJ....131.1163S} {131, 1163}

\bibitem[\protect\citeauthoryear{{Taylor}}{{Taylor}}{2005}]{Taylor:2005}
{Taylor} M.~B.,  2005, in {Shopbell} P.,  {Britton} M.,   {Ebert} R.,  eds,
  ~ASPC Vol. 347, Astronomical Data Analysis Software and Systems XIV. p.~29

\bibitem[\protect\citeauthoryear{{Zasowski} et~al.,}{{Zasowski}
  et~al.}{2013}]{2013AJ....146...64Z}
{Zasowski} G.,  et~al., 2013, \mn@doi [\aj] {10.1088/0004-6256/146/3/64}, \href
  {https://ui.adsabs.harvard.edu/abs/2013AJ....146...64Z} {146, 64}

\bibitem[\protect\citeauthoryear{{Zhang}, {Chen}  \& {Zhao}}{{Zhang}
  et~al.}{2021}]{2021ApJ...919...52Z}
{Zhang} H.,  {Chen} Y.,   {Zhao} G.,  2021, \mn@doi [\apj]
  {10.3847/1538-4357/ac0e92}, \href
  {https://ui.adsabs.harvard.edu/abs/2021ApJ...919...52Z} {919, 52}

\bibitem[\protect\citeauthoryear{{van Helshoecht} \& {Groenewegen}}{{van
  Helshoecht} \& {Groenewegen}}{2007}]{2007A&A...463..559V}
{van Helshoecht} V.,  {Groenewegen} M.~A.~T.,  2007, \mn@doi [\aap]
  {10.1051/0004-6361:20052721}, \href
  {https://ui.adsabs.harvard.edu/abs/2007A&A...463..559V} {463, 559}

\makeatother
\end{thebibliography}

%\bibliography{ref_doutorado.bib}

% Alternatively you could enter them by hand, like this:
% This method is tedious and prone to error if you have lots of references
%\begin{thebibliography}{99}
%\bibitem[\protect\citeauthoryear{Author}{2012}]{Author2012}
%Author A.~N., 2013, Journal of Improbable Astronomy, 1, 1
%\bibitem[\protect\citeauthoryear{Others}{2013}]{Others2013}
%Others S., 2012, Journal of Interesting Stuff, 17, 198
%\end{thebibliography}

%%%%%%%%%%%%%%%%%%%%%%%%%%%%%%%%%%%%%%%%%%%%%%%%%%

%%%%%%%%%%%%%%%%% APPENDICES %%%%%%%%%%%%%%%%%%%%%
\appendix
\section{Tables}
\label{sect:apendiz_tables}

This appendix contains informations from OCs CMDs properties. In Table \ref{Tab:ocs_mags} the average RC apparent magnitudes ($G$ and $K$) and colour indices ($(G_{BP}-G_{RP})$ and $\sigma_{(J-K)}$) and their uncertainties are presented. The number of RC stars $N_{G}$ and $N_{K}$ filtered by the box-shaped filter for \textit{Gaia} and 2MASS CMDs is also presented. In Table \ref{Tab:ocs_deltas} the morphological age indices 
$\Delta G$, $\Delta BR$, $\Delta K$ and $\Delta JK$ and their uncertainties are presented. A review of the astrophysical parameters for the cluster NGC2354 taken from the literature is presented in Table \ref{Tab:NGC2354_parametros}.

\begin{table*}
\small
%\centering
\caption{Average RC apparent magnitudes and colour indices and their uncertainties. The number of RC stars  filtered by the box-shaped filter for \textit{Gaia} and 2MASS CMDs is also presented.}
\label{Tab:clusters_prop3}
\begin{tabular}{|l|r|r|r|r|r|r|r|r|r|r|}
\hline
  \multicolumn{1}{|c|}{OC} &
  \multicolumn{1}{c|}{$G$} &
  \multicolumn{1}{c|}{$\sigma_{G}$} &
  \multicolumn{1}{c|}{$(G_{BP}-G_{RP})$} &
  \multicolumn{1}{c|}{$\sigma_{(G_{BP}-G_{RP})}$} &
  \multicolumn{1}{c|}{$K$} &
  \multicolumn{1}{c|}{$\sigma_{K}$} &
  \multicolumn{1}{c|}{$(J-K)$} &
  \multicolumn{1}{c|}{$\sigma_{(J-K)}$} &
  \multicolumn{1}{c|}{$N_{G}$} &
  \multicolumn{1}{c|}{$N_{K}$} \\
\hline
  NGC 188 & 12.13 & 0.07 & 1.37 & 0.02 & 9.69 & 0.08 & 0.72 & 0.01 & 14.0 & 15.0\\
  NGC 752 & 8.90 & 0.06 & 1.16 & 0.01 & 6.77 & 0.06 & 0.60 & 0.01 & 14.0 & 14.0\\
  NGC 1245 & 13.65 & 0.03 & 1.38 & 0.01 & 11.14 & 0.05 & 0.65 & 0.01 & 41.0 & 44.0\\
  NGC 1817 & 12.15 & 0.05 & 1.35 & 0.01 & 9.73 & 0.05 & 0.66 & 0.01 & 26.0 & 28.0\\
  NGC 2099 & 10.96 & 0.05 & 1.42 & 0.01 & 8.44 & 0.06 & 0.69 & 0.01 & 27.0 & 26.0\\
  Trumpler 5 & 14.30 & 0.02 & 1.88 & 0.01 & 11.01 & 0.03 & 0.91 & 0.01 & 92.0 & 110.0\\
  Collinder 110 & 13.20 & 0.04 & 1.72 & 0.01 & 10.25 & 0.04 & 0.83 & 0.01 & 43.0 & 45.0\\
  NGC 2354 & 11.29 & 0.07 & 1.25 & 0.03 & 9.15 & 0.11 & 0.64 & 0.02 & 12.0 & 12.0\\
  NGC 2355 & 12.16 & 0.09 & 1.22 & 0.01 & 10.01 & 0.09 & 0.61 & 0.01 & 8.0 & 10.0\\
  NGC 2360 & 10.92 & 0.05 & 1.21 & 0.01 & 8.76 & 0.06 & 0.62 & 0.01 & 12.0 & 13.0\\
  NGC 2423 & 10.70 & 0.05 & 1.19 & 0.0 & 8.59 & 0.06 & 0.60 & 0.01 & 11.0 & 11.0\\
  NGC 2420 & 12.32 & 0.02 & 1.16 & 0.01 & 10.25 & 0.08 & 0.61 & 0.01 & 10.0 & 12.0\\
  NGC 2447 & 9.90 & 0.06 & 1.08 & 0.01 & 8.01 & 0.07 & 0.53 & 0.02 & 12.0 & 14.0\\
  NGC 2477 & 12.09 & 0.03 & 1.48 & 0.01 & 9.48 & 0.02 & 0.70 & 0.01 & 77.0 & 78.0\\
  NGC 2527 & 9.23 & 0.06 & 1.12 & 0.01 & 7.21 & 0.06 & 0.58 & 0.01 & 4.0 & 4.0\\
  NGC 2539 & 10.63 & 0.08 & 1.14 & 0.01 & 8.61 & 0.1 & 0.57 & 0.01 & 9.0 & 9.0\\
  NGC 2660 & 13.93 & 0.03 & 1.60 & 0.01 & 11.05 & 0.04 & 0.77 & 0.01 & 31.0 & 31.0\\
  NGC 2682 & 10.16 & 0.07 & 1.26 & 0.03 & 7.93 & 0.08 & 0.65 & 0.01 & 10.0 & 9.0\\
  IC 2714 & 10.86 & 0.08 & 1.54 & 0.02 & 8.21 & 0.08 & 0.75 & 0.01 & 14.0 & 14.0\\
  NGC 3960 & 12.92 & 0.07 & 1.47 & 0.02 & 10.37 & 0.08 & 0.67 & 0.01 & 12.0 & 12.0\\
  NGC 4337 & 13.50 & 0.02 & 1.54 & 0.01 & 10.73 & 0.03 & 0.76 & 0.01 & 21.0 & 21.0\\
  NGC 4349 & 11.10 & 0.07 & 1.57 & 0.02 & 8.30 & 0.09 & 0.78 & 0.01 & 9.0 & 9.0\\
  Collinder 261 & 13.51 & 0.04 & 1.63 & 0.02 & 10.66 & 0.05 & 0.86 & 0.01 & 28.0 & 34.0\\
  NGC 5822 & 10.43 & 0.04 & 1.25 & 0.01 & 8.19 & 0.05 & 0.62 & 0.01 & 17.0 & 17.0\\
  NGC 6134 & 11.75 & 0.04 & 1.53 & 0.01 & 8.98 & 0.05 & 0.75 & 0.01 & 24.0 & 22.0\\
  NGC 6253 & 12.32 & 0.05 & 1.58 & 0.01 & 9.52 & 0.05 & 0.82 & 0.01 & 20.0 & 20.0\\
  IC 4651 & 10.57 & 0.03 & 1.31 & 0.01 & 8.15 & 0.06 & 0.67 & 0.01 & 10.0 & 12.0\\
  NGC 6583 & 13.45 & 0.04 & 1.82 & 0.02 & 10.24 & 0.06 & 0.86 & 0.01 & 20.0 & 20.0\\
  IC 4756 & 9.05 & 0.08 & 1.31 & 0.02 & 6.72 & 0.09 & 0.64 & 0.01 & 10.0 & 10.0\\
  NGC 6705 & 11.24 & 0.04 & 1.68 & 0.01 & 8.27 & 0.06 & 0.84 & 0.01 & 27.0 & 26.0\\
  Ruprecht 147 & 8.10 & 0.04 & 1.31 & 0.02 & 5.74 & 0.03 & 0.70 & 0.02 & 5.0 & 5.0\\
  NGC 6811 & 10.99 & 0.03 & 1.12 & 0.01 & 9.00 & 0.03 & 0.56 & 0.01 & 6.0 & 6.0\\
  NGC 6819 & 12.71 & 0.03 & 1.38 & 0.01 & 10.26 & 0.03 & 0.70 & 0.01 & 39.0 & 39.0\\
  NGC 7789 & 12.64 & 0.02 & 1.47 & 0.01 & 10.03 & 0.02 & 0.72 & 0.01 & 105.0 & 104.0\\
\hline\end{tabular}
   \label{Tab:ocs_mags}
\end{table*}

\begin{table*}
\caption{Morphological age indices and their uncertainties calculated for our OC sample.}
   \label{Tab:ocs_deltas}
\begin{tabular}{|l|r|r|r|r|r|r|r|r|}
\hline
  \multicolumn{1}{|c|}{$Oc$} &
  \multicolumn{1}{c|}{$\Delta G$} &
  \multicolumn{1}{c|}{$\sigma_{\Delta G}$} &
  \multicolumn{1}{c|}{$\Delta BR$} &
  \multicolumn{1}{c|}{$\sigma_{\Delta BR}$} &
  \multicolumn{1}{c|}{$\Delta K$} &
  \multicolumn{1}{c|}{$\sigma_{\Delta K}$} &
  \multicolumn{1}{c|}{$\Delta JK$} &
  \multicolumn{1}{c|}{$\sigma_{\Delta JK}$} \\
\hline
  NGC 188 & 3.21 & 0.15 & 0.53 & 0.02 & 4.55 & 0.24 & 0.44 & 0.05\\
  NGC 752 & 1.91 & 0.27 & 0.65 & 0.01 & 2.76 & 0.24 & 0.42 & 0.03\\
  NGC 1245 & 0.87 & 0.26 & 0.76 & 0.01 & 2.01 & 0.2 & 0.46 & 0.04\\
  NGC 1817 & 1.13 & 0.15 & 0.81 & 0.01 & 2.85 & 0.52 & 0.48 & 0.03\\
  NGC 2099 & 0.99 & 0.30 & 1.03 & 0.01 & 3.36 & 0.21 & 0.58 & 0.04\\
  Trumpler 5 & 2.88 & 0.14 & 0.68 & 0.01 & - & - & - & - \\
  Collinder 110 & 2.12 & 0.23 & 0.71 & 0.02 & 3.44 & 0.27 & 0.48 & 0.04\\
  NGC 2354 & 1.95 & 0.16 & 0.68 & 0.03 & 3.11 & 0.23 & 0.46 & 0.06\\
  NGC 2355 & 1.33 & 0.23 & 0.82 & 0.03 & 2.81 & 0.45 & 0.48 & 0.05\\
  NGC 2360 & 1.38 & 0.40 & 0.77 & 0.01 & 2.87 & 0.25 & 0.49 & 0.04\\
  NGC 2423 & 1.00 & 0.23 & 0.76 & 0.01 & 2.53 & 0.29 & 0.46 & 0.02\\
  NGC 2420 & 2.42 & 0.05 & 0.60 & 0.01 & 3.49 & 0.08 & 0.41 & 0.03\\
  NGC 2447 & 0.96 & 0.27 & 1.01 & 0.02 & 2.42 & 0.22 & 0.55 & 0.06\\
  NGC 2477 & 0.70 & 0.13 & 0.85 & 0.01 & 2.37 & 0.05 & 0.55 & 0.05\\
  NGC 2527 & 0.65 & 0.13 & 0.90 & 0.01 & 2.54 & 0.49 & 0.56 & 0.02\\
  NGC 2539 & 0.96 & 0.22 & 0.97 & 0.02 & 2.51 & 0.37 & 0.54 & 0.02\\
  NGC 2660 & 1.03 & 0.19 & 0.70 & 0.02 & 2.85 & 0.25 & 0.51 & 0.05\\
  NGC 2682 & 3.05 & 0.2 & 0.55 & 0.03 & 4.12 & 0.22 & 0.39 & 0.05\\
  IC 2714 & 1.60 & 0.36 & 1.07 & 0.02 & 3.85 & 0.23 & 0.65 & 0.04\\
  NGC 3960 & 0.86 & 0.41 & 0.84 & 0.03 & 2.77 & 0.38 & 0.53 & 0.04\\
  NGC 4337 & 1.51 & 0.22 & 0.61 & 0.02 & 2.60 & 0.15 & 0.48 & 0.04\\
  NGC 4349 & 1.53 & 0.34 & 1.10 & 0.02 & 4.23 & 0.30 & 0.68 & 0.05\\
  Collinder 261 & 3.19 & 0.17 & 0.54 & 0.02 & - & - &-  & -\\
  NGC 5822 & 0.57 & 0.38 & 0.83 & 0.02 & 2.48 & 0.43 & 0.49 & 0.04\\
  NGC 6134 & 1.49 & 0.16 & 0.66 & 0.01 & 2.83 & 0.07 & 0.48 & 0.03\\
  NGC 6253 & 2.82 & 0.17 & 0.53 & 0.01 & 4.05 & 0.23 & 0.48 & 0.03\\
  IC 4651 & 2.11 & 0.19 & 0.62 & 0.01 & 3.62 & 0.16 & 0.45 & 0.05\\
  NGC 6583 & 1.31 & 0.26 & 0.82 & 0.02 & 2.78 & 0.38 & 0.55 & 0.07\\
  IC 4756 & 0.59 & 0.31 & 0.84 & 0.02 & 2.77 & 0.31 & 0.49 & 0.03\\
  NGC 6705 & 1.56 & 0.41 & 1.22 & 0.02 & 4.4 & 0.49 & 0.73 & 0.04\\
  Ruprecht 147 & 2.81 & 0.24 & 0.56 & 0.02 & 3.94 & 0.29 & 0.46 & 0.05\\
  NGC 6811 & 0.90 & 0.26 & 0.80 & 0.01 & 2.51 & 0.15 & 0.49 & 0.02\\
  NGC 6819 & 2.59 & 0.13 & 0.62 & 0.01 & - & - & - & -\\
  NGC 7789 & 2.13 & 0.20 & 0.70 & 0.02 & 3.23 & 0.24 & 0.45 & 0.04\\
\hline\end{tabular}
\end{table*}

\subsection{The NGC2354 age}
\label{sect:ngc_2354}

\begin{figure}
\includegraphics[width=0.48\linewidth]{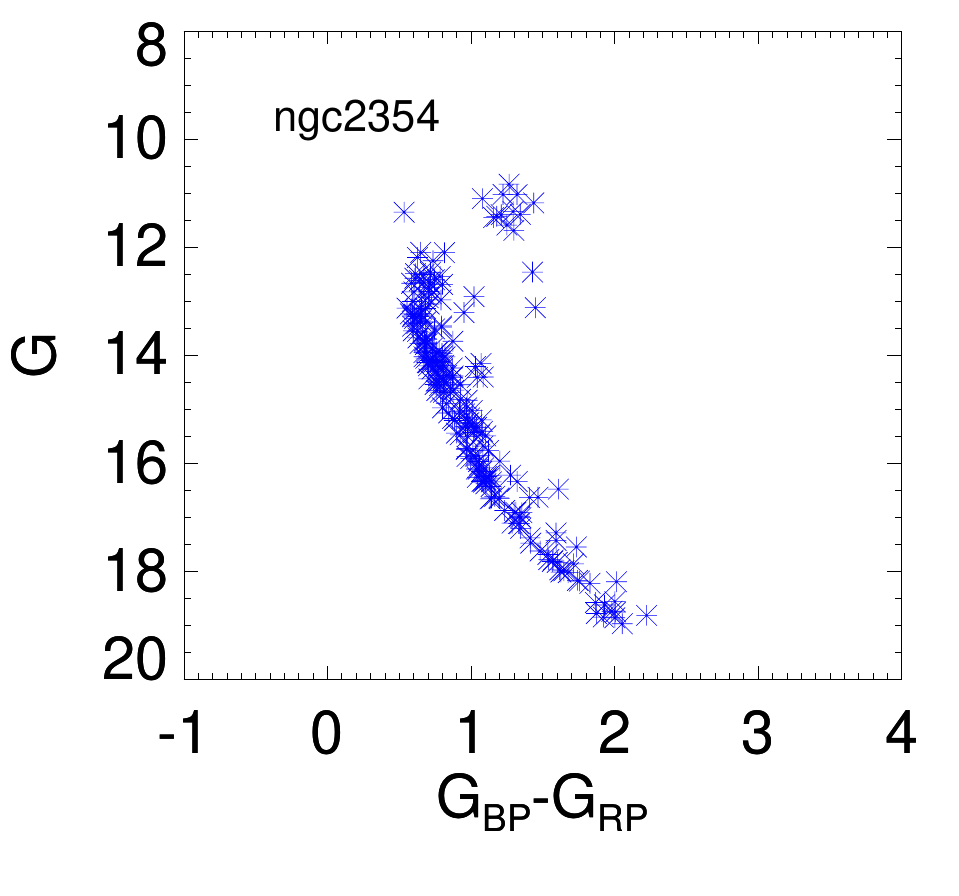}
\includegraphics[width=0.50\linewidth]{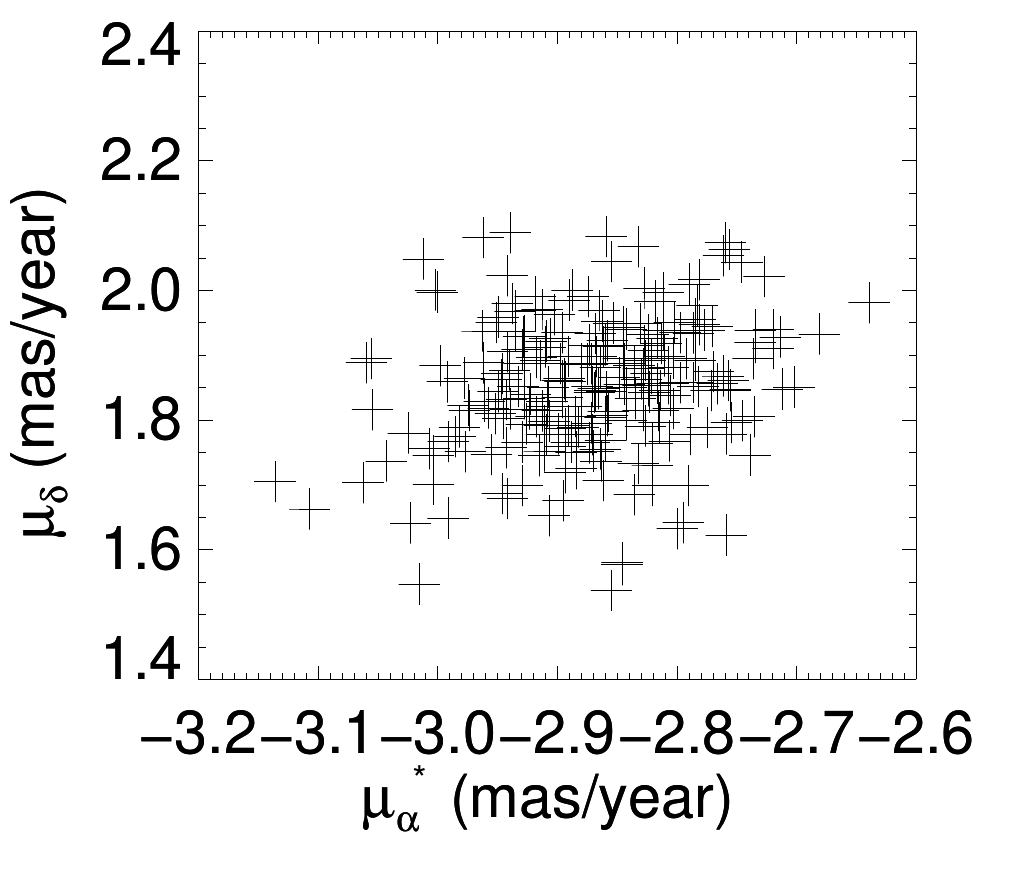}
\caption{Results of the membership assignment procedure for NGC 2354. Left: CMD of the most probable members. Right: VPD of the most probable members.}

\label{fig:NGC2354_parametros}
\end{figure}

NGC2354 is an OC located in the constellation of Canis Major which, according to our analysis from \textit{Gaia} data, has an apparent diameter of 25 arcmin and 234 members. Our analysis proved to be efficient in obtaining reliable members of this cluster, as its CMD clearly presents an evolutionary sequence without many \textit{outliers} (see Fig. \ref{fig:NGC2354_parametros}) and the dispersion values of the members in astrometric space were compatible with those of the other clusters with similar distances (see Table \ref{Tab:members}).

\begin{figure}
\includegraphics[width=0.95\linewidth]{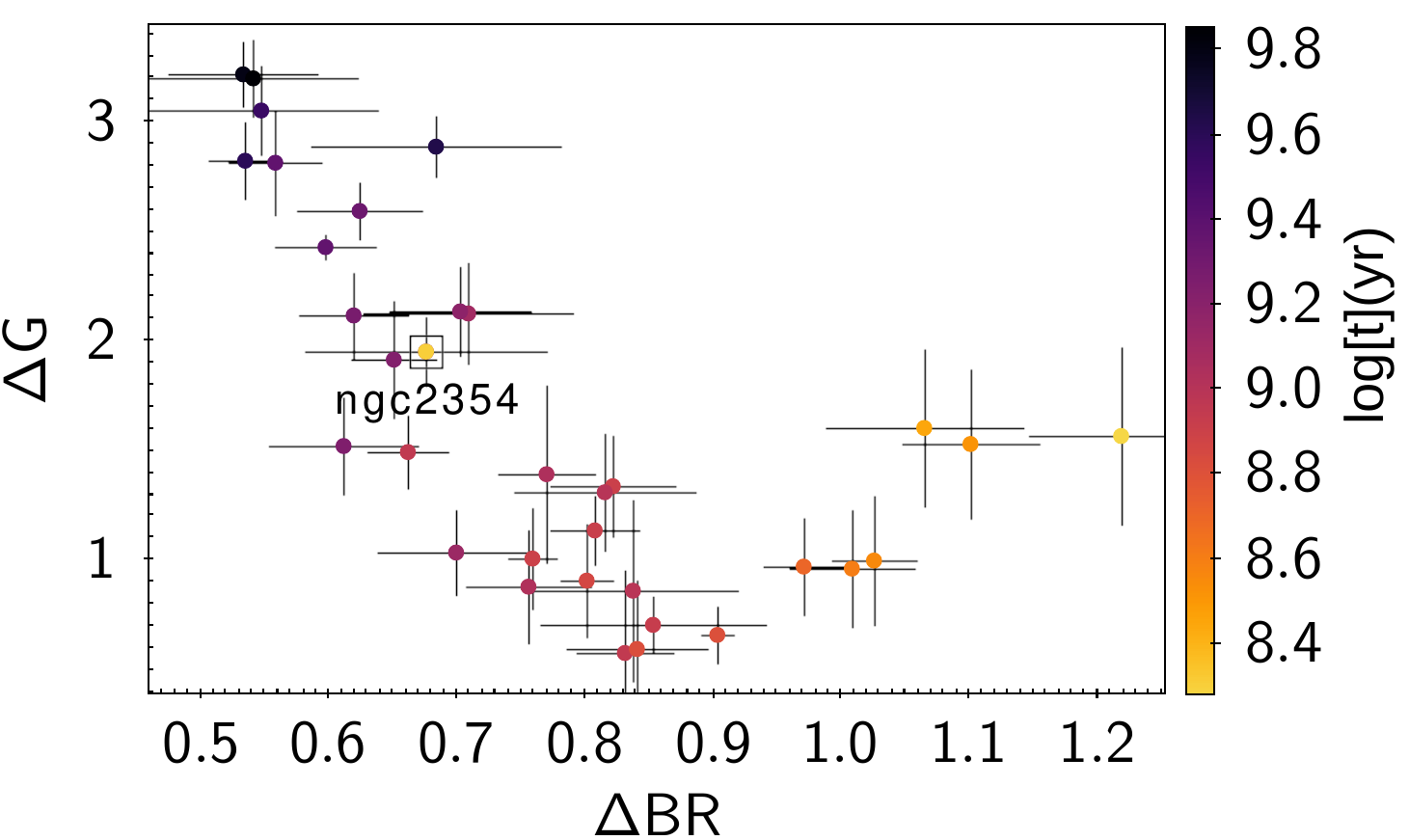}
\includegraphics[width=0.95\linewidth]{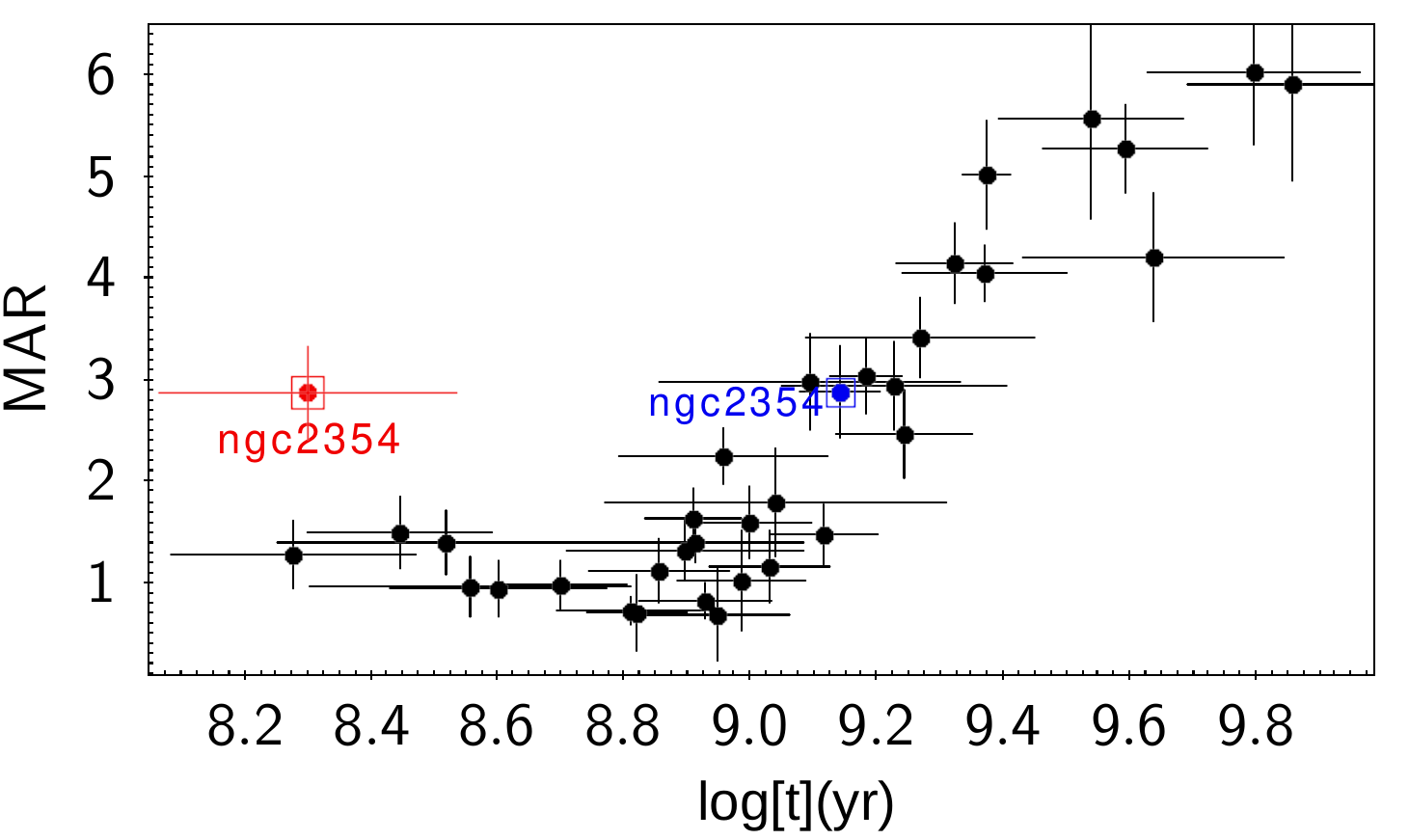}

\caption{Top: Relation $\Delta BR$ versus $\Delta G$. NGC 2354 highlight by a black square. The colour bar represents the age values. Bottom: Relation log[t] versus MAR. NGC 2354 is plotted with two different age values and represented by two different colours: red (age value from N2016) and blue (age from average \textit{Gaia} values).}

\label{fig:NGC2354_clump}
\end{figure}

However, we note that the age value used in this work (established in N2016) proved to be underestimated in relation to recent age values established in the literature, based on determinations with data from \textit{Gaia}. We also noticed discrepancies in the properties presented by this cluster when compared to others. This object presented very discrepant values on the plane $\Delta G$ versus $\Delta BR$ and in the age relation of the MAR index, when compared to objects with the same age, as can be seen in Figure \ref{fig:NGC2354_clump}. This reinforces the importance of such indices in charaterizing OCs morphologically.

We carried out a review of the astrophysical parameters of this object in the literature and noticed that in studies where the authors did not use precise proper motion data, low age values were assigned to this object (see Table \ref{Tab:NGC2354_parametros}). We note that prior to the availability of data from \textit{Gaia}, the determinations of age, distance and colour excess, on average, were: $\overline{log[t]}=8.35$ and $\sigma_{\log[t({\rm yr})]}=0.27$, $\overline{d}=3027$ pc and $\sigma_{d}=915$ pc and $\overline{E(B-V)}=0.31$ and $\sigma_{E(B-V )}=0.17$. It is evident that the age value is compatible with the average established in N2016 of $\overline{\log[t({\rm yr})]}= 8.30 \pm 0.23$. With the exception of \cite{kharchenko2013}, the other authors who characterized this cluster before the \textit{Gaia} era, used UBV photometry and interpreted its CMD as a much younger and more distant object.

After the availability of \textit{Gaia}, the same determinations led to the following average values: $\overline{\log[t({\rm yr})]}=9.14$ and $\sigma_{\log[t({\rm yr})]}=0.06$, $\overline{d}=1259$ pc and $\sigma_{d}=84$ pc and $\overline{E(B-V)}=0.16$ and $\sigma_{E(B-V)}=0.06$. A smaller fluctuation of the values and a 
better convergence of the distance estimate ($d=1279^{+188}_{-145}$ pc) was obtained through the average parallax of the cluster members in \cite{cjv18}. For this particular object, we replaced its age from N2016 by the averaged \textit{Gaia} determinations and assumed the corresponding dispersion as its uncertainty: $\overline{\log[t({\rm yr})]}=9.14 \pm 0.06$.

  \begin{table}
  \caption[Astrophysical parameters of the NGC2354 cluster from the literature]{Astrophysical parameters of the NGC2354 cluster from the literature: \citeauthor{lp19} (\citeyear{lp19}, LP2019), \citeauthor{cjv18} (\citeyear{cjv18}, CG2018), \citeauthor{2017AstBu..72..257L} (\citeyear{2017AstBu..72..257L}, L2017), \citeauthor{2005A&A...438.1163K} (\citeyear{2005A&A...438.1163K}, K2005), \citeauthor{1960ZA.....49..214D} (\citeyear{1960ZA.....49..214D}, Du1960), \citeauthor{1994A&AS..104..379B} (\citeyear{1994A&AS..104..379B}, Ba1994).}
  \small
\begin{tabular}{|l|r|r|r|r|r|}
\hline
   \multicolumn{1}{|c|}{$d(pc)$} &
   \multicolumn{1}{c|}{$log[t](yr)$} &
  \multicolumn{1}{c|}{$E(B-V)$} &
   \multicolumn{1}{c|}{$ref$} \\
\hline
  1850 & 8.84 & 0.14 & Du1960 \\
  \hline
   1837 & 8.26 & 0.14 & Ba1994 \\
  \hline
  4085 & 8.13 & 0.31 & webda \\
  \hline
  3794 & 8.10 & 0.29 & K2005 \\
  \hline
   2865.0 & 8.61 & 0.666 & K2013 \\
  \hline
   $3732 \pm1100 $& $8,174 \pm 0.242$ & $0.286 \pm 0.057 $ & L2017 \\
  \hline
   $1279^{+188}_{-145}$ & - & - & CG2018 \\
  \hline
   1132 & $9.07 \pm 0.02$ & 0.26 & LP2019 \\
  \hline
   1370.0 & 9.15 & 0.11 & CG2020 \\
  \hline
    $1258 \pm 42$ & $9.21 \pm 0.03$ & $0.17 \pm 0.02$ & D2021 \\
  \hline
\hline\end{tabular}

   \label{Tab:NGC2354_parametros}
\end{table}

% Don't change these lines
\bsp	% typesetting comment
\label{lastpage}
\end{document}